\RequirePackage{amsthm}
\documentclass[referee,pdflatex]{sn-jnl}

\usepackage{graphicx}%
\usepackage{multirow}%
\usepackage{amsmath,amssymb,amsfonts}%
\usepackage{amsthm}%
\usepackage{mathrsfs}%
\usepackage[title]{appendix}%
\usepackage{xcolor}%
\usepackage{textcomp}%
\usepackage{manyfoot}%
\usepackage{booktabs}%
\usepackage{algorithm}%
\usepackage{algorithmicx}%
\usepackage{algpseudocode}%
\usepackage{listings}%
\usepackage{geometry}
\usepackage{multicol}
\usepackage{multirow}
\usepackage{anyfontsize}

\usepackage{subcaption}
\usepackage{booktabs}\let\cline\cmidrule
\theoremstyle{thmstyleone}%
\theoremstyle{thmstyletwo}%

\theoremstyle{thmstylethree}%

\begin{document}

\title[Article Title]{A magnetised Galactic halo from inner Galaxy outflows}


\author*[1]{\fnm{He-Shou} \sur{Zhang}}\email{heshouzhang.astrophy@gmail.com; heshou.zhang@inaf.it}

\author*[1,2]{\fnm{Gabriele} \sur{Ponti}}\email{gabriele.ponti@inaf.it}

\author*[3]{\fnm{Ettore} \sur{Carretti}}\email{carretti@ira.inaf.it}

\author*[4,5]{\fnm{Ruo-Yu} \sur{Liu}}\email{ryliu@nju.edu.cn}

\author*[6]{\fnm{Mark R.} \sur{Morris}}\email{morris@astro.ucla.edu}

\author[7]{\fnm{Marijke} \sur{Haverkorn}}

\author[1,2]{\fnm{Nicola} \sur{Locatelli}}

\author[2]{\fnm{Xueying} \sur{Zheng}}

\author[8,9,10]{\fnm{Felix} \sur{Aharonian}}

\author[4,11]{\fnm{Hai-Ming} \sur{Zhang}}

\author[2]{\fnm{Yi} \sur{Zhang}}

\author[1,12]{\fnm{Giovanni} \sur{Stel}}

\author[2]{\fnm{Andrew} \sur{Strong}}

\author[2]{\fnm{Michael C. H.} \sur{Yeung}}

\author[2]{\fnm{Andrea} \sur{Merloni}}

\affil*[1]{\orgname{INAF - Osservatorio Astronomico di Brera}, \orgaddress{\street{Via E. Bianchi 46}, \city{Merate}, \postcode{23807}, \state{Lecco}, \country{Italy}}}

\affil[2]{\orgname{Max-Planck-Institut f\"ur Extraterrestrische Physik (MPE)}, \orgaddress{\street{Giessenbachstrasse 1}, \city{Garching bei M\"unchen}, \postcode{85748}, \country{Germany}}}

\affil[3]{\orgname{INAF - Istituto di Radioastronomia}, \orgaddress{\street{Via Gobetti 101}, \city{Bologna}, \postcode{40129}, \country{Italy}}}

\affil[4]{\orgdiv{School of Astronomy and Space Science}, \orgname{Nanjing University}, \orgaddress{\street{Xianlin Avenue 163}, \city{Nanjing}, \postcode{210023}, \country{China}}}

\affil[5]{\orgdiv{Key Laboratory of Modern Astronomy and Astrophysics (Nanjing University)}, \orgname{Ministry of Education}, \orgaddress{\city{Nanjing}, \postcode{210023}, \country{China}}}

\affil[6]{\orgdiv{Department of Physics and Astronomy}, \orgname{ University of California}, \orgaddress{\city{Los Angeles}, \postcode{90095-1547}, \state{CA}, \country{USA}}}

\affil[7]{\orgdiv{Department of Astrophysics/IMAPP}, \orgname{Radboud University Nijmegen}, \orgaddress{\street{P.O. Box 9010}, \city{Nijmegen}, \postcode{6500}, \state{GL}, \country{The Netherlands}}}

\affil[8]{\orgname{Dublin Institute for Advanced Studies}, \orgaddress{\street{31 Fitzwilliam Place}, \city{Dublin}, \postcode{2}, \country{Ireland}}}

\affil[9]{\orgname{Max-Planck-Institut f\"ur Kernphysik}, \orgaddress{\street{P.O. Box 103980}, \city{Heidelberg}, \postcode{69029}, \country{Germany}}}

\affil[10]{\orgname{Yerevan State University}, \orgaddress{\street{1 Alek Manukyan St}, \city{Yerevan}, \postcode{0025}, \country{Armenia}}}

\affil[11]{\orgdiv{Guangxi Key Laboratory for Relativistic Astrophysics, School of Physical Science and Technology}, \orgname{Guangxi University}, \orgaddress{\city{Nanning}, \postcode{530004}, \country{China}}}

\affil[12]{\orgdiv{DiSAT}, \orgname{Universit\`a degli Studi dell’Insubria}, \orgaddress{\street{Via Valleggio 11}, \city{Como}, \postcode{22100}, \country{Italy}}}


\abstract{Magnetic halos of galaxies are crucial {for understanding} galaxy evolution, galactic-scale outflows, and feedback from star formation activity. Identifying the magnetised halo of the Milky Way is challenging because of the potential contamination from foreground emission arising in local spiral arms. Additionally, it is unclear how our magnetic halo is influenced by recently revealed large-scale structures such as the X-ray emitting eROSITA Bubbles. Here we report the identification of several kpc-scale magnetised structures based on their polarized radio emission and their gamma-ray counterparts, which can be interpreted as the radiation of relativistic electrons in the Galactic magnetic halo. These non-thermal structures extend far above and below the Galactic plane and are spatially coincident with the thermal X-ray emission from the eROSITA Bubbles. The morphological consistency of these structures suggests a common origin, which can be sustained by Galactic outflows driven by active star-forming regions located {in the Galactic Disc} at $3-5$~kpc from the Galactic Centre. These results reveal how X-ray-emitting and magnetised halos of spiral galaxies can be related to intense star formation activities and suggest that the X-shaped coherent magnetic structures observed in their halos can stem from galaxy outflows.\\
\textbf{Comments}: Submitted \textbf{2 Mar. 2024}; Accepted \textbf{12 Aug. 2024}.
}

\maketitle

\textbf{Main Text}

X-ray-emitting galactic halos have been discovered in star-forming galaxies\cite{Strickland04xHalo,StricklandHeckman07xHalo}{}. Several of them are accompanied by large-scale coherent magnetic structures revealed by radio data\cite{Krause20ChangesXXII,Krause19halo}{}. These large-scale magnetic fields are observed off the galaxy mid-planes, indicating that these galaxies harbour magnetised halos. However, the relationship between the X-ray-emitting and magnetised galactic halos is unclear, and similarly for their nature and origins. The recent discovery of the X-ray emitting large structures {within} the Milky Way halo, the eROSITA Bubbles\cite{Predehl20}{}, provides important physical insights that enhance our understanding of galactic halos.

\begin{figure}
\begin{center}
\includegraphics[width=0.85\textwidth]{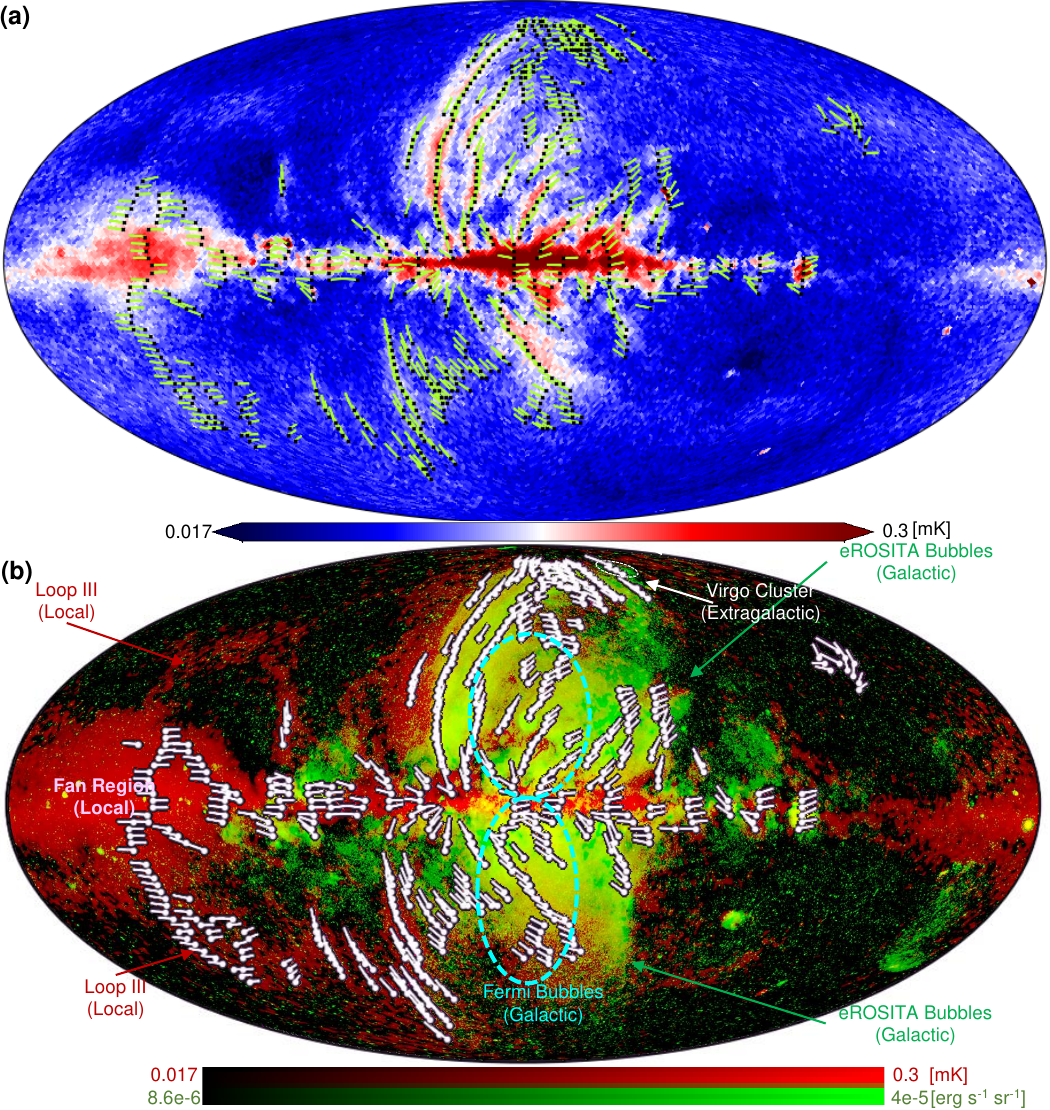}
\end{center}
\caption{{\bf Figure 1 $\mid$ Polarized radio counterpart of the eROSITA Bubbles.} (a) Background image: polarized synchrotron intensity (PI$_{syn}$) from WMAP data  at 22.8~GHz; Black dots: local maxima  of the polarized synchrotron intensity estimated from constant latitude profiles; Green bars: magnetic field direction inferred from polarized synchrotron emission. Several coherent magnetised ridges rise from the Galactic disc and are progressively bent toward the west with increasing latitude. (b)  Magnetised  ridges (PI$_{syn,max}$,  white dots) and polarized synchrotron emission at 22.8 GHz  (red) compared to the eROSITA X-ray emission at 0.6-1.0~keV (green). The large structure close to the  east end of the map  with no X-ray counterpart is the Fan Region (local emission, see Extended Data Figure~4). The magnetised  ridges  in the inner Galaxy ($\mid l\mid\lesssim60^\circ$) are spatially correlated with  Galactic  structures: the ridges associated spatially with the Fermi Bubbles appear to emanate from the CMZ\cite{Carretti13Nat,Lacki14,Crocker15FermiMultiCMZ}{}, while the  ridges in the {\it outer region} appear to originate in the disc a few  kpc  from  the Galactic Centre. The  ridges show an approximate north-south symmetry.
The polarized intensity is enhanced at the edges of the eROSITA Bubbles. The eastern edges of these Bubbles are parallel to their magnetic field. The western roots of the eROSITA Bubbles' edges are limb-brightened in polarized synchrotron emission.
}
\label{fig:outflows}
\end{figure}

Figure \ref{fig:outflows} compares the eROSITA all-sky emission at 0.6-1.0~keV with the magnetic field determined from the polarized synchrotron emission at 22.8~GHz from WMAP\cite{DataWMAP1}{}, for which Faraday rotation effects are marginal. Several magnetic structures revealed through their polarized emission and coherent field line direction, here denoted as magnetic ridges, appear in the inner Galaxy, emerging from the Galactic plane\cite{Vidal15}{} and stretching for more than $\pm20^\circ$. The polarized intensity is enhanced at the edges of the eROSITA Bubbles (see Extended Data Figure~1). The magnetic field directions are parallel to the Bubbles' edges in the east. The magnetic ridges show a general tilt westwards, starting from the disc and rising to high latitudes. This implies a potential connection between magnetic ridges and the eROSITA Bubbles, {as the eROSITA Bubbles also show a westward tilt}.

\begin{figure}
\begin{center}
\includegraphics[width=1\textwidth]{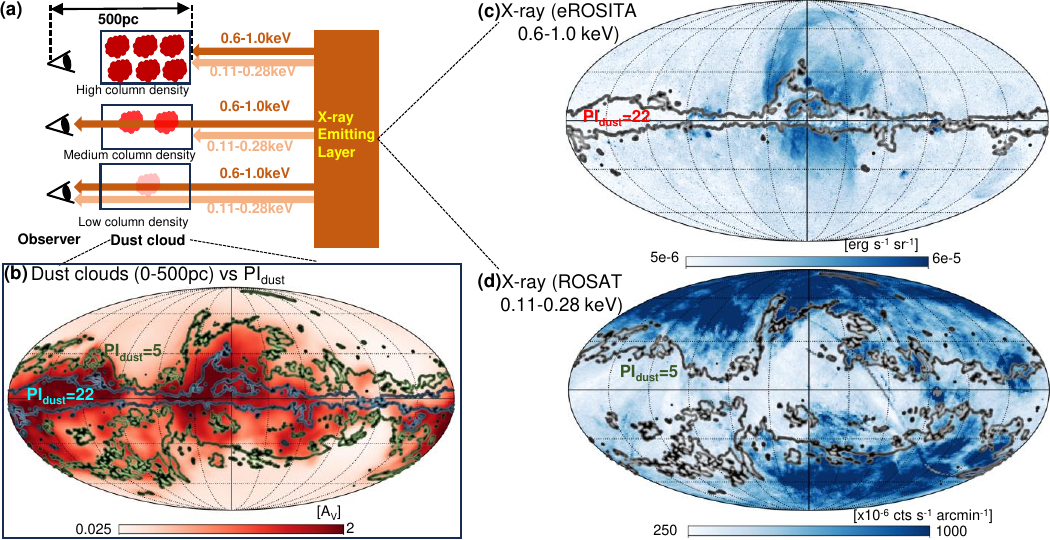}
\end{center}
\caption{{\bf Figure 2 $\mid$ distance measurements to the eROSITA Bubbles.} The schematics in panel (a) illustrate how X-ray absorption depends on the column density of the foreground medium. Panel b: The extinction map for the layer at 0-500~pc, as obtained from the 3D dust distribution by the Gaia-2MASS-Apogee dataset\cite{Lallement18} (background image), matches the polarized dust intensity map  at 353~GHz from Planck\cite{DataPLANCKCommander} (Contours are two levels of polarized intensity, $\mathrm{PI_{dust}}$). An  anti-correlation is obvious between  high dust column density (panel c, contours,  $\mathrm{PI_{dust}}=22 \mu$K) with high-energy X-ray emission (0.6-1.0~keV), and between low  dust  column density (panel d, contour, $\mathrm{PI_{dust}}=5\mu$K) with low energy X-ray emissions (0.11-0.28~keV).
}
\label{fig:depoldust}
\end{figure}

It is difficult to study the halos of the Milky Way because radiation coming from the Galactic halo is mixed with {foreground emission} arising in local spiral arms\cite{Carretti13Nat,Wolleben21}{}. Hence, a key issue for these extended structures is whether they are local objects within the Local hot Bubble (LB)\cite{Liu17LHB}{}, or distant Galactic structures. Thus far, they have mostly been modeled as shells of old supernova remnants in the LB\cite{Berkhuijsen71SNR,Vidal15}{}. Figure~\ref{fig:depoldust} reveals an anti-correlation between the X-ray maps of eROSITA\cite{Predehl20}{} (0.6--1.0~keV) at mid/low Galactic latitudes and the dust column density based on the dust distribution within 500~pc from the Sun by Ref.\cite{Lallement18}{}. Therefore, the local dust within 500~pc is responsible for the X-ray absorption, implying that the X-ray emitting eROSITA Bubbles are not local and the bulk of the emitting structures must originate from a distance beyond {the 500-pc line-of-sight extent of our local arm}. This is consistent with the comparison between dust emission and ROSAT data from Ref.\cite{Sofue15mn}.

\begin{figure}
\begin{center}
\includegraphics[width=0.7\textwidth]{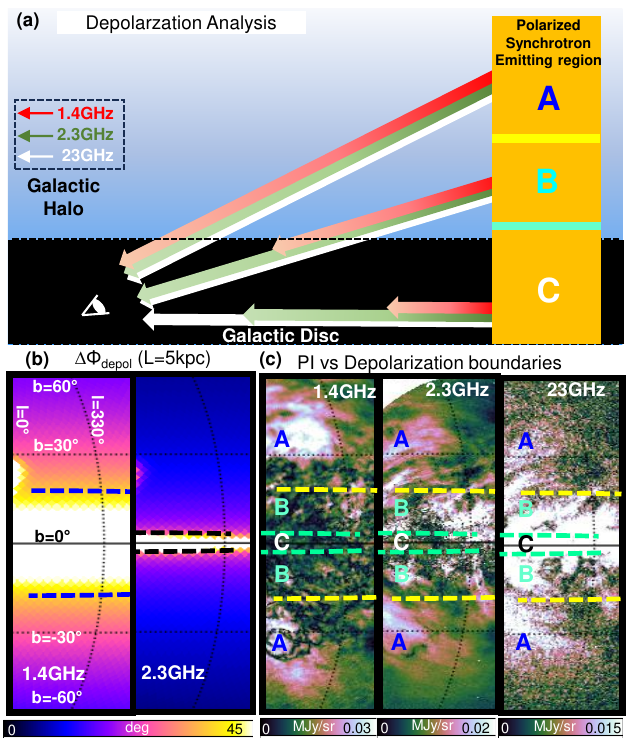}
\end{center}
\caption{{\bf Figure 3 $\mid$ distance measurements for magnetised ridges by Faraday rotation depolarization.} The schematic in panel (a) shows that: For the polarized synchrotron emission, assuming the Galactic  turbulent  magnetic field out to 5-kpc,  Faraday rotation depolarization is negligible at 22.8~GHz (white, WMAP map). The depolarization occurs up to a few degrees of Galactic latitude at 2.3~GHz (green, S-PASS), and further extends up to a Galactic latitude of $|b|\approx 20^\circ$  at 1.4~GHz (red, GMIMS) (see Extended Data Figure~3). {The structures A-C at different heights hence show different depolarization.} The depolarization screen at 5~kpc is shown in (b) for the two lowest  frequencies as the latitude at which the depolarization is 0.5 and $\Delta\Phi\simeq40^\circ$. The observed  polarized intensity maps  are compared in (c). The observations are consistent with an origin of  the polarized synchrotron emission at mid/high Galactic latitudes from  beyond 5~kpc.
}
\label{fig:depolsyn}
\end{figure}

Extended Data Figure~2 shows that these magnetic ridges corresponding to the X-ray emitting outer halo either have no dust counterparts or their directions are not correlated with {the magnetic field inferred from} the polarized dust emission. Hence the magnetic ridges are also non-local structures. We estimate our distance from the polarized synchrotron emission of the magnetised ridges in Figure~\ref{fig:depolsyn} thanks to the Faraday-rotation depolarization of the foreground turbulent magnetised medium\cite{Burn66,Sokoloff98}{}. As demonstrated in Figure~\ref{fig:depolsyn}(a), the depolarization is largest at lower frequencies and decreases at higher latitudes because of a decline with latitude of the magnetic field strength and electron density in the foreground medium, {as well as a shorter column through the foreground medium}. Using the depolarization expected from the magneto-ionic medium out to different depths, the comparison between Figures~\ref{fig:depolsyn}(b) and (c) demonstrates that the observed depolarization is consistent with that produced by the medium out to distances of several~kpc (see Methods and Extended Data Figures~3--4 for details). The polarized magnetic ridges are thus {Galactic structures of several kiloparsec scales}. This indicates that the bulk of the emission associated with these extended structures (including the North Polar Spur\cite{Lallement22}) is beyond several kpc from us, however, {this analysis} does not exclude a smaller contribution from some local features.

\begin{figure}
\begin{center}
\includegraphics[width=0.95\textwidth]{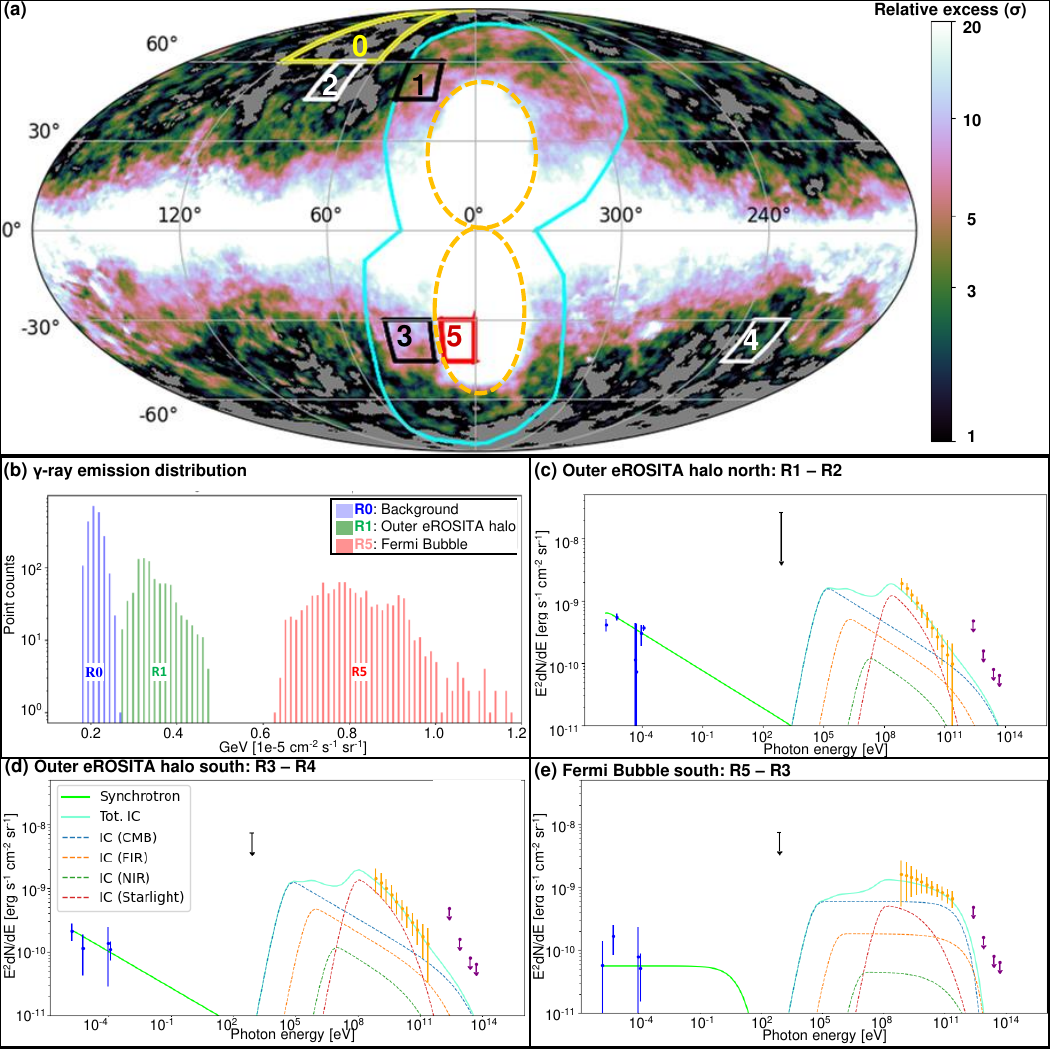}
\end{center}
\caption{{\bf Figure 4 $\mid$  Radio--$\gamma$-ray analysis and SED.} (a) Intensity map (defined in Methods) of the Fermi all-sky diffuse map at $E_\gamma\gtrsim100$~GeV expressed in units of the noise level from the reference background (patch R0 in the northeastern high galactic latitude sector, see \S~2.2 in the Methods section for details). Patches R1-R5 are used for the SED study. {The two lines are the edges of the eROSITA (solid) and Fermi (dashed) Bubbles.} (b) the distribution of the pixel counts for the gamma-ray fluxes in the background (R0), the {\it outer region} of the eROSITA Bubbles (R1), and the Fermi Bubble (R5). The three distributions are well separated. (c-e) SED best-fit of leptonic radiation in: (c) the {\it outer region} north (R1-R2) with electron energy index $-3.40\pm0.06$ and magnetic field $1.97\pm0.20$~$\mu$G; (d) south (R3-R4) with electron index $-3.38\pm0.10$, magnetic field $1.40\pm0.20$~$\mu$G; (e) southeastern cap of the Fermi Bubble (R5-R3) with electron index $-3.00^{+0.30}_{-0.13}$, magnetic field $1.0^{+0.5}_{-0.4}$~$\mu$G. {The error-bars reported here for the gamma-ray data are based on the statistical uncertainties (see \S~3.1 in the Methods section for details). The reference frequencies are listed in the Supplementary Table 2. The error-bars reported for the radio fluxes are calculated based on the flux density calibration accuracy and beam sensitivity of the corresponding surveys, as defined in the Supplementary.} The fitting results for the electron distribution indices for the {\it outer regions} are consistent, but the Fermi Bubble has a harder electron energy distribution index. The detailed results are summarized in the Supplementary.
}
\label{fig:sed}
\end{figure}

The fact that synchrotron radiation is enhanced at the edges of the X-ray emitting {halo structures} indicates the presence of relativistic electrons. Those electrons can also give rise to gamma-ray emission via inverse Compton (IC) scattering of photons from interstellar radiation and the Cosmic Microwave Background (CMB). We investigate the potential gamma-ray counterparts of the eROSITA Bubbles using the Fermi-LAT data of the diffuse all-sky gamma-ray intensity from Ref.\cite{Platz22}. The relative excess of the gamma-ray intensity above the background is presented in Figure~\ref{fig:sed}(a) and Extended Data Figure~5. The horizontal cuts at north and south high Galactic latitudes above and below the Fermi Bubbles are shown in Extended Data Figure~6. We observe that extended structures with $\gamma-$ray enhancements show agreement with a large part of the edges of the X-ray emitting outer halo. The consistency with the eROSITA Bubble in the north can be observed in all three gamma-ray maps ($E_\gamma\gtrsim1$, $10$, and $100$~GeV) at mid/high latitude ($b>30^\circ$). The consistency in the south is observed for $E_\gamma\gtrsim100$~GeV and for $b\gtrsim-60^\circ$ (Extended Data Figure~6), while no clear structure is observed at the cap of the southern eROSITA Bubble.

The eROSITA Bubbles contain the Fermi Bubbles in projection, but it is extremely difficult to imagine a scenario in which the two Galaxy-scale features are centered at different distances. It is worth noting that the origin of the Fermi Bubbles and their potential low-latitude radio counterparts (the so-called ``radio haze'')\cite{Dobler10haze,Planck13haze}{} can either be the outflows from the star formation activity of the Central Molecular Zone (CMZ)\cite{Lacki14,Crocker15FermiMultiCMZ}{} or the past activity of the central Supermassive Black hole (SMBH) Sgr A$^\ast$\cite{Su10Fermi,Ackermann14FermiSpect}{}.
Investigating the origin of the Fermi Bubbles is beyond the scope of this work.

Our analysis in this work focuses on the non-thermal emission in the region outside the Fermi Bubbles but inside the edges of the eROSITA Bubbles (hereafter the {\it outer region}). The similarity in the morphology between radio and $\gamma$-ray bands implies a common origin of the emission in these bands and we study the spectral energy distribution (SED).
In Figure~\ref{fig:sed}, two patches within the {\it outer region} are selected for further investigation: patch R1 in the north ($20^\circ<l<40^\circ$, $45^\circ<b<60^\circ$) and patch R3 in the south ($20^\circ<l<40^\circ$, $-45^\circ<b<-30^\circ$). The average flux densities of the two patches outside the eROSITA Bubbles at the same Galactic latitudes (patches R2 and R4, respectively) are subtracted from the two selected patches within {\it the outer region} to remove fore/background emission from the Galactic halo. The radio data\cite{Remazeilles15Haslam408,DataDRAO1,DataQUIJOTE,WMAP14synchro,DataPLANCKCommander} from 0.408--30~GHz are used to characterize the synchrotron emission. The effective area of Fermi-LAT drops quickly with decreasing energy below 1~GeV, leading to poor statistics for diffuse emission in the chosen patches for $E_\gamma\lesssim0.3$~GeV\cite{Fermi21DataInstrument}{}. Therefore, we use only the data above 0.3~GeV in our study. We fit the spectrum with a single power-law in each individual band (radio or gamma-ray, see Extended Data Figure~7), and note that the gamma-ray flux density exhibits a softer spectrum compared to the radio flux in the {\it outer region} due to the Klein-Nishina effect.

Assuming the same electrons within a given patch are responsible for both the synchrotron emission and IC scattering, we fit the multi-wavelength SED in different patches of the eROSITA Bubbles to study the cosmic rays (CRs) and magnetic fields therein (Methods). The best-fit results of the SED fitting are presented in Figure~\ref{fig:sed}(c-e). The SED fitting results demonstrate a {\it north/south symmetry of the outer region of the eROSITA Bubbles}. The derived electron distributions $N\propto E^{\alpha}$ have shown consistent, very steep slopes, with $\alpha=-3.40\pm0.06$ in the northern patch R1 and $-3.38\pm0.10$ in the southern patch R3. The magnetic field directions are largely symmetric about the Galactic disc, and the average magnetic field strengths are $1.97\pm0.20$~$\mu$G in the north and $1.40\pm0.20$~$\mu$G in the south.
{We calculate the plasma-beta by $\beta\equiv p_{th}/p_{B}$, where the magnetic pressure is $p_B=B^2/8\pi$ and the thermal pressure is $p_{th}=n_e k_B T$.} We adopt a temperature of 0.3~keV\cite{kataoka13,Predehl20}, a halo electron density of $3\times10^{-3}$~cm$^{-3}$ calculated from Ref.\cite{Locatelli24neDensityMW}, and magnetic field strength obtained through our SED fitting. In patch R1: $\beta_{r1}\simeq9$, and in patch R3: $\beta_{r3}\simeq18$.

Plausible origins of the CRs responsible for the non-thermal radiation in the {\it outer region} of the eROSITA Bubbles include acceleration processes in the inner Fermi Bubbles or Galactic outflows from the disc. As a comparison, we perform an SED fitting for the southeastern edge of the Fermi Bubbles in Figure~\ref{fig:sed}(e) and find an electron energy index $-3.00^{+0.3}_{-0.13}$ for patch R5. This electron index is harder than that in the patch R3 of the {\it outer region} at the same latitude. We compare the synchrotron flux densities for the outer patch R3 and the inner patch R5 after foreground subtraction, for 0.408~GHz: $F_{\nu,R3}=0.0523\pm0.0078$~MJy/sr, $F_{\nu,R5}=0.0140\pm0.0099$~MJy/sr; for 1.4~GHz: $F_{\nu,R3}=0.0082\pm0.052$~MJy/sr, $F_{\nu,R5}=0.0118\pm0.0059$~MJy/sr; for 23~GHz: $F_{\nu,R3}=(5.9\pm4.6)\times10^{-4}$~MJy/sr, $F_{\nu,R5}=(3.4\pm6.7)\times10^{-4}$~MJy/sr; for 30~GHz: $F_{\nu,R3}=(3.3\pm1.3)\times10^{-4}$~MJy/sr, $F_{\nu,R5}=(1.5\pm1.4)\times10^{-4}$~MJy/sr.
The synchrotron flux densities in the outer region are equvalent or higher compared to the foreground-subtracted values inside the Fermi Bubbles at a similar Galactic height.
Therefore, diffusion from the Fermi Bubbles cannot be the primary process for injecting the relativistic electrons into the {\it outer region}. Additionally, the magnetic field is parallel to the shell of the Bubbles and the diffusion from the Fermi Bubbles into the outer halo requires cross-field transport of relativistic electrons, which is very inefficient.

\begin{figure}
\begin{center}
\includegraphics[width=0.9\textwidth]{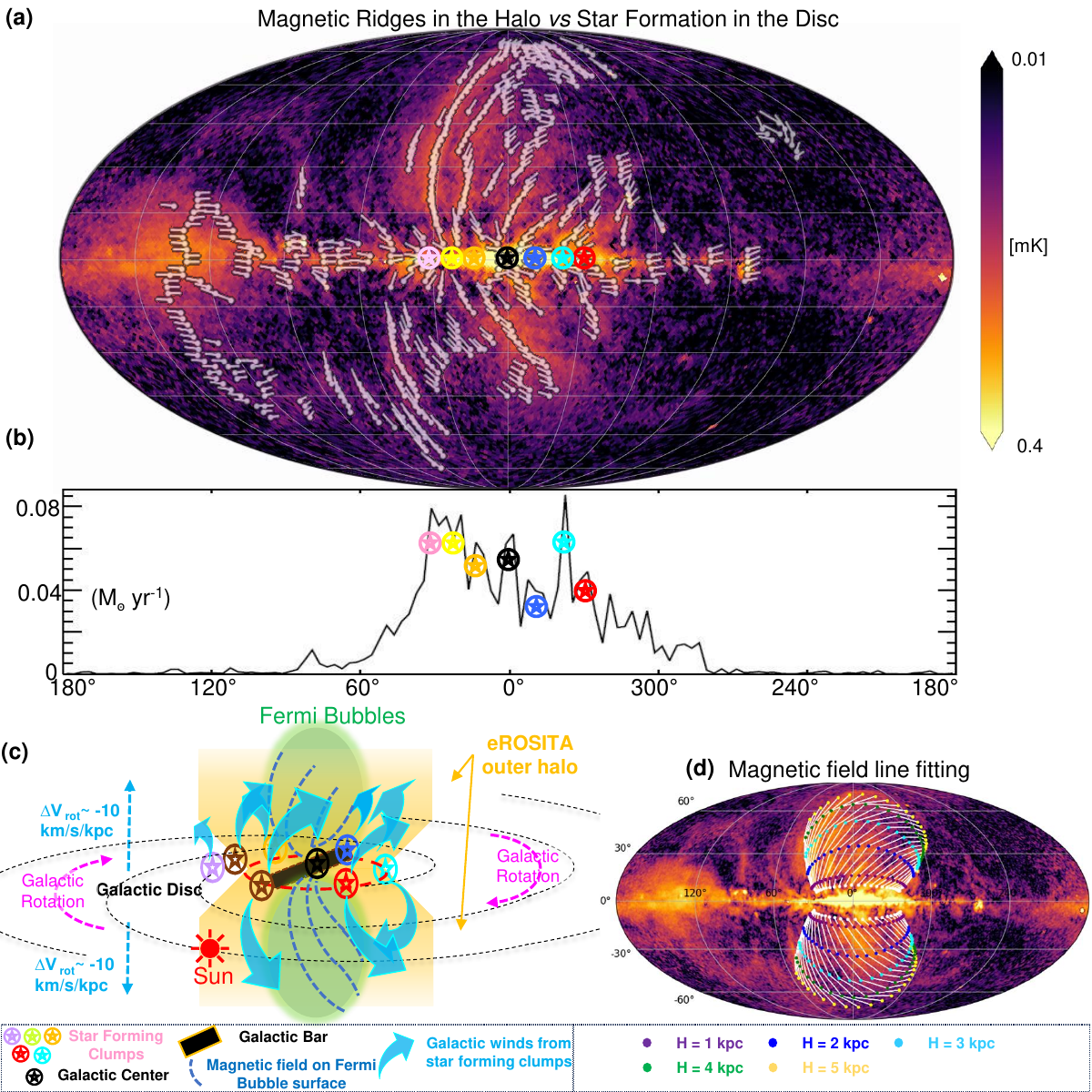}
\end{center}
\caption{{\bf Figure 5 $\mid$ Energy sources of the  Galactic outflows.} (a) the polarized intensity map at 22.8~GHz by WMAP with the magnetic ridges (white) overlaid. The footpoints of the magnetic ridges correspond to the marked clumps with a high star-formation rate on the Galactic plane in (b) measured by Ref.\cite{Elia22}. A north polar view of the active star-forming clumps is shown in the Extended Data Figure~8. (c) Proposed scenario for the observed multi-wavelength Milky Way outflows. The 3--5~kpc star-forming ring powers the outer outflows, showing a ``bouquet'' morphology. We assume the presence of a central outflow that generates the Fermi Bubbles, as discussed in previous works\cite{Su10Fermi,Lacki14}{}. We note that  some magnetised ridges appear to be wrapped around the Fermi Bubbles. The outflows show a lag as the plasma is transported to higher Galactic latitudes because of the decrease of rotational speed at higher heights  from the plane. (d) Modelling of  the   field line flow wrapping around the surface of the outer outflows. An  angular  lag of $4.5^\circ\frac{\mathrm{H}}{\mathrm{kpc}}$ with an increasing height H is used. The east-to-west tilted orientation of the magnetic ridges can be reproduced by such lagging. The resulting magnetic field is parallel to the surface of the Bubbles, providing confinement to the CRs.
}
\label{fig:origin}
\end{figure}

Figure~\ref{fig:origin} shows that the magnetic ridges of the {\it outer region} connected to the locations having high star-formation rates in the disc, corresponding to the star-forming ring {located 3--5~kpc from the Galactic Center (GC) at the end of the Galactic Bar}.
The magnetised ridges related to the Fermi Bubbles, on the other hand, appear to originate from the few-hundred parsec Central Molecular Zone (CMZ) and to wrap around the surface of the Fermi Bubbles, which is consistent with previous works\cite{Carretti13Nat,Su10Fermi}{}.
This distinction is also consistent with the conclusion above that there are different origins of the relativistic electrons in the Fermi Bubbles and {those in} the {\it outer region} (the region between the outer boundaries of the Fermi and eROSITA Bubbles). In search of the sources of relativistic electrons, we show in Extended Data Figure~8 a polar view of the specific star-formation rate distribution in the Milky Way, as measured by Herschel\cite{Elia22}{}, displaying several clumps in the 3--5~kpc star-forming ring with rates $\Sigma_{SFR}\gtrsim0.02~\mathrm{M_\odot yr^{-1} kpc^{-2}}$. This distribution aligns with the footprints of the magnetised ridges, which gives us a clue to the origin of the relativistic electrons in the {\it outer region}. The collective effect of merging supernova explosions can generate a Galactic wind by expelling material at speeds ranging from $100$ to $1000$~km~s$^{-1}$ out of the Galactic disc\cite{Chevalier85SF,Heckman00,Veilleux05,Strickland09}{}. Consequently, a wind termination shock is anticipated at the high-altitude extent of the wind, where particles are accelerated and heated. Hence, the primary source of CRs responsible for the {\it outer region} is likely to be Galactic outflows from the 3--5~kpc star-forming ring (hereafter ``outer outflows'').
The radio-gamma flux densities in the outer outflows can be effectively modeled {with electrons having} energies higher than 2~GeV, following a single power-law distribution. This supports the hypothesis of a shared origin for the multi-wavelength radiations. The electron index of $\alpha\simeq-3.4$ observed in the outer outflows is too soft with respect to the expectation from strong shock acceleration (the classical value for the index in a strong shock is $\alpha=-2$)\cite{Vink14_shockMA,Guo18_shockMA}{}. Instead, the observed soft spectrum arises from a significant cooling above 2~GeV within the investigated patches. This indicates that the dynamic timescale of the outer outflows is longer than the cooling timescale of the relativistic electrons, $10^8$ yr (see Methods).
In this scenario, the thermal X-ray emission of the eROSITA Bubbles comes from the shock-heated plasma. Applying the Rankine-Hugoniot relation, we find that the wind velocity is approximately $v_w \sim 400$~km/s (Methods), falling within the anticipated range for galactic winds driven by collective supernova explosions. Our calculations indicate that sustaining the outer outflows up to 10~kpc height requires less than $21\%$ of the energy released from supernova explosions in the 3--5~kpc star-forming ring and the mass loss rate from the star forming ring would be 0.3-1.3~M$_\odot$/yr (see calculations in \S~4 in Methods and Extended Data Table~1). Our results are consistent with {prior} hydrodynamical simulations\cite{Nguyen22SFringHDsimu} focusing on the thermal emission from the eROSITA Bubbles.

We present a detailed 3D picture for the eROSITA Bubbles in Figure~\ref{fig:origin}c, where the outflows form a ``bouquet'' shape in the Galactic halo.
Our geometric check in the Supplementary material shows that a ``bouquet'' outer halo can be projected into a bubble shape (see Methods, Extended Data Figure~9). In our model, magnetic ridges appear as coherent structures emerging from the active star-forming regions in the Galactic disc. {Previous research \cite{Carretti13Nat,MarascoFraternali11,Faerman17}{} revealed} that the Galactic halo medium rotates clockwise as seen from Galactic north, similar to the the Galactic disc, and that the azimuthal angular speed decreases with height above the disc. {This is a natural consequence of the conservation of angular momentum as the outflowing winds expand into the halo.} Indeed, external spiral galaxies show a similar behavior and have been found to have a lag in the halo, with a reduction of the gas rotational speed on the order of $10~\mathrm{km/s}\, \frac{H}{~\mathrm{kpc}}$\cite{Sancisi2001,Marasco13}{}, where $H$ is the height above the galactic plane.
Figure~\ref{fig:origin}d shows that the angular lag of outflowing gas with increasing height from the Galactic plane is compatible with the general orientation of the magnetic field, aligned with the magnetic ridges {showing a global westwards-oriented tilt from the Galactic disc into the halo}. Recent simulations\cite{Meliani24BhaloSimulation}{} of the magnetic halo of a galaxy found that ordered magnetic fields are associated with free gas winds in galactic outflows, whereas more turbulent fields appear in the shocked region.
{Thus, these magnetic ridges in the Galactic halo is likely to trace the outflowing winds.} This is consistent with what was previously found for the magnetic ridges wrapping the Fermi Bubbles\cite{Carretti13Nat}{}, thus suggesting similar gas dynamics for the inner and outer outflows.

Our data analysis and modelling {lead us to propose} that the 3 -- 5~kpc star-forming ring powers the eROSITA Galactic outflows, which result in both thermal and nonthermal large-scale emission. As shown in Figure~\ref{fig:origin}a, magnetic ridges are connected to the active star-forming regions on the Galactic Disc rather than the GC, which supports our proposed scenario. Earlier {investigations} have also discussed the possibility that the thermal emission from the {\it outer region} results from the Fermi Bubbles' expansion because of either previous AGN-activities\cite{Sofue16mn,Sofue21,Yang2022NatAs,Mou23NC} of the SMBH Sgr A$^\ast$ or outflows from starburst regions at the GC\cite{Sarkar19}. Whether these GC-outflow scenarios can reproduce such magnetic structures in Figure~\ref{fig:origin}a is unclear, and dedicated simulations are needed to answer this question. With our findings, future research focusing on the circumgalactic medium of the Milky Way should take the magnetic halo and the non-thermal fluxes in the corresponding regions into account while analysing the X-ray halo. Additionally, future studies of the Milky Way's outflows should consider the 3--5~kpc star-forming ring as another potential energy source other than the SMBH and the star-burst activities at the GC.

Our results indicate a connection between the X-ray-emitting and magnetised Galactic halos, and offer insights into the origin of these halos in other galaxies. {We have shown that galactic-scale star-forming activities play a strong role in the formation of magnetically organized structures within galactic halos.} Notably, observations of edge-on galaxies have revealed a distinctive X-shaped magnetic halo at radio frequencies, featuring kpc-scale anisotropic magnetic ridges emerging from the galaxies' inner region\cite{Krause20ChangesXXII}{}. Such features can be attributed to galactic outflows launched from active star-forming regions, which can regulate the gas ecosystem of galaxies and have a fundamental impact on galaxy evolution.

\clearpage
\newpage

\section*{Methods}
\setcounter{figure}{0}
\setcounter{section}{0}

\section{Distance measurements through multi-wavelength observations}

A key issue that we want to answer is whether those gigantic structures projected in the sky are within the Local Arm or stemming out of the Galactic disc at much further distance from us. In this section, we will present our distance estimates through multi-wavelength comparison: \S~1.1 X-ray vs dust observation; \S~1.2 dust vs synchrotron polarization; \S~1.3 Faraday rotation depolarization analysis based on multi-frequency synchrotron data.

\subsection{X-ray vs dust observations}

We estimate the distance to the X-ray emitting eROSITA Bubbles in Figure~\ref{fig:depoldust}. The X-ray photons emitted by hot plasma can be absorbed by foreground medium. Figur~\ref{fig:depoldust}(b) shows that the polarized dust intensity from Planck survey is originated mainly from the local medium. We take the measurements from Ref.\cite{Lallement18}, and integrate the extinction for distances lower than $500$~pc around the Sun, as obtained from the 3D dust distribution by the Gaia-2MASS-Apogee dataset. Figure~\ref{fig:depoldust}(c) shows that an anti-correlation is observed between the eROSITA Bubbles (0.6--1.0~keV) and the contour of polarized intensity from the dust at $\mathrm{PI_{dust}}=22 \mu$K. On the other hand, Figure~\ref{fig:depoldust}(d) reports the anti-correlation between softer X-ray emission from ROSAT\cite{DataROSAT}{} (0.11--0.28~keV) and the contour of a lower polarized intensity from the dust ( $\mathrm{PI_{dust}}=5 \mu$K).
The X-ray absorber for softer X-ray emission has extended to a higher Galactic latitude. This is consistent with the picture that the bulk of the X-ray emitting structure is behind the dust within 500~pc as illustrated in Figure~\ref{fig:depoldust}(a). Therefore we conclude that the eROSITA Bubbles stand behind the dust emission within 500~pc from the Sun, beyond the Local Arm. Hence, they should extend above and below the Galactic disc.

\subsection{Dust vs synchrotron polarization}

We obtain the magnetic ridges by performing same-latitude cuts across the all-sky polarized synchrotron emission map (PI$_{syn}$) from WMAP K band\cite{DataWMAP1}{} to find the local maximum points (PI$_{syn,max}$) along each cut. The points with PI$_{syn,max}>0.04$~K are preserved in Figure~\ref{fig:outflows}, and the corresponding magnetic field directions are overlaid. Several coherent magnetic structures extending more than $15^\circ$ are found and we define them as magnetic ridges. Figure~\ref{fig:depoldust}(b) demonstrates that the polarized dust intensity from Planck survey are originated mainly from the thermal dust within 500~pc from us. This agrees with the previous simulations\cite{Maconi23DustSimulation} that the dust emission observed at 353~GHz agrees with the radiation of dust from the Local Bubble, which is within 200~pc from the Sun\cite{LocalismFrisch11araa,Yeung23LHB}{}.
Extended Data Figure~2 presents the comparison between polarization angles of dust (353~GHz) and the magnetic field inferred from the synchrotron polarized emission. The polarization angle of dust emission at mm-wavelengths is perpendicular to the magnetic field.
As shown in Extended Data Figure~2(b), the known local structures, the Fan Region, and most of the Galactic plane ($\mid b\mid <5^\circ$) apply to the polarized E-vector of dust emission perpendicular to the magnetic field. There is no uniform correlation between the magnetic field and the polarized dust emission in the Serpens-Aquila Rift.
Furthermore, most of the magnetic ridges corresponding to the outer region of the eROSITA Bubbles and the southern Fermi Bubbles have no dust counterparts. Hence these ridges are Galactic structures which are not contaminated with other components along the line-of-sight.
Because the polarized synchrotron emission also does not have the distance information for the emitting layer, we will introduce the Faraday rotation depolarization analysis in \S~1.3 to measure the distances to those magnetic ridges.

\subsection{Faraday rotation depolarization analysis}

The depolarization is the ratio between the polarization fraction at a frequency and that at a reference frequency, assumed  not depolarized.
The polarized intensity and polarization angles of the magnetic ridges show wavelength-dependent depolarization. Based on the derivations from Refs.\cite{Burn66,Sokoloff98,Beck15}{}, the wavelength-dependent Faraday depolarization for a synchrotron-emitting and Faraday rotating turbulent magneto-ionic plasma is:
\begin{equation}\label{eq:fdepol}
f_{depol}=\frac{1-\exp(-S)}{S},
\end{equation}
where $S\equiv2\sigma_{RM}^2\lambda^4$. In our calculations, we pick  the $22.8$~GHz of the WMAP\cite{WMAPdata}{} data  as the reference frequency, at which the Faraday rotation depolarization can be  assumed negligible.
The RM dispersion $\sigma_{\rm RM}$ is:
\begin{equation}\label{eq:simgaRM}
    \sigma_{RM} =0.81 \, \sigma_{B\parallel }\, n_e \, d\, N_{\mid\mid}^{1/2}
\end{equation}
where  $n_e$ [cm$^{-3}$] is the electron number density of the plasma, $ \sigma_{B\parallel }$ [$\mu$G] is the component along the line.-of-sight of the isotropic, turbulent magnetic field, and  $N_{\mid\mid}$ is the number of  random-walk cells of  length $d$ [pc] along the line-of-sight. The latter is defined as $N_{\mid\mid}\equiv L\,f/d$ where  $L$ [pc] is the distance from us and $f\equiv<n_e>^2/<n_e^2>$ is the electron volume filling factor.  The RM dispersion can thus be written as\cite{Ehle93}{}:
\begin{equation}\label{eq:simgaRMsquare}
\sigma_{RM}^2=\left(0.81\, \sigma_{B\parallel}\right)^2\, L\, d\, <n_e^2>
\end{equation}
We adopt $d=100~\mathrm{pc}$ following Ref.\cite{Armstrong95,Chep2010,Beck15}. We use the electron density  model by Ref.\cite{ymw16ne}.
We assume a distance of  the Sun from the Galactic Centre of  $D_\odot=8.5~\mathrm{kpc}$. Given a position on the  line-of-sight of Galactic coordinates ($l$, $b$) and a distance $L$ from the Sun, the position in the Galaxy can be expressed as:
\begin{equation}\label{eq:lbcoord}
\begin{array}{ll}
r&=\sqrt{D_\odot^2+L^2\cos^2(b)-2D_\odot L\cos(b)\cos(l)},      \\
z&=L\sin(b),                              \\
\end{array}
\end{equation}
where $r$ is the separation from the Galactic Centre on the Galactic plane and $z$ is the height from the plane. {Along each line of sight, we perform the integration with a discrete  step  ${ \delta L}=5~\mathrm{pc}$, which yields an RM dispersion of:
\begin{equation}\label{eq:intsigmaRM}
\begin{split}
\sigma_{RM}^2&=\int_{L}^{O}\left(0.81\, B_\parallel(l,b,L)\right)^2\, d\, n_e^2(l,b,L) {\bf \delta L}   \\
&=\int_{L}^{O}\left(0.81\, B^{JF12}_{\parallel}(l,b,L)\right)^2\, d\, n_e^2(l,b,L) {\bf \delta L}   \\
\end{split}
\end{equation}
We take the magnetic field model of the Milky Way by Jansson\&Farrar12 (hereafter ``JF12'', Ref.\cite{JF12b_turbB}) modified to fit Planck results\cite{PLANCK16XLIIBfield}{}. Here $ B^{JF12}_{\parallel}=B^{JF12}_{regular,\parallel}+1/\sqrt{3}B^{JF12}_{turb}$ is the magnetic field strength adopted from Ref.\cite{JF12b_turbB,PLANCK16XLIIBfield}{}, $B^{JF12}_{regular,\parallel}$ is the LOS projection of the regular magnetic component, and $B^{JF12}_{turb}$ is the strength of the turbulent component.}
Additionally, the wavelength-dependent dispersion caused by the foreground medium with $\sigma_{\rm RM}$ is:
\begin{equation}\label{eq:AngleRM}
    \Delta\Phi=\sigma_{RM}\lambda^2
\end{equation}

The results from these equations are shown in Figure \ref{fig:depolsyn}. Extended Data Figures~3 demonstrates the Faraday rotation depolarization for synchrotron emitting structures at different distances. We estimate the depolarization from the observations at three different frequencies (22.8, 2.3, and 1.4~GHz using data from WMAP\cite{DataWMAP1,DataWMAP2}{}, S-PASS\cite{Carretti13Nat,Carretti19MN}{}, and DRAO/Villa-Elisa surveys\cite{DataDRAO1,DataDRAO2}{}, respectively). The depolarization gets smaller moving to high latitudes and higher  frequencies. The  depolarization screen generated by the magneto-ionic medium out to 5-kpc  is reported in Figure~\ref{fig:depolsyn}. Our calculations tells that the depolarization depends on the frequency and the latitude that determines  how much of the disc the polarized radiation goes through. Specifically: as shown in Figure~\ref{fig:depolsyn}(a), for Zone A (high latitudes $\mid b\mid\gtrsim20^\circ$), both 1.4 and 2.3~GHz can be observed; for Zone B (mid latitudes $5^\circ\lesssim\mid b\mid\lesssim20^\circ$), 1.4~GHz is depolarized whilst  2.3~GHz is observable ; for Zone C  (low-latitudes $\mid b\mid\lesssim5^\circ$) both 1.4~GHz and 2.3~GHz are depolarized. Negligible depolarization occurs at 22.8~GHz at any latitudes and the magnetic coherence are preserved from the Galactic disc up to the Galactic poles. Figure~\ref{fig:depolsyn}, panel c, shows the observed polarized emission  at 1.4~GHz (DRAO/Villa-Elisa)\cite{DataDRAO1,DataDRAO2}{},  2.3~GHz (S-PASS)\cite{Carretti13Nat,Carretti19MN}{}, and at 22.8~GHz (WMAP)\cite{DataWMAP1,DataWMAP2}{}. The image shows that our calculations of the depolarization broadly match the regions of high depolarization. Extended Data Figures~4 presents a detailed analysis for the comparison between all-sky polarized synchrotron emitting structures at 1.4~GHz and 22.8~GHz, and the depolarization screen at the distance $L=5$~kpc.
In the eastern sky of Extended Data Figure 4b, the regions with high polarization intensity at 1.4~GHz are observed at low Galactic latitudes (i.e., the Fan Region and Loop III). Different from the central magnetic ridges, these regions can be interpreted as local, or at a distance closer than some 1-kpc. This is consistent with  previous measurements at  $150$~MHz\cite{Iacobelli13LOFAR}{}. Instead, the Galactic magnetised ridges coincident to the  eROSITA Bubbles are clearly depolarized at 1.4-GHz at low and mid Galactic latitudes, consistently  with the prediction of a depolarization screen at a distance of at least 5~kpc.  The frequency sensitive depolarization  does not affect  the local structures.

\section{Consistency of the X-ray eROSITA Bubbles’ edges with emission at other wavelengths}

In this section, we will discuss the comparison between the edges of the X-ray emitting eROSITA Bubbles and the enhancements at other wavelengths. While the radiations from radio and gamma-ray bands result from different radiation processes than the X-ray, large scale enhancements from these bands are observed to be coincident with the edges of the eROSITA Bubbles by only an offset of a few degrees.

\subsection{Comparison with synchrotron emission}

We compare the magnetised ridges, as we defined them  using synchrotron polarization data from WMAP\cite{DataWMAP1,DataWMAP2}{}, with the X-ray surface brightness at  0.6-1.0~keV  in  Extended Data Figure~1. The polarized intensity peaks (PI$_{syn,max}$) are observed close to the edges of the eROSITA Bubbles with an offset of only a few degrees. Therefore, these Galactic magnetic ridges are enhanced close to the edges of the eROSITA Bubbles, with the only exception of the south-west Bubble's edge. The enhancements of the PI$_{syn}$ at all of the four roots of the eROSITA Bubbles suggests  that the eROSITA Bubbles are limb-brightened in synchrotron polarized emission, similarly to the roots of the Fermi Bubbles\cite{Carretti13Nat,Su10Fermi}{}. The magnetic field directions are parallel to the eastern edges of the eROSITA Bubbles in both the north and the south. Instead, there is no such alignment in the west. A possible explanation is given by  the modelling presented in the main text and  Figure~\ref{fig:origin}(d). The points at the same height ($H=1\sim5$~kpc) on the eROSITA Bubbles's surface are projected on the polarized synchrotron intensity map. An anticlockwise lag of $4.5^\circ$ is introduced between points differing in height by $\Delta H=1$~kpc, which is a natural outcome of the conservation of angular momentum as the outflowing winds expand into the halo. {The tracks of the gas in Galactic outflows have a global east-to-west tilt from the Galactic disc towards higher latitude, and they are consistent with the observed magnetic ridges.}

\subsection{Comparison with Diffuse gamma-ray radiation}

In our $\gamma-$ray intensity map, we calculate the relative excess of the $\gamma-$ray flux along different lines-of-sight comparing to the standard deviation of the selected background area located at high Galactic latitude towards the northeast (patch R0 in Figure~\ref{fig:sed}a). The relative excess is defined by $\sigma\equiv(I_f-\overline{I}_{f0})/std(I_{f0})$, where $I_f$ is the flux at a given line-of-sight, $\overline{I}_{f0}$ and $std(I_{f0})$ are the average value and standard deviation of the background patch. The results are presented in Figure~\ref{fig:sed}a and Extended Data Figure~5.  The northern edges of the eROSITA Bubble are consistent with $\sigma\gtrsim5$ for all three energy bands. For the Galactic south, the diffuse gamma-ray emission for $E_\gamma\gtrsim100$~GeV has shown an enhancement at the edges of the southern eROSITA Bubble upto $l\simeq-60^\circ$ in latitude. But it is less clear for the other two energy bands.
The more detailed comparison between X-ray and $\gamma-$ray at $E_\gamma\gtrsim100$~GeV for eROSITA Bubbles is performed in Extended Data Figure~6. We perform cuts in high Galactic latitudes in the north ($b=+70^\circ$ and $+65^\circ$) and south ($b=-60^\circ$) to avoid the potential influence by the emission from the  foreground Galactic disc or the Fermi Bubbles. In the northern cuts, both X-ray and $\gamma-$ray radiations have shown central enhancements ($-70^\circ\lesssim l\lesssim40^\circ$) beyond the background within the edges of the eROSITA Bubbles with a clear ``plateau'' shape. Additionally, the edges of the enhancements are in agreement with an offset of only a few degrees. In the southern cut, the central enhancements are observed for both X-ray and $\gamma-$ray bands, but they are less evident as compared to the cuts in the Galactic north. Below $b=-60^\circ$, there is no clear edge of $\gamma-$ray emitting structures.

We intend to compare the correlation between different extended structures at large scales (several tens of degrees). Indeed, the edges of the eROSITA Bubbles have shown a consistency with enhancements of polarized synchrotron intensity/gamma-ray intensity within a separation of a few degrees. However, at smaller scales down to a few degrees, the emission could be subject to the local physical conditions, which may vary from place to place. Especially, X-ray emission comes from the thermal electrons, radio continuum comes from synchrotron radiation of nonthermal electrons in the magnetic field, and gamma-ray photons comes from the Inverse Compton scatterings of non-thermal electrons. Their radiation efficiencies do not rely on the same physical quantities. Therefore, we do not expect a perfect correlation down to a few degrees.

\section{SED analysis}

In this section, we will provide details for our SED analysis. We will discuss data and errors in \S~3.1, fitting for the photons of Inverse Compton (IC) in \S~3.2, MCMC fitting in \S~3.3, and fitting results in \S~3.4.

\subsection{data and errors for SED}

{The spectral energy is computed as:
\begin{equation}
E^2d^4N/(dEdAd\Omega dt),
\label{eq:SEDenergy}
\end{equation}
which we will introduce for different bands of data we use below.}

For $\gamma-$ray, Equation~\eqref{eq:SEDenergy} is: the energy of the photons received at the receiver (area A) at a given band width (log-spaced) per solid angle per time (using the notation E$^2$dN/dE for simplicity).

For radio data, spectral flux density ($\mathcal{F}(\nu)$) is the quantity that describes the rate at which energy is transferred by electromagnetic radiation through a surface, per unit surface area and per unit frequency.
$\mathcal{F}(\nu)=\frac{\partial F}{\partial\nu}$, where F is the flux density. The SED quantifies the energy emitted by a radiation source in the log energy band, hence Equation~\eqref{eq:SEDenergy} is (in the unit of ~erg~s$^{-1}$~cm$^{-2}$~sr$^{-1}$):
\begin{equation}
\begin{array}{ll}
\nu\mathcal{F}(\nu)&=\nu\frac{\partial F}{\partial\nu}=\frac{\partial F}{\partial(\log(\nu))}\\
&=\frac{S[{\rm MJy/sr}]}{10^{17}{\rm MJy}/[{\rm erg\cdot s^{-1}\cdot cm^{-2}\cdot Hz^{-1}]}}*\nu[{\rm Hz}] \\
\end{array}
\label{eq:radioSEDdata}
\end{equation}

For the $\gamma-$ray flux density, we use the data of diffuse emission from Fermi-LAT. The HAWC sensitivity is used as upper limits for $E_\gamma\gtrsim1$~TeV as the Bubbles are not observed by HAWC\cite{HAWC2017}{}. We use two methodologies to extract the diffuse gamma-ray emission of the given patches:

1) We use the diffuse radiation after the point sources subtraction from Platz+\cite{Platz22} using a Bayesian analysis on the 12 year Fermi data (hereafter Platz+). Platz+ separated the data in the range $0.316$~GeV$\lesssim E_\gamma \lesssim316$~GeV into 11 energy bins. In order to compute data uncertainties, we consider:
(1a) The standard deviation of the flux intensity among all pixels in the patch ($n_{st}=std(I_{R})$); (1b) The Poisson noise of the observed photons ($n_{pois}=\frac{\sqrt{N}}{N}\bar{I}_{R}$, where N is the total number of photons and $\bar{I}_{R}$ is the average flux intensity in the patch). Hence, the total error of the patch is defined by $err_{R}=\sqrt{n_{st}^2+n_{pois}^2}$. We then use the normal error propagation rules in order to compute the uncertainty of the $gamma-$ray maps. Take patch R1 subtracted by patch R2 as an example:
\begin{equation}
\begin{array}{ll}
Signal: & I_{signal}=I_{R1}-I_{R2}\\
Noise: & s_{noise}=\sqrt{(err_{R1})^2+(err_{R2})^2} \\
\end{array}
\label{eq:error_gam}
\end{equation}
where $err_{R1}$ and $err_{R2}$ are the total errors in patch R1 and patch R2.

2) Complementary, we take the diffuse $\gamma-$ray data of the same patches from the 14 year Fermi-LAT data (hereafter Fermi14yr). Fermi14yr uses the most recent Fermi data and the 14 year source catalog 4FGL-DR3 \cite{FermiCatalog4FGLDR3}. Fermi14yr separates the data in the range $0.1$~GeV$\lesssim E_\gamma \lesssim500$~GeV into 14 energy bins (the low energy bins with poor statistics due to a small effective area will not be used for further calculations). We mask the influence from the point sources within a radius of $14^\circ$ from the boundaries and find the best fit for the diffuse emission through Fermipy\cite{Fermipy_Wood17}{}. The outcome indices ($dN/dE\propto E_\gamma^{\beta_{F14}}$) for the emission from the 5 patches by a power-law fitting are (listing here only the statistical errors): patch R1: $-2.180\pm0.002$; patch R2: $-2.235\pm0.003$; patch R3: $-2.205\pm0.002$; patch R4: $-2.167\pm0.002$; patch R5: $-2.092\pm0.002$. We can see that these fitting indices are rather similar to each other. Therefore, it is necessary to exclude the Galactic foreground influences in our analyses.

To perform fore-/back-ground subtraction in our fitting, we remove the average flux outside the bubbles' edge at the same latitude.
In the southeastern sky ($b>30^\circ, 60\lesssim l\lesssim150^\circ$), there are several known local structures which might influence the estimate of the total flux (see Extended Data Figure~2a and~4). Therefore we select the patch R4 in the southwestern sky to represent the foreground.
As demonstrated in Figure~\ref{fig:sed}a, we choose the patches in the mid-latitude: R1 and R3 for northern/southern {\it outer region} of the eROSITA Bubbles, R5 for southern Fermi Bubble cap (northern Fermi Bubble is not selected because it is overlapped with the Serpens-Aquila Rift, see Extended Data Figure~2b). In order to exclude the influence of the emission of the fore-/back-ground, we subtract the emission at the same Galactic latitude outside the considered patches (i.e., patch R1-patch R2; patch R3-patch R4; patch R5-patch R3). We plot the flux density of 0.6-1.0 keV at the corresponding patches from Ref.\cite{Predehl20} in the SED for reference.

\subsection{Fitting of the photon field for IC}
Based on our analysis, the patches that we study would be only a few~kpc away from the Galactic disc, hence the seed photons in IC radiations of the SED fitting are mainly from the Interstellar radiation field (ISRF) plus the CMB. We neglect the slight anisotropy in the starlight radiation field and only considers the fitting of radiation energy density based on the radiation model proposed by Ref.\cite{Popescu17ISRF} and simplify the seed photon field by fitting the radiation spectrum with 4 blackbody radiation fields at  ``CMB'' (at 2.725~K), ``FIR'' (far-infrared), ``NIR'' (near-infrared), ``Star-light'' (scattered star light from the Spiral Arms of the Galaxy). We present the 4-blackbody modellings in the Supplementary Figure 1-3(a) and summarize the results in Supplementary Table~1(a).

\subsection{MCMC fitting for SED}
We use the package ``naima''\cite{naima}{} to model the multi-wavelength results. We choose to see if the multi-wavelength emission fits purely leptonic processes (Synchrotron + IC).
We assume that the synchrotron emission and IC are from the same electron distributions of the same patches. The synchrotron emission spectrum of the ``naima'' package is calculated from magnetic field strength and electron distribution based on Ref.\cite{Aharonian10syn}. The IC emission spectrum of the ``naima'' package is calculated from seed photon fields and electron distribution based on Ref.\cite{Khangulyan14IC}.

We presume the electron distribution to be in a power-law following the equation defined in ``naima'':
\begin{equation}
f(E_{e})= A_{e}(\frac{E_{e}}{1~\rm TeV})^\alpha,
\label{eq:naima_powerlaw}
\end{equation}
where $A_{e}$ [eV$^{-1}$] is the amplitude of the electron spectrum, $\alpha$ is the electron index, and $E_{e}$ is the electron index.

The non-detection of the Bubbles from the HAWC survey provides the upper limit in our SED fitting in $E_\gamma\gtrsim$1~TeV bands. As a result, the SED data cannot be fitted with a single power-law distribution of electrons.
We also test the electron distribution with an exponential cut-off at the high-energy end of the electron spectrum following the equation defined in ``naima'':
\begin{equation}
f(E_{e}) = A_{e}(\frac{E_{e}}{1~{\rm TeV}})^\alpha\exp(-(\frac{E_{e}}{E_{\rm cutoff}})^{\beta_e}),
\label{eq:naima_expcutoffpowerlaw}
\end{equation}
where we take $\beta_e=2$ as the cut-off power index based on Ref.\cite{Zirakashvili07AAexpcutoff}.

In our fitting, linear priors are used for all the parameters and the first 500 steps are discarded as the burn-in phase. We run $5\times10^3$~steps to get the best fit and errors of the amplitude $A_{e}$, the index $\alpha$, and the magnetic field strength $B$. The MCMC processes and corner maps are presented in Supplementary, showing that the multi-wavelength data dispersion is within 2-$\sigma$ to the best fit. The fit results are summarized in Supplementary Table~1(b).

\subsection{Fitting results}

We first show the power-law fit for the fluxes of individual energy bands in Extended Data Figure~7: the radio flux ($F_\nu\propto E_\gamma^{\alpha_r}$), and the gamma-ray flux ($EdN/dE\propto E_\gamma^{\alpha_{Pl}}$ for Platz+\cite{Platz22} and $EdN/dE\propto E_\gamma^{\alpha_{F14}}$ for Fermi 14 year diffuse map\cite{FermiCatalog4FGLDR3}).

The gamma-ray spectral indices obtained from the two tested-methods have small differences ($\Delta\alpha_{\gamma}\lesssim0.1$, e.g., for northeastern outer halo, $\alpha_{Pl}=-1.499\pm0.004$, and $\alpha_{F14}=-1.422\pm0.008$). They are consistent with each other considering the systematic uncertainties of the Fermi-LAT\cite{Fermi_Instrument12}. For reference, the power-law index obtained in the southeastern cap of the Fermi Bubbles at the same latitude is significantly harder ($\alpha_{Pl}=-1.239\pm0.002$).

The radio fluxes in the outer halo show a harder spectrum compared to the gamma-ray flux for the outer outflows (for northeastern outer halo: $\alpha_r=-1.07\pm0.04$, $\alpha_{Pl}=-1.499\pm0.004$; for southeastern outer halo: $\alpha_r=-1.14\pm0.11$, $\alpha_{Pl}=-1.415\pm0.006$).
But we need to note that the $\gamma-$ray emission is influenced by the Klein-Nishina (KN) effect at higher energy which would result in a softer spectrum because of the suppressed cross section for the IC process\cite{Blumenthal70}. Therefore we need to check if the radio and $\gamma-$ray flux densities could be fitted with one single power-law electron distribution in the SED fitting.
As is shown in the Supplementary Table~1(b), the SED fitted magnetic field strength and electron indices based on Platz+\cite{Platz22} and Fermi14yr\cite{FermiCatalog4FGLDR3} are consistent for all the fittings within error range.

We can verify the obtained magnetic field based on the SED fitting as follows:
The typical IC photon energy radiated by an electron with the energy $E_e$ up-scattering a photon of energy $\epsilon\,$ can be given by $E_{\rm IC}\approx 3(E_e/40\,{\rm GeV})^2(\epsilon/0.4\rm eV)\,$GeV, given that the KN effect is not important. The same electron radiates {synchrotron photons} in the magnetic field $B$ at a typical energy of $ E_{\rm syn}=10^{-4}(E_e/40{\rm GeV})^2(B/1\mu \rm G)\,$eV. Combing this two formulae, we get
\begin{equation}\label{eq:syn-ic}
E_{\rm IC}\approx 3(E_{\rm syn}/10^{-4}{\rm eV})(B/1\mu \rm G)^{-1}(\epsilon/0.4\rm eV)\,{\rm GeV}.
\end{equation}
On the other hand, the synchrotron-to-IC flux ratio is $F_{\rm syn}/F_{\rm IC}=u_{\rm B}/u_{\rm ph}$. Take the north outer outflow for instance, the $\gamma$-ray flux at 3\,GeV is measured to be about $10^{-9}\,\rm erg~s^{-1}cm^{-2}sr^{-1}$, and the radio flux at $10^{-4}\,$eV is about $4\times 10^{-10}\,\rm erg~s^{-1}cm^{-2}sr^{-1}$. For a soft electron spectrum, the optical radiation field is the dominant target radiation field for the IC radiation at 3\,GeV, at which energy the KN effect is not important. Thus, for $u_{\rm ph}=0.23\,{\rm eV/cm^3}$ and $\epsilon\approx 0.4\,$eV, we obtain $B\approx 1.9\,\mu$G via Eq.~(\ref{eq:syn-ic}), which is consistent with the fitting result.

The magnetic field strength we obtained for the southern cap of Fermi Bubbles of $B\approx 1\,\mu$G is smaller than the magnetic field measurements for the Fermi Bubbles at lower galactic latitude\cite{Su10Fermi,kataoka13,Ackermann14FermiSpect}. This is expected because the magnetic field strength is expected to decrease when we measure medium higher in the Galactic halo.

We calculate the cooling time for the non-thermal radiations from the electron at the energy $E_{e}$ Ref.\cite{Heesen16} based on the following equation considering the synchrotron cooling time $\tau_{syn}$ and the Inverse Compton cooling time $\tau_{IC}$:
\begin{equation}
\begin{array}{ll}
\tau_{cool}&=(\tau_{syn}^{-1}+\tau_{IC}^{-1})^{-1} \\
& \simeq 5\times10^8\left(\frac{E_{e}}{1~\mathrm{GeV}}\right)^{-1}\left(\frac{U_B+U_{\rm ph}}{10^{-12}~\mathrm{erg}~\mathrm{cm^{-3}}}\right)^{-1}~\mathrm{yrs}\\
\label{eq:tcoolrad}
\end{array}
\end{equation}
where $U_B\equiv\frac{B^2}{8\pi}$ is the magnetic energy and $U_{ph}$ is the energy density for the radiation field relevant to the IC process. The relativistic Bremsstahlung is negligible in our analysis because the gas density in the halo is too low and the corresponding cooling time is more than $10$~Gyr.

\section{Outer outflow modelling}

Regions with a star formation rate surface  density larger than 0.01~$\mathrm{M_\odot yr^{-1} kpc^{-2}}$  drive superwinds that can become galaxy  outflows of speed of $\sim$ 100--1000~km/s\cite{Chevalier85SF,Heckman00,Strickland09,Heesen21halorev}{}. Extended Data Figure~8 shows the star-formation rate density of the Milky Way's disc as measured with Herschel telescope data\cite{Elia22}{}. We find  that there are several clumps that have sufficient star formation rate to drive galactic winds and that are at and about the star-forming ring located at 3-5~kpc  from the Galactic Center (GC).

The energy injection rate in the outer halo can be estimated by:
\begin{equation}\label{eq:E_tot}
E_{tot}\sim\dot{E}_{inj}\min\{t_{dyn},t_{therm,cool}\}
\end{equation}
where $E_{tot}$ is the total energy in the outer halo, $t_{dyn}$ is the dynamical timescale of the outer halo and $t_{therm,cool}$ is the cooling time of the hot plasma. The quantity $\dot{E}_{inj}$ is the energy injection rate into the system. The system energy depends on  the system time before the cooling is dominant, whilst it depends on the cooling time when it is shorter than age of the outer halo ($t_{dyn}>t_{therm,cool}$). The cooling time for the hot plasma in the eROSITA Bubbles is estimated of approximately $2\times10^8$yrs \cite{Predehl20}{}.

The total energy injection  is made up   of 1) the thermal energy of the hot plasma ($\dot{E}_{therm}$), 2) the energy of non-thermal electrons ($\dot{E}_{CR}$), and 3) the magnetic field energy ($\dot{E}_B$). The energy in the hot thermal plasma that emits the X-ray halo is summarized in the Extended Data Table~1 and is estimated depending on the  Bubbles'  height (see Supplementary for more details). The CR energy can be derived from the electron energy distribution of the patch R1 SED fitting, which gives $2.2\times10^{53}$~erg. If we assume the same electron energy density resides in the rest of the  outer eROSITA Bubbles, the total relativistic CR energy would be $1.2\times10^{55}$~erg. The magnetic energy can be estimated by $E_B=VB^2/8\pi$, where $V$ is the volume of the outer outflows. In our calculations for the energy of the halos below, we adopt a reasonable height ranging from 4 to 10~kpc. From our SED fitting, the magnetic strength at an height of 3~kpc is 1-2~$\mu$G. Assuming an average magnetic field strength of 3~$\mu$G across the entire outer outflows, the magnetic field energy is reported in Extended Data Table~1.

The injection rate of the dynamical energy in the wind can be expressed by:
\begin{equation}\label{eq:Edot_inj}
\dot{E}_{inj}=\frac{1}{2}\dot{M}_{inj} v_w^2
\end{equation}
where $\dot{M}$ is the mass injection rate due to the Galactic outflows from the star forming clumps, and $v_w$ is the velocity of the Galactic wind. We consider the ions and electrons downstream have reached the same temperature $T_g$. The value of $v_w$ can be estimated using Rankine-Hugoniot relation:
\begin{equation}\label{eq:RHrelation}
T_g=\frac{3}{16}\frac{m_H}{k_B}(v_w-v_g)^2,
\end{equation}
where {the quantity $m_H$ is the mass of hydrogen} and the quantity $v_g$ is the velocity of the shock heated gas. For a temperature of 0.3~keV\cite{Predehl20}{} and $v_g\ll v_w$ for the outflows, the wind velocity is around $v_w\simeq400$~km/s. The wind velocity significantly exceeds the sound speed in the hot wind, which is $c_s\simeq180$~km/s. Thus, a termination shock is  expected in the outer outflows.

A number of previous works  reported a range of   the supernova  rate  in the Milky Way of 2--6 per century\cite{SNRrate1Reed05,SNRrate2Diehl06,SNRrate3Li11,SNRrate4Rozwadowska21,SNRrate5Adams13}{}. The Herschel measurements for the star-forming rate of the Milky Way\cite{Elia22} show  that considerable amount of star forming activity occurs in the 3--5~kpc star-forming ring of the Galaxy. Hence, the rate of supernovae at and about the 3--5~kpc star forming ring can be approximated as 1  per century. The ejected energy of a supernova explosion is $\sim10^{51}$~erg\cite{Poznanski13SNRenergy}{}. This corresponds to an energy injection rate from the 3--5~kpc star forming ring of $\dot{E}_{SFR}\simeq3.2\times10^{41}$~erg/s.

Based on our estimate of  the height of magnetic ridges and non-thermal electron cooling time scale (see Main Text), we test outer outflows with  height of 4--10~kpc and system time of  $10^8$~yrs and $10^9$~yrs, and summarize the results in Extended Data Table~1.
Our calculations show that the total energy in the outer halo is 8--20$\times10^{55}$~erg, similar to what found in previous work\cite{Predehl20}{}.
The energy injection rate required for the outer outflows is a few times $10^{40}$~erg/s, which corresponds to only 5--21$\%$ of that produced by  core collapse supernova explosions in the 3--5~kpc star forming ring, which hence can  amply  supply the outer outflows.
The mass loss rate from the 3--5~kpc star forming ring is 0.3--1.3~M$_\odot$/yr.
Hence, our conclusion here is that the bulk of multi-wavelength emission related to the outer halo can result from the galactic outflows powered by the star-forming ring of our Milky Way located 3 -- 5~kpc from the GC.

Our model provides an explanation for the bulk of the extended thermal and non-thermal emitting structures, as well as the magnetic field direction we measured. Nevertheless, there are some other phenomena of the Milky Way's outflows that need further investigation. For example, the NPS is observed to have enhanced brightness compared to its southern counterpart. As demonstrated in Figure~\ref{fig:origin}, the magnetic ridges in NPS are connected to the near end of the Galactic Bar, and hence the density of the plasma can be more influenced in NPS by violent star-forming activities. The X-ray surface brightness is proportional to the density squared of the emitting hot plasma, and hence a slight intrinsic variation to the density results in a substantial change to the surface brightness. Moreover, previous research\cite{Inoue15} proposed that the metallicity of galactic outflows can vary due to either previous AGN events or starburst activities at the GC. Earlier X-ray research have studied the metallicity of the hot plasma in various patches of the Fermi and eROSITA Bubbles and drawn different conclusions (sub-solar\cite{kataoka13,LaRocca20} or super-solar\cite{Gupta23NA_SFerass}). Indeed, at such low X-ray temperatures it is basically impossible to detect the intrinsic continuum (bremsstrahlung) at CCD resolution, therefore abundances are unreliable. Additionally, the abundances will be characteristic of the ambient medium that has been shock heated, therefore it is logical that the abundances are assumed to be as low as the one of the Galactic halo. More studies on detailed simulations and observations for galactic outflows are needed to answer these questions.

\section{Three-dimensional geometry of outer outflows}

The eROSITA Bubbles appear as two gigantic spheres extending up to more than $80^\circ$ in latitude. We argue that the caps of the Bubbles at such a high latitude can be reproduced by different intrinsic three-dimensional geometries. We have tested the X-ray emitting structures with a height from 4 to 100~kpc using an open ``bouquet'' shape in Extended Data Figure~9. The bubble shape can even be reproduced with open outer outflows at an unrealistic height (H$=100$~kpc, Extended Data Figure~9d).
Therefore, our modelling is consistent with the observation (``bubble-shape'') because of projection effect.

\clearpage
\newpage

\begin{table}
\small
\centering
\begin{tabular}{|c|c|c|c|c|c|c|}
\hline \hline
\multicolumn{7}{|c|}{{\bf Assumptions}}\\
\hline
$H_{sys}$ [~kpc]  & \multicolumn{2}{||c||}{4} & \multicolumn{2}{c||}{7} & \multicolumn{2}{c|}{10}\\
\hline
$t_{dyn}$ [~yrs] & \multicolumn{1}{||c|}{$10^8$} & \multicolumn{1}{c||}{$10^9$} & $10^8$ & \multicolumn{1}{c||}{$10^9$} & $10^8$ & \multicolumn{1}{c|}{$10^9$} \\
\hline \hline
\multicolumn{7}{|c|}{{\bf Results}}\\
\hline
$E_{therm}$ [$\times10^{55}$~erg] & \multicolumn{2}{||c||}{5.9}  & \multicolumn{2}{c||}{10} & \multicolumn{2}{c|}{14}\\
\hline
$E_{B}$ [$\times10^{55}$~erg] & \multicolumn{2}{||c||}{1.5} & \multicolumn{2}{c||}{3.3} & \multicolumn{2}{c|}{5.3}\\
\hline
$E_{tot}$ [$\times10^{55}$~erg] & \multicolumn{2}{||c||}{8.6} & \multicolumn{2}{c||}{15} & \multicolumn{2}{c|}{21}\\
\hline
$\dot{E}_{inj}$ [$\times10^{40}$~erg/s] & \multicolumn{1}{||c|}{$2.7$} & \multicolumn{1}{c||}{$1.4$} & $4.7$ & \multicolumn{1}{c||}{$2.5$} & $6.6$ & \multicolumn{1}{c|}{$3.5$} \\
\hline
$\chi_{inj}$ [$\%$] & \multicolumn{1}{||c|}{$8.5$} & \multicolumn{1}{c||}{$4.5$} & $14.6$ & \multicolumn{1}{c||}{$7.7$} & $20.5$ & \multicolumn{1}{c|}{$10.8$} \\
\hline
$\dot{M}_{inj}$ [~M$_\odot$/yr] & \multicolumn{1}{||c|}{$0.54$} & \multicolumn{1}{c||}{$0.28$} & $0.92$ & \multicolumn{1}{c||}{$0.49$} & $1.3$ & \multicolumn{1}{c|}{$0.68$} \\
\hline \hline
\end{tabular}
\caption{{\bf Extended Data Table 1 $\mid$ Calculations for the outer outflows.} The quantities $H_{sys}$ and $t_{dyn}$ represent the assumed height and dynamical timescale of the outer halo, respectively. $E_{therm}$ and $E_B$ are the obtained energy of the outer halo for hot plasma and magnetic fields. The CR energy is obtained to be $1.2\times10^{55}$~erg from SED analysis (Methods). $E_{tot}$ is the total energy of the outer halo, which is at similar order of magnitude with a factor of a few difference compared to the energy estimated from Ref.\cite{Predehl20}. The geometries of the system are adopted from Figure~5c: plasma filled evenly in the outer shell with a vacant inner cylinder having 3~kpc bottoms radius. $\dot{E}_{inj}$ is the energy injection rate obtained from Equation~\eqref{eq:E_tot} and $\chi_{inj}\equiv\dot{E}_{inj}/\dot{E}_{SFR}$ is the required percentage of the energy injection rate due to the supernova explosions from the star forming ring (see Methods). $\dot{M}_{inj}$ is the mass injection rate obtained from Equation~\eqref{eq:Edot_inj}. {\it Our results show that the supernova explosions from the star forming ring are enough to fuel all the tested scenarios, and the mass injection rate is consistent with the measurements from Ref.\cite{FoxRichter19massoutflow}.}}
\label{tab:checkbubble}
\end{table}

\begin{figure}
\begin{center}
\includegraphics[width=0.9\textwidth]{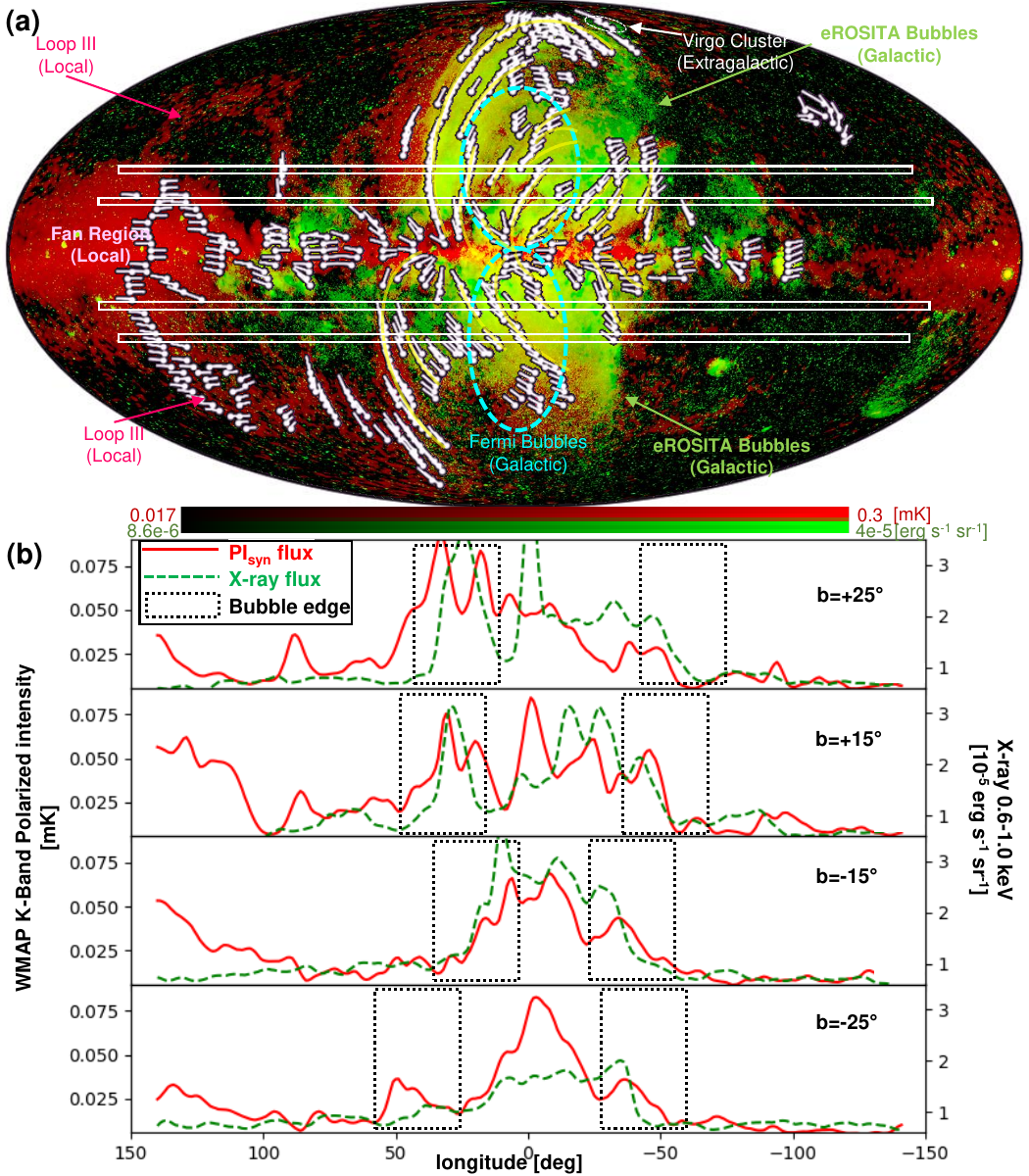}
\end{center}
\caption{{\bf Extended Data Figure 1 $\mid$  Comparison between the X-ray surface brightness of the eROSITA all-sky map (0.6-1.0~keV) and the magnetic ridges.} Panel\, (a) presents the magnetic ridges from Figure~1(b), with four constant latitude cuts at the roots of the X-ray outer halos. The comparisons between the polarized intensity WMAP-K Band (Ref.[55]) (PI$_{syn}$, red) and the 0.6-1.0~keV eROSITA (Ref.[5]) X-ray emission (green) for the four cuts are presented in panel\, (b). The detected magnetic ridges are clear peaks of the red curves in (b). At the edges of the X-ray outer halo (highlighted in dotted boxes), the enhancements of the polarized synchrotron intensity (PI$_{syn,max}$) are observed with an offset of only a few degrees. While the large-scale magnetic structures surrounding the northern cap of the X-ray outer halo appear to largely enclose the Bubble, the same is not obviously true for the southern halo;  there, only the southeastern magnetic ridge approaches the cap of the southern X-ray outer halo.
}
\label{figext:xrayvsradio}
\end{figure}

\begin{figure}
\begin{center}
\includegraphics[width=1.0\textwidth]{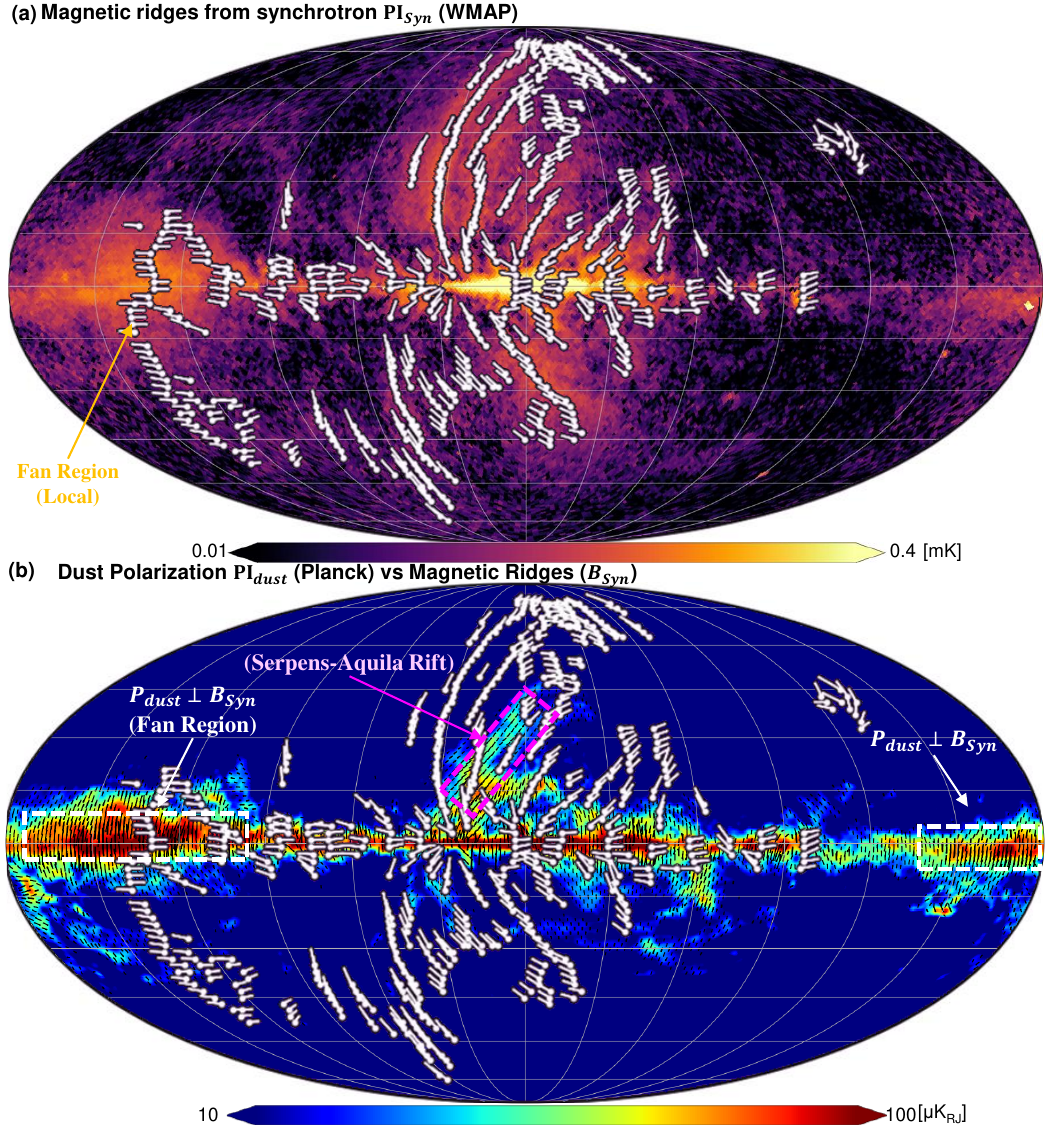}
\end{center}
\caption{{\bf Extended Data Figure 2 $\mid$ Comparison between dust  and synchrotron polarization.} (a) Magnetic field ridges detected by synchrotron polarization ($B_{syn}$,  WMAP at 22.8~GHz). The coherent ridges are enhanced in their polarized synchrotron intensity and connected by the magnetic field lines. (b) Comparison between the polarized emission of thermal dust ($P_{Dust}$, background black bars for direction and filled color for polarized intensity by Planck at 353~GHz from Ref.[28]) and the magnetic ridges deduced from synchrotron (white lines adapted from panel\, a). The polarized E-vectors of the dust emission shows a general perpendicular direction to the magnetic field from synchrotron in the Galactic plane ($\mid b\mid <5^\circ$), while some of the polarized E-vectors of dust emission is parallel to the magnetic field in the known local structure within the Serpens-Aquila Rift. However, most of the magnetic ridges presented in (a) have no dust counterparts, hence they are Galactic structures.
}
\label{figext:dustpol}
\end{figure}

\begin{figure}
\begin{center}
\includegraphics[width=1.0\textwidth]{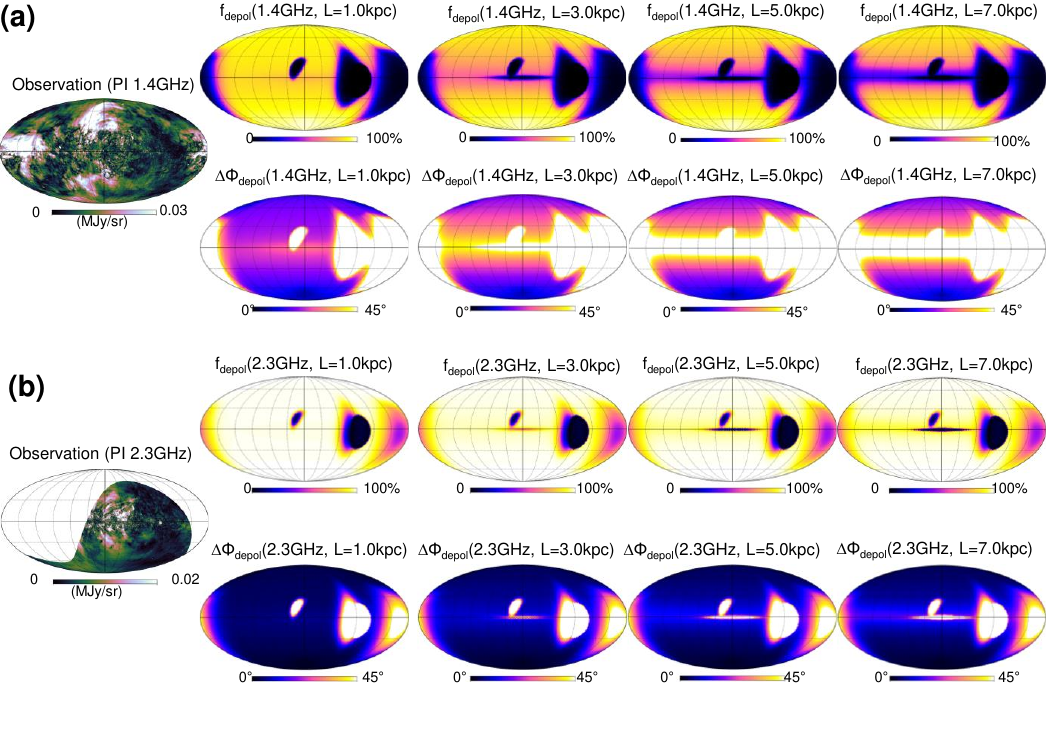}
\end{center}
\caption{{\bf Extended Data Figure 3 $\mid$  Faraday depolarization by the turbulent  Galactic magnetic field out to different distances from the Sun.} $\mid$ The first column is the observed polarized  emission  at 1.4~GHz (a) and 2.3~GHz (b). Columns 2--5 show the Faraday depolarization $f_{depol}$ and angle dispersion $\Delta\Phi$ due to the turbulent Galactic magnetic field, estimated as described in the Methods section. These maps represent the depolarization effects for polarized synchrotron radiations at distances of 1, 3, 5, and 7~kpc from us, shown at frequencies of 1.4~GHz (top two rows) and 2.3~GHz (bottom two rows), respectively. The white color in $\Delta\Phi$ maps indicates where the dispersion of polarization angles is expected to be more than $45^\circ$. At 1.4~GHz frequency, the depolarization screen shows significant growth in latitude between $L=1$--5~kpc.
The $L=5$-kpc case matches well the observed depolarization and the emission at mid and high Galactic latitudes, and therefore the radiation from the magnetic ridges must arise at a distance beyond 5~kpc.
}
\label{figext:depol_dist}
\end{figure}

\begin{figure}
\begin{center}
\includegraphics[width=1.0\textwidth]{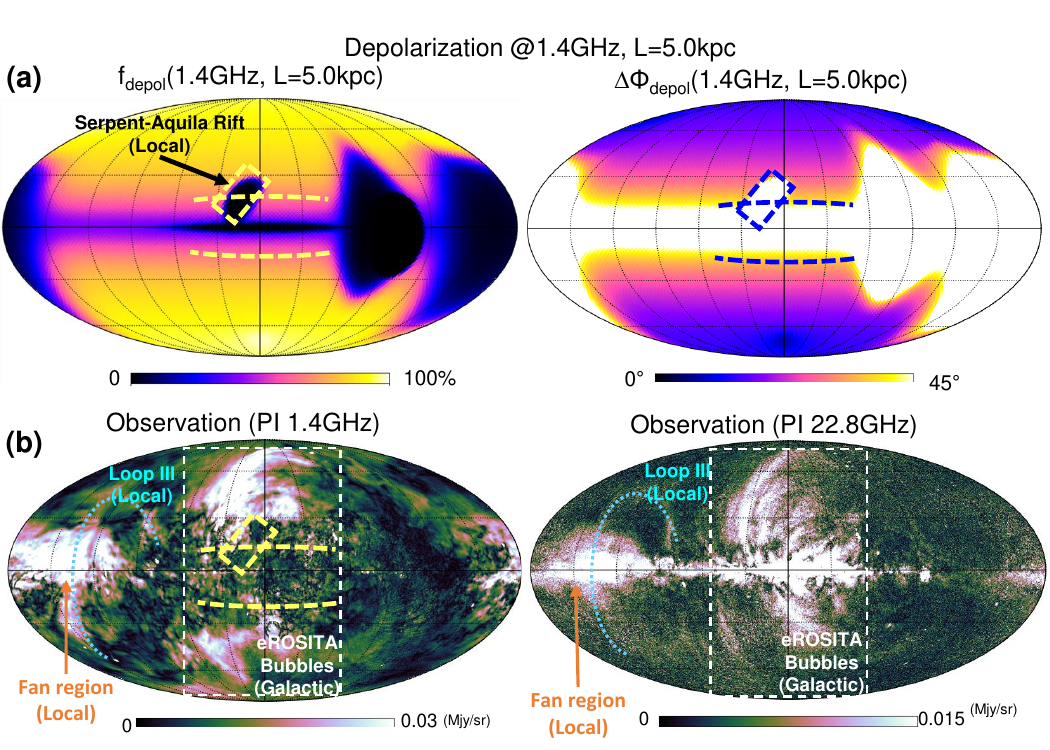}
\end{center}
\caption{{\bf Extended Data Figure 4 $\mid$ Depolarization of the polarized radio emission ridges and  local structures, such as the Fan region and Loop III.} (a) Depolarization analyses ($f_{depol}$, $\Delta\Phi_{depol}$) at 1.4~GHz; (b) observations for polarized intensity at 1.4~GHz (Refs.[25,64]) and 22.8~GHz (Refs.[6,62]). No Faraday depolarization  is expected for local emission down to the Galactic disc. At 1.4~GHz, the radio counterpart of the  eROSITA Bubbles is  depolarized at  Galactic latitudes $|b|\lesssim 20^\circ$, whilst  no depolarization is observed for the Fan region or Loop III.
}
\label{figext:localvsfar}
\end{figure}

\begin{figure}
\begin{center}
\includegraphics[width=1.0\textwidth]{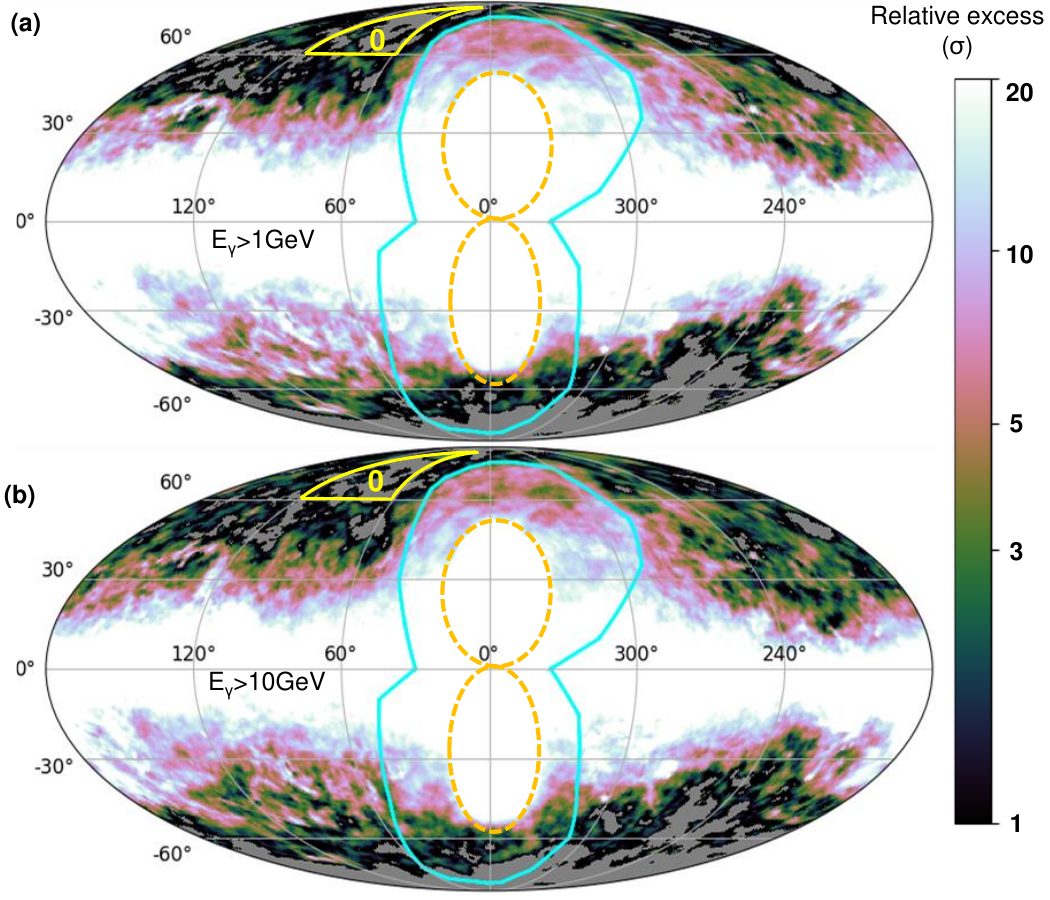}
\end{center}
\caption{{\bf Extended Data Figure 5 $\mid$ $\gamma-$ray diffuse emission intensity maps.} Following Figure~4a and the calculations in Methods, we calculate the relative excess of the $\gamma-$ray flux density compared to patch R0 for $\gamma-$ray photons with (a) $E_\gamma\gtrsim1$~GeV; (b) $E_\gamma\gtrsim10$~GeV. The background area is selected in the yellow triangle in the northeast (the same as Patch R0 in Figure~4a). The two lines are the edges of the eROSITA (solid) and Fermi (dashed) Bubbles.
}
\label{figext:diffusesigma}
\end{figure}

\begin{figure}
\begin{center}
\includegraphics[width=1.0\textwidth]{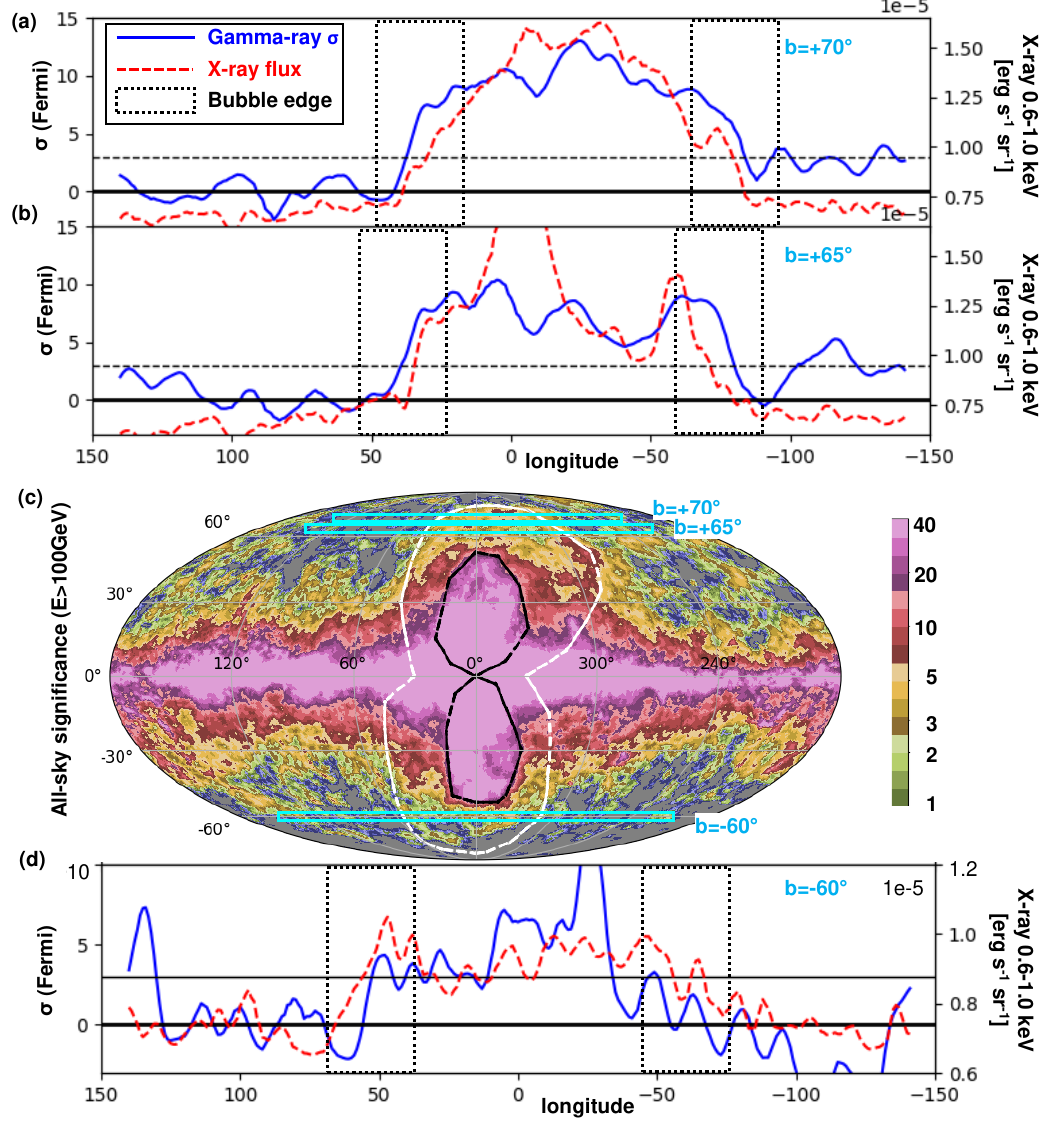}
\end{center}
\caption{{\bf Extended Data Figure 6 $\mid$  Comparison between the X-ray surface brightness (0.6-1.0~keV) and gamma-ray intensity ($E_\gamma\gtrsim100$~GeV) at high Galactic latitudes.} Two cuts in the Galactic north (a: $l=+70^\circ$, b: $l=+65^\circ$) and one cut in the Galactic south (d: $l=-60^\circ$) are considered for X-ray (red dashed lines) and gamma-ray (blue lines). Lower latitudes are not considered to avoid the influence of foreground structures or the Fermi Bubbles. The two energy bands have shown enhancements beyond the background within the edges of the X-ray outer halo, and the edges of the enhancements are in agreement with a separation of only a few degrees. The consistencies are observed in the southern cut, but the enhanced plateau is less evident for the southern Bubble.
}
\label{figext:xrayvsgammaCuts}
\end{figure}

\begin{figure}
\begin{center}
\includegraphics[width=1.0\textwidth]{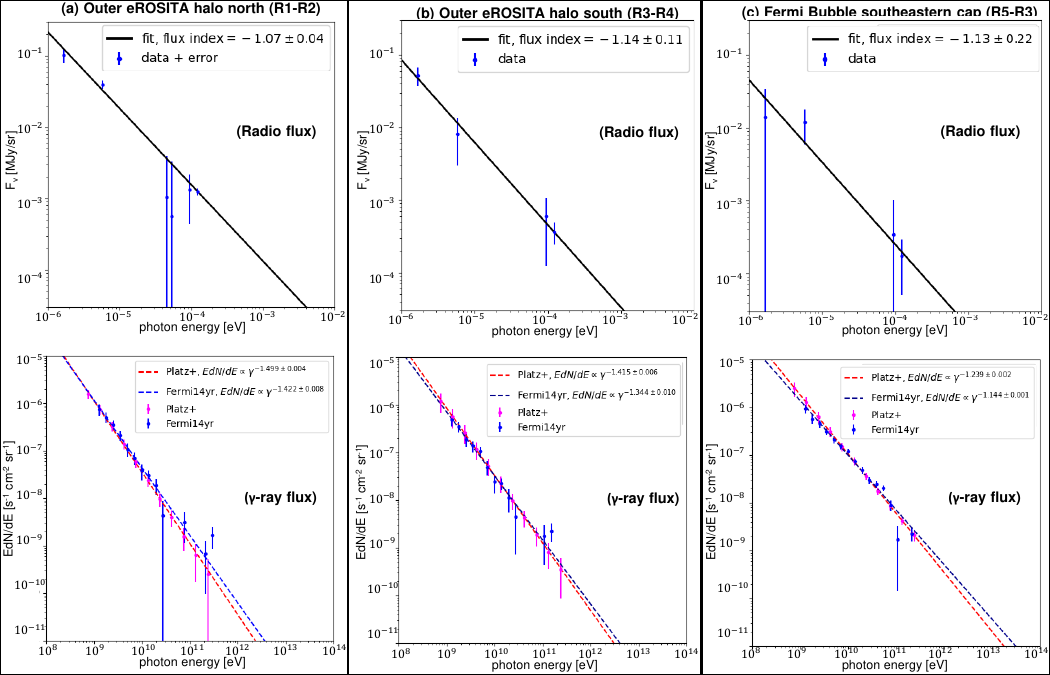}
\end{center}
\caption{{\bf Extended Data Figure 7 $\mid$ Power-law fits for observed fluxes in radio and $\gamma-$ray bands.} Upper plots fit the radio fluxes with respect to the energy $F_\nu\propto E_\gamma^{\alpha_r}$. Lower plots fit the gamma-ray fluxes with respect to the energy for Platz+ (Ref.[17]) --$EdN/dE\propto E_\gamma^{\alpha_{Platz}}$, Fermi14yr -- $EdN/dE\propto E_\gamma^{\alpha_{F14}}$. The error-bars reported here for the gamma-ray data are based on the statistical uncertainties (see \S~3.1 in the Methods section for details). The reference frequencies are listed in the Supplementary Table 2. The error-bars reported for the radio fluxes are calculated based on the flux density calibration accuracy and beam sensitivity of the corresponding surveys, as defined in the Supplementary. The definitions of the fitting parameters are described \S~3.4 in the Methods section. (a) northeastern outer outflows (R1-R2). $\alpha_r=-1.07\pm0.04$, $\alpha_{Pl}=-1.499\pm0.004$, and $\alpha_{F14}=-1.422\pm0.008$. (b) southeastern outer outflows (R3-R4). $\alpha_r=-1.14\pm0.11$, $\alpha_{Pl}=-1.415\pm0.006$, and $\alpha_{F14}=-1.344\pm0.010$. (c) southeastern Fermi Bubble cap (R5-R3). $\alpha_r=-1.13\pm0.22$, $\alpha_{Pl}=-1.239\pm0.002$, and $\alpha_{F14}=-1.144\pm0.001$.
}
\label{figext:powerlawfit}
\end{figure}

\begin{figure}
\begin{center}
\includegraphics[width=0.87\textwidth]{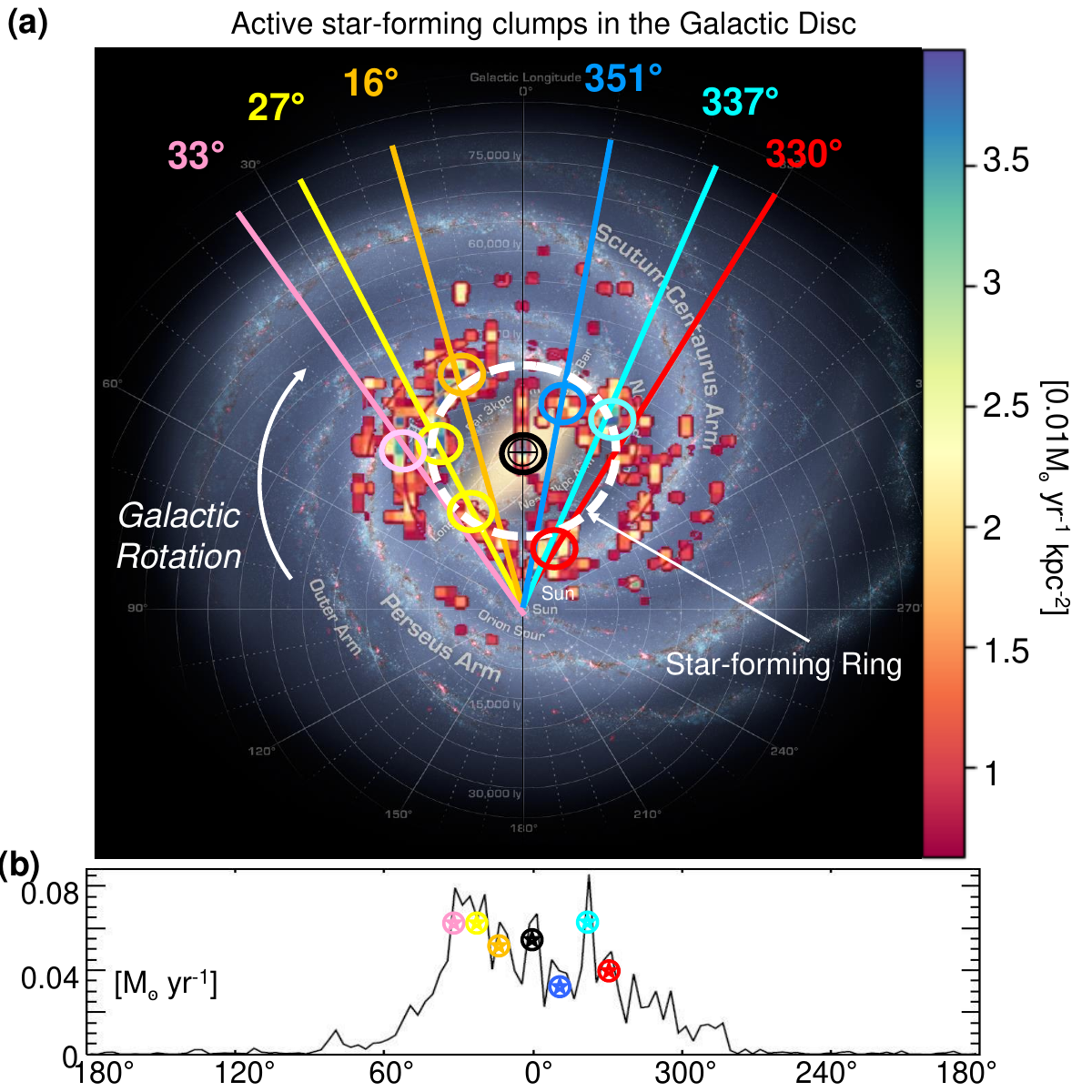}
\end{center}
\caption{{\bf Extended Data Figure 8 $\mid$ Active star forming clumps.} (a) an artist's view of the Galaxy (NASA/JPL-Caltech/R. Hurt) with the active star-forming clumps and their Galactic longitude overlaid. The specific star formation rates are measured from Ref.[32] binned by a resolution of $0.5\times0.5~\mathrm{kpc^2}$, and the clumps with $\Sigma_{SFR}\gtrsim0.02~\mathrm{M_\odot yr^{-1} kpc^{-2}/bin}$ are considered. (b) The footpoints of the magnetic ridges correspond to the marked clumps with a high star-formation rate on the Galactic plane (from measurements in the Figure 4 of Ref.[32]).}
\label{figext:SFR}
\end{figure}

\begin{figure}
\begin{center}
\includegraphics[width=0.9\textwidth]{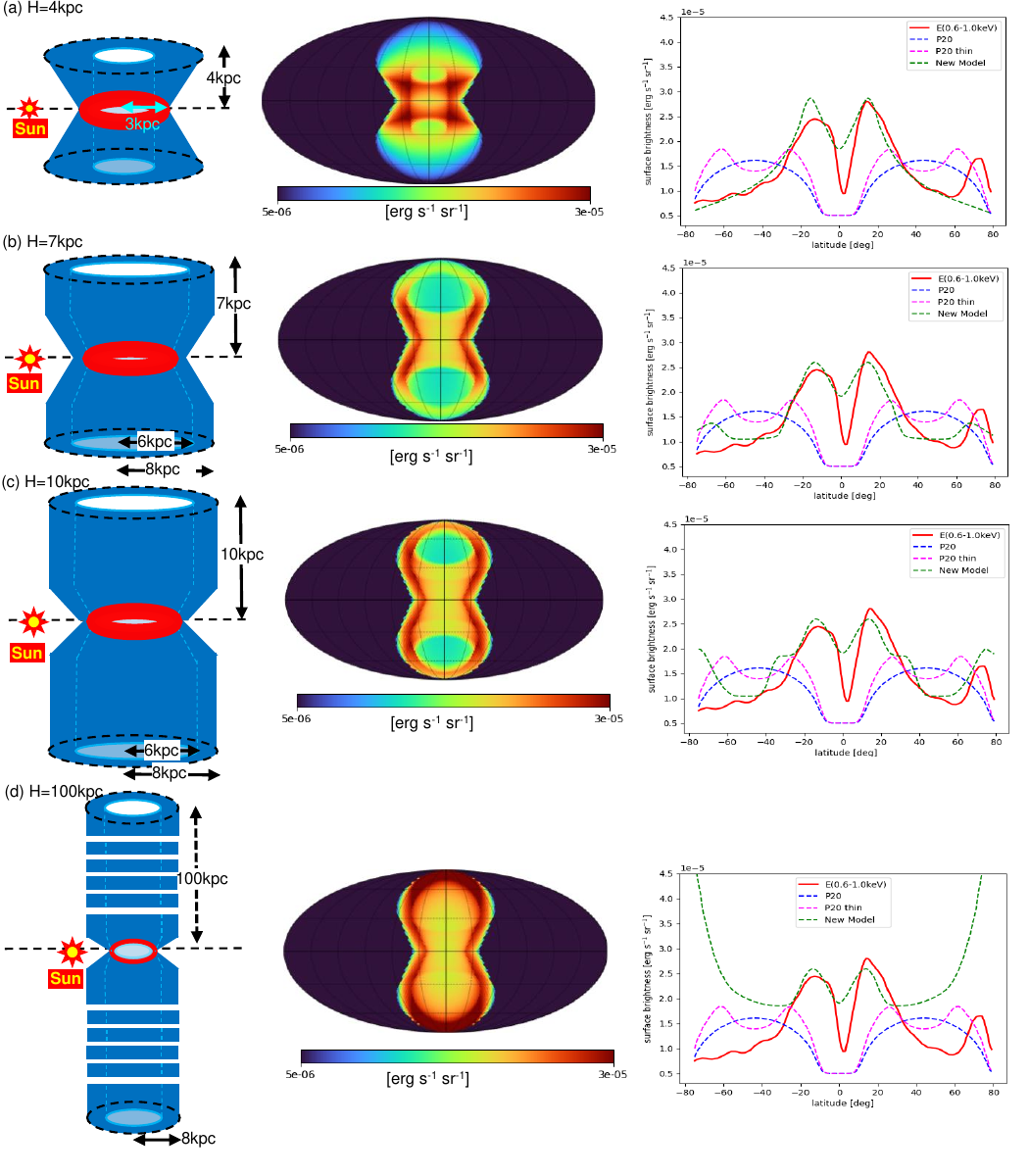}
\end{center}
\caption{{\bf Extended Data Figure 9 $\mid$ Projection effect of the outer halo.} The geometric check for the projection of an open ``bouquet'' outer halo is presented with four different heights: (a) H$=4$~kpc (having an inner vacant cylinder with bottom radius of $3$~kpc); (b) H$=7$~kpc (with a 2-kpc thickness); (c) H$=10$~kpc (with a 2-kpc thickness); (d) H$=100$~kpc (with a 2-kpc thickness). We assume that the density in the outer halo is uniform, the temperature and the metallicity in the outer outflow are T=0.3~keV, metallicity is 0.2 solar (Refs.[30, 5]). Their corresponding projections are presented in the middle column. In the right column, cuts at western sky at $l=330^\circ$ for the three projections (green line) are compared with the observation (red, Ref.[5]), the ``{\bf P20}'' model (blue) and ``{\bf P20} thin'' model (magenta). This figure shows that a ``bubble-shape'' with cap can be reproduced by an open halo because of the projection effect. The intrinsic 3D structure of the X-ray emitting outer halo cannot be inferred from only a 2D projection.}
\label{figext:XrayProjection}
\end{figure}

\clearpage
\newpage

\paragraph*{Data availability}
We use the following surveys in our paper to analyze the magnetic halo: the synchrotron and dust data (\url{https://lambda.gsfc.nasa.gov/}), the Fermi gamma-ray data (\url{https://fermi.gsfc.nasa.gov/ssc/data/access/lat/14yr_catalog/}), the 3D dust extinction map (\url{https://astro.acri-st.fr/gaia_dev/about}), the ROSAT allsky survey (\url{https://cade.irap.omp.eu/dokuwiki/doku.php?id=rass}). The other data are taken from the maps of published papers which we will provide references in the Supplementary material.

\paragraph*{Code availability}
The following software and code packages have been used in our analysis: Python (\url{https://www.python.org/}) with the package Numpy (\url{https://numpy.org/}), Healpy (\url{https://healpy.readthedocs.io/}), Astropy (\url{https://www.astropy.org/}), Fermipy (\url{https://fermipy.readthedocs.io/}); Jupyter Notebook (\url{https://jupyter-notebook.readthedocs.io/}), Matplotlib (\url{https://matplotlib.org/}); and DS9 (\url{https://sites.google.com/cfa.harvard.edu/saoimageds9}). The package ``naima'' (\url{https://naima.readthedocs.io/}) is used to model the multi-wavelength results.
the electron distribution is from the package ``ymw16'' (\url{https://www.atnf.csiro.au/research/pulsar/ymw16/}). The magnetic field model of ``JF12'' comes from the package ``CRPropa'' (\url{https://crpropa.github.io/CRPropa3/api/classcrpropa_1_1PlanckJF12bField.html}).

\noindent{\bf Correspondence and requests for materials} should be addressed to He-Shou Zhang, Gabriele Ponti, Ettore Carretti, Ruo-Yu Liu, or Mark R. Morris.

\paragraph*{Acknowledgements}
HSZ acknowledges support by the X-riStMAs project (Seal of Excellence n. [0000153]) under the National Recovery and Resilience Plan (PNRR), Mission 4, Component 2, Investment 1.2 - Italian Ministry of University and Research, funded by the European Union – NextGenerationEU. HSZ also acknowledges the computing support from PLEIADI supercomputer from INAF. HSZ, GP, NL, XZ, YZ, GS acknowledge financial support from the European Research Council (ERC) under the European Union’s Horizon 2020 research and innovation program HotMilk (grant agreement No. [865637]). GP also acknowledges support from Bando per il Finanziamento della Ricerca Fondamentale 2022 dell’Istituto Nazionale di Astrofisica (INAF): GO Large program and from the Framework per l’Attrazione e il Rafforzamento delle Eccellenze (FARE) per la ricerca in Italia (R20L5S39T9). MM acknowledges support from NASA under ADAP grant 80NSSC24K0639. The authors acknowledge X.Y. Li, X. Wu, M. Sasaki, G. Pareschi, and G. Ghisellini.

\paragraph*{Author Contributions Statement}
H.S.Z. and G.P. led the project. H.S.Z. performed the analysis. E.C, H.S.Z., and M.H. led the radio data analysis and magnetic field measurement. G.P., H.S.Z., N.L., and X.Z. led the X-ray study. R.Y.L., H.S.Z., F.A., H.M.Z., M.R.M., Y.Z., and G.S. led the gamma-ray study. H.S.Z., G.P., E.C., R.Y.L., and M.R.M. led the multi-wavelength comparison and wrote the manuscript. All authors contributed to improving the analysis and the manuscript.

\paragraph*{Competing Interests Statement}
The authors declare no competing interests.

\clearpage
\newpage

\section*{Supplementary}
\setcounter{figure}{0}

\paragraph*{References for data and software}
We use the following surveys in our paper to analyze the magnetic halo: the polarized synchrotron emission surveys at different frequencies -- Haslam408 (0.408~GHz)\cite{Haslam74,Haslam82,Remazeilles15Haslam408}{}; DRAO/Villa-Elisa (1.4~GHz)\cite{DataDRAO1,DataDRAO2}{}, S-PASS (2.3GHz)\cite{Carretti13Nat,SPASS_Carretti19}{}, QUIJOTE (11, 13~GHz)\cite{DataQUIJOTE}{}, WMAP K-Band (22.8~GHz)\cite{DataWMAP1,DataWMAP2}{}, and Planck (30 GHz)\cite{PLANCK18diffuseSepa}{}. To exclude the local  contribution we use the dust  polarized emission at 353~GHz  from Planck \cite{PLANCK18dustPol}{}. The SED analysis made use of the diffuse $\gamma-$ray emission measured by Fermi-LAT with sources subtracted off\cite{Platz22}{}. The diffuse Fermi emission is compared with data from Fermi 14 year data subtracting the 14-year Fermi point source catalog\cite{FermiCatalog4FGLDR3}{}. The HAWC\cite{HAWC2017}{} also provides constraints to the SED. The Gaia-2MASS survey uses the star dust absorption of  the Gaia survey to study the dust structures at different distances\cite{Lallement18}{}. The X-ray emission the energy band 0.6--1.0~keV by eROSITA are used \cite{Predehl20}. The  data of the  energy band 0.11--0.28~keV are from ROSAT\cite{DataROSAT}{}. The star-forming rates of  the Galactic disc are  from the infrared Herschel  Legacy survey\cite{Elia22}{}. The following softwares packages are used for calculations and observational data analysis:
Python\cite{Python} with the package Numpy\cite{numpy1,numpy2}{}, Healpy\cite{healpy1,healpy2}{}, Astropy\cite{Astropy13,Astropy18,Astropy22}{}, Fermipy\cite{Fermipy_Wood17}{}, naima\cite{naima}{}, CRPropa\cite{CRPropa}; Jupyter Notebook\cite{Jupyter}{}, Matplotlib\cite{Matplotlib}{}; and DS9\cite{ds9Joye03}{}.

{In the Supplementary material, we provide the reference list at the end of the document following the numbering of the main text.}

\paragraph*{Radio data in SED analysis}

We calculate the noise through the flux density calibration accuracy by (taking patch R1 subtracted by patch R2 as an example):
\begin{equation}
\begin{array}{ll}
Signal: & I_{signal}=I_{R1}-I_{R2}\\
Noise: & s_{noise,accu}=R_{accuracy}\cdot\sqrt{(I_{R1})^2+(I_{R2})^2} \\
\end{array}
\label{eq:snoise_accu}
\end{equation}

If the flux density calibration accuracy is high, we get the error from the beam sensitivity ($S_{beam}$) by:
\begin{equation}
\begin{array}{ll}
s_{noise,sens}&=\frac{S_{beam}}{\sqrt{N_{beam}}}\\
&=\frac{S_{beam}}{\sqrt{A_{all}/A_{beam}}}\\
&=\frac{S_{beam}}{\sqrt{(N_{pix,all}*A_{pix})/(1.13*FWHM^2)}}\\
\end{array}
\label{eq:snoise_sens}
\end{equation}

We list the data and two types of errors in the Supplementary Table 2 from all 5 patches, and R1-R2, R3-R4, R5-R3 subtractions.

We convert the radio data from temperature T[K] to the spectral flux density $\mathcal{F}(\nu)$[MJy/sr] by
\begin{equation}
\mathcal{F}(\nu)[{\rm MJy/sr}]=\frac{T[{\rm K}] \cdot r_{fac}[{\rm Jy/K}]}{10^6 {\rm Jy/MJy}}\frac{3283 {\rm deg^2/sr}}{1.13\cdot(FWHM)^2 {\rm deg^2}},
\label{eq:radiofluxConvert}
\end{equation}
where $r_{fac}$ is the conversion factor.

For radio data, we summarize the conversion methodologies below:

For 0.408~GHz, we use the 2014-Reprocessed Haslam 408 MHz\cite{Remazeilles15Haslam408}{} from Ref.\cite{Haslam74,Haslam82}{}, $r_{fac}=5114$~Jy/sr~K$^{-1}$, accuracy $10\%$.

For 1.4~GHz, we use the Stockert/DRAO survey\cite{DatastockertReich86,DatastockertReich01}{}, $r_{fac}=11.25$~Jy/K, accuracy $5\%$.

For 11~GHz, we use the Quijote survey\cite{DataQUIJOTE}{},  $r_{fac}=\frac{1}{961.9\mu{\rm K}/{\rm Jy}}$, accuracy $5\%$.

For 13~GHz, we use the Quijote survey\cite{DataQUIJOTE}{}, $r_{fac}=\frac{1}{703.8\mu{\rm K}/{\rm Jy}}$, accuracy $5\%$.

For 23~GHz, we use the synchrotron separation from WMAP\cite{DataWMAP2}{} for data and error, $r_{fac}=\frac{1}{250.6\mu{\rm K}/{\rm Jy}}$.

For 30~GHz, we use the synchrotron separation from Planck survey\cite{DataPLANCKCommander}{}. $r_{fac}=24.33$~MJy/sr$\cdot $K$^{-1}$.

Note that for 23~GHz and 30~GHz, we use the synchrotron component from WMAP\cite{WMAPdata}{} and Planck\cite{DataPLANCKCommander}{} surveys, respectively.

\paragraph*{X-ray-emitting halo re-analysis}

As a pure geometric check, we try to test the expected X-ray emission maps from the ``Bouquet Model''. We take the surface brightness measured from the paper of Predehl+2020\cite{Predehl20} (here after {\bf P20}) by taking the all-sky map in 0.6-1.0~keV energy band.
The observed X-ray flux is assumed to be produced by the emission of thermal collision of hot plasma with the temperature of $T$ (with the emissivity of $\epsilon(T)$), and the observed surface brightness along the direction (l,b) can be expressed by (from Ref.\cite{MillerBregman15}):
\begin{equation}\label{eq:surfBright}
I_{X}(l,b)\propto \int dl_{depth} n_H^2 \epsilon(T).
\end{equation}
And the averaged surface brightness of the Bubbles is:
\begin{equation}\label{eq:brightX}
B_{X} \propto \frac{\int \int dl_{depth} d\theta\cdot n_H^2\epsilon(T)}{\Omega_{Bubble}},\\
\end{equation}
where $l_{depth}$ is the emitting depth, $d\theta$ is the solid angle for the size of a pixel, $k_B$ is Boltzmann's constant, and $\Omega_{Bubble}$ is the solid angle of the emitting region observed in the sky.
In our work, the outer outflows is modelled by the ``Bouquet Model'', which we simplify into the following geometry (Supplementary Figure~\ref{figSup:XrayEmission}a) as an example:
we assume the outer outflow in each side of the Galactic disc has the shape of an up-side-down truncated cone with a vacant cylindrical center. The radius at the bottom is 5~kpc and at the top is 8~kpc. The vacant cylinder has a radius of 3~kpc. The geometry in the disc is an annulus as extended as the star-forming ring of the Milky Way extending approximately at 3-5~kpc Galactic Centric.
The contribution of the Fermi Bubbles are not considered in the modelling.

As a comparison, we also rebuild the {\bf P20} model, where the X-ray outer halo are modelled as two spherical bubble shells. The outer radius in {\bf P20} model is set at 7~kpc and different parameters for the inner radius (3 and 5~kpc) were tested in Predehl+\cite{Predehl20}. Here we reproduce the {\bf P20} modelling by two cases: 1) inner radius at 3~kpc (denoted as {\bf P20} model); 2) inner radius at 5~kpc (denoted as {\bf P20 thin} model).
In Supplementary Figure~\ref{figSup:XrayEmission}, we show the geometry of the modellings and their projected surface brightness in the all-sky map.
In comparison, we compare the modelling the X-ray observation from 0.6-1.0~keV in the longitude cut at $l=10^\circ$ (In Supplementary Figure~\ref{figSup:XrayEmission}g) and $l=340^\circ$ (In Supplementary Figure~\ref{figSup:XrayEmission}h).

We consider that the averaged surface brightness resulting from  the ``Bouquet Model'' is the same as what measured for the eROSITA Bubbles but the emitting geometry is defined by the new modelling ($B_{X,new}=B_{X,P20}$). We assume that in the ``Bouquet Model'', the density is uniform, the temperature and the metallicity in the outer outflow are T=0.3~keV, metallicity is 0.2 solar\cite{kataoka13,Predehl20}.
The total energy in the emitting plasma can be estimated by $E_{tot}=n_H k_B T V$, where $V$ is the volume of the emitting plasma.
Hence, we obtain the total energy in the outer outflow ($E_{tot,new}$) based on the total energy calculation based on the {\bf P20} modelling:
\begin{equation}\label{eq:E_ratio}
\begin{array}{ll}
\frac{\int \int dl_{depth} d\theta_{new}}{\int d\theta }\cdot&\frac{E_{tot,new}^2}{V_{new}^2}=\frac{\int \int dl_{depth} d\theta_{old}}{\int d\theta }\cdot\frac{E_{tot,P20}^2}{V_{old}^2}\\
&\frac{E_{tot,new}}{E_{tot,P20}}=\frac{ V_{new}/\sqrt{\int \int dl_{depth} d\theta_{new}/\Omega_{new}}}{V_{old}/\sqrt{\int \int dl_{depth} d\theta_{old}/\Omega_{old}}}\\
\end{array}
\end{equation}
Here, $\int dl_{depth}$ is the length depth along different lines-of-sight, and $\int d\theta$ is the sky area that the projected model occupies. The footnote ``new'' is for the ``Bouquet Model'' and ``P20'' is for the {\bf P20} model. The energy for different Bouquet heights are summarized in Extended Data Table 1.

We make comparison for the observed surface brightness in Supplementary Figure~\ref{figSup:XrayEmission}(d-h), with the three modellings: 1) the ``Bouquet Model'' (height at 4~kpc); 2) {\bf P20} model (inner radius 3~kpc); 3) {\bf P20 thin} model (inner radius 5~kpc).
It is not surprising that the inner part ($l\lesssim15^\circ$) shows lower surface brightness in our model when compared with the data. Indeed, we attribute such mismatch to the contribution from the Fermi Bubbles which have been observed to have enhanced X-ray emission at the roots\cite{Bland-Hawthorn03,Su10Fermi,Ponti19}{}.
For the cut of the outer outflow ($l=-30^\circ$), the lower-/mid-latitudes ($\mid b\mid<45^\circ$) are better reproduced by the ``Bouquet Model''. Also, the southern sky is better reproduced by the ``Bouquet Model'' comparing to the previous models. On the other hand, the enhancement in the northern sky at high latitude (in NPS) is better reproduced by {\bf P20 thin} model and is not present in the ``Bouquet Model''. This indicates that either part of X-ray emission from the NPS is not associated with the Galactic outflows or that the NPS requires additional explanation in the ``Bouquet Model''.

\paragraph*{Remarks on the electron density model selections} We note that the electron distribution from ``ymw16''\cite{ymw16ne}{} uses the dispersion measure of pulsars and fast radio bursts and includes the components such as the spiral arms, thin and thick disc. Hence, the ``ymw16'' model is advantageous to represent the electron distribution within the Galactic disc. This is suitable for the calculations in this section of the Faraday rotation depolarization analysis due to the Galactic disc as foreground. However, the ``ymw16'' model does not contain the component of X-ray-emitting warm-hot plasma in the circumgalactic medium beyond the Galactic disc. In our plasma-beta calculations for the halo presented in the main text, we employed the latest electron distribution data Ref.\cite{Locatelli24neDensityMW}, derived from an analysis of X-ray observations conducted with eROSITA. Ref.\cite{Locatelli24neDensityMW}{} model the warm-hot plasma density at moderate-high Galactic latitudes and towards the anti-center direction. We extrapolate their analytic model towards the central regions of the Galaxy. We note that the Galactic outflows possibly introduce a deviation in the central regions of the Milky Way. If we adopt a smaller electron density in the X-ray halo following Ref.\cite{Predehl20}{}($2\times10^{-3}$~cm$^{-3}$), the plasma betas are $\beta_{r1}\simeq6$ and $\beta_{r3}\simeq12$, respectively. Patches R1 and R3 in the outer outflows are still in high plasma beta regime.

\newpage

\begin{figure}
\begin{center}
\includegraphics[width=1.0\textwidth]{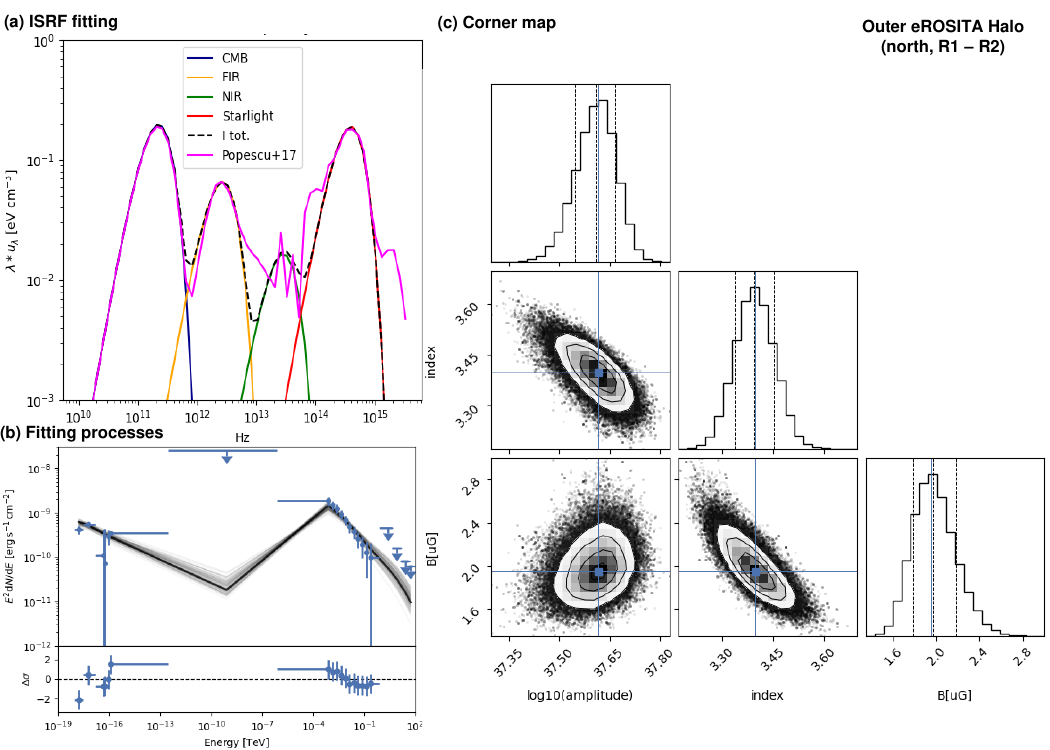}
\end{center}
\caption{{\bf Supplementary Figure 1 $\mid$ The fitting details for northeastern outer outflows.} These calculations correspond to the SED fitting in Figure 4c for the patch R1 - patch R2. (a) the seed photon field through the fitting of the ISRF+CMB from Ref. \cite{Popescu17ISRF}. (b) The fitting processes for the MCMC fitting (different grey lines) and the data deviation ($\Delta\sigma$) compared to the best fit.  {The error-bars reported here for the gamma-ray data are based on the statistical uncertainties (see \S~3.1 in the Methods section for details). The reference frequencies are listed in the Supplementary Table 2. The error-bars reported for the radio fluxes are calculated based on the flux density calibration accuracy and beam sensitivity of the corresponding surveys, as defined in the Supplementary. The measure center of the bottom sub-figure is the observational data deviation (y-axis) with respect to the reference energy (x-axis) compared to the best fit (the black fitting curve above). The grey shadowed lines represent fitting curves with different investigated parameters within the MCMC process.} The data dispersion is within 2-$\sigma$ to the best fit. (c) the corner map for the MCMC fitting.
}
\label{figSup:NEoutereRASS}
\end{figure}

\begin{figure}
\begin{center}
\includegraphics[width=1.0\textwidth]{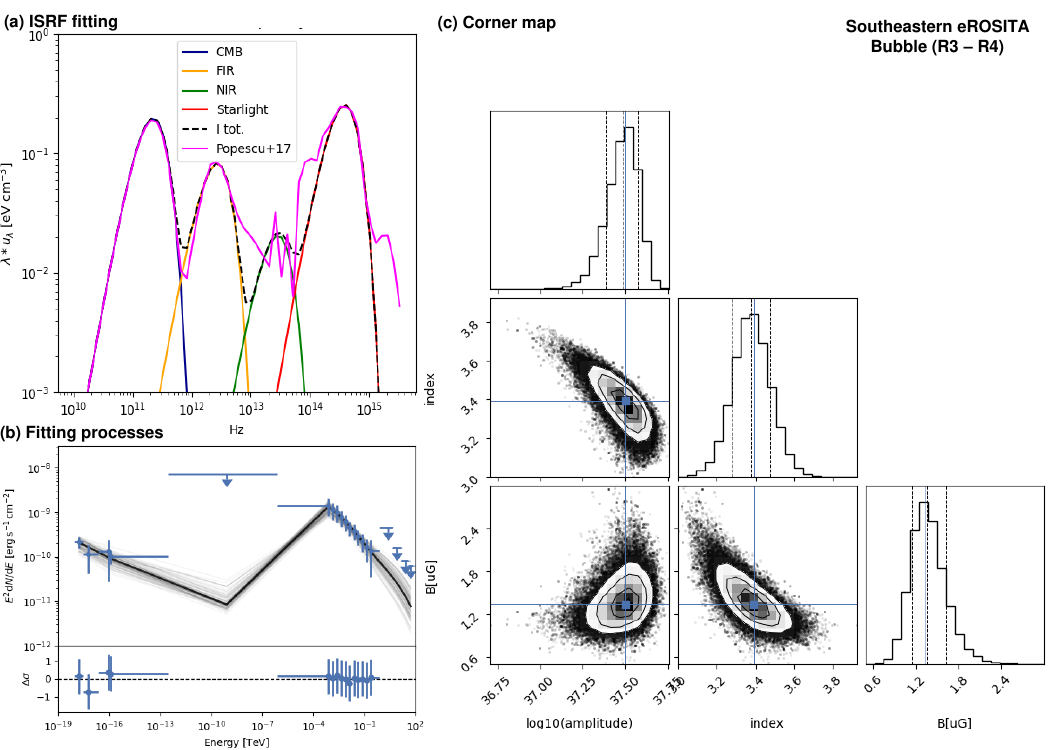}
\end{center}
\caption{{\bf Supplementary Figure 2 $\mid$ The fitting details for southeastern outer outflows.} Same as Supplementary Figure 1 but for the SED fitting in Figure 4d for the patch R3 - patch R4. The data dispersion is within 1-$\sigma$ to the best fit.
}
\label{figSup:SEoutereRASS}
\end{figure}

\begin{figure}
\begin{center}
\includegraphics[width=1.0\textwidth]{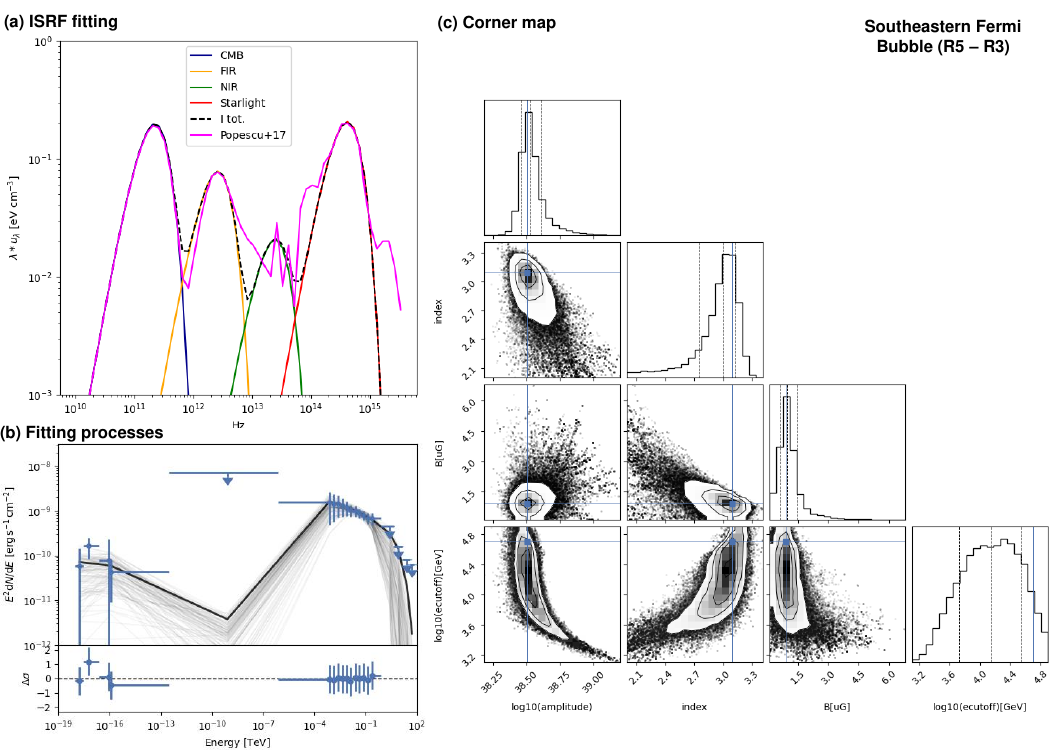}
\end{center}
\caption{{\bf Supplementary Figure 3 $\mid$ The fitting details for southeastern cap of the Fermi Bubble.} Same as Supplementary Figure 1 but for the SED fitting in Figure 4e for the patch R5 - patch R3. The data dispersion is within 2-$\sigma$ to the best fit.
}
\label{figSup:SEFermiBubble}
\end{figure}

\begin{figure}
\begin{center}
\includegraphics[width=0.9\textwidth]{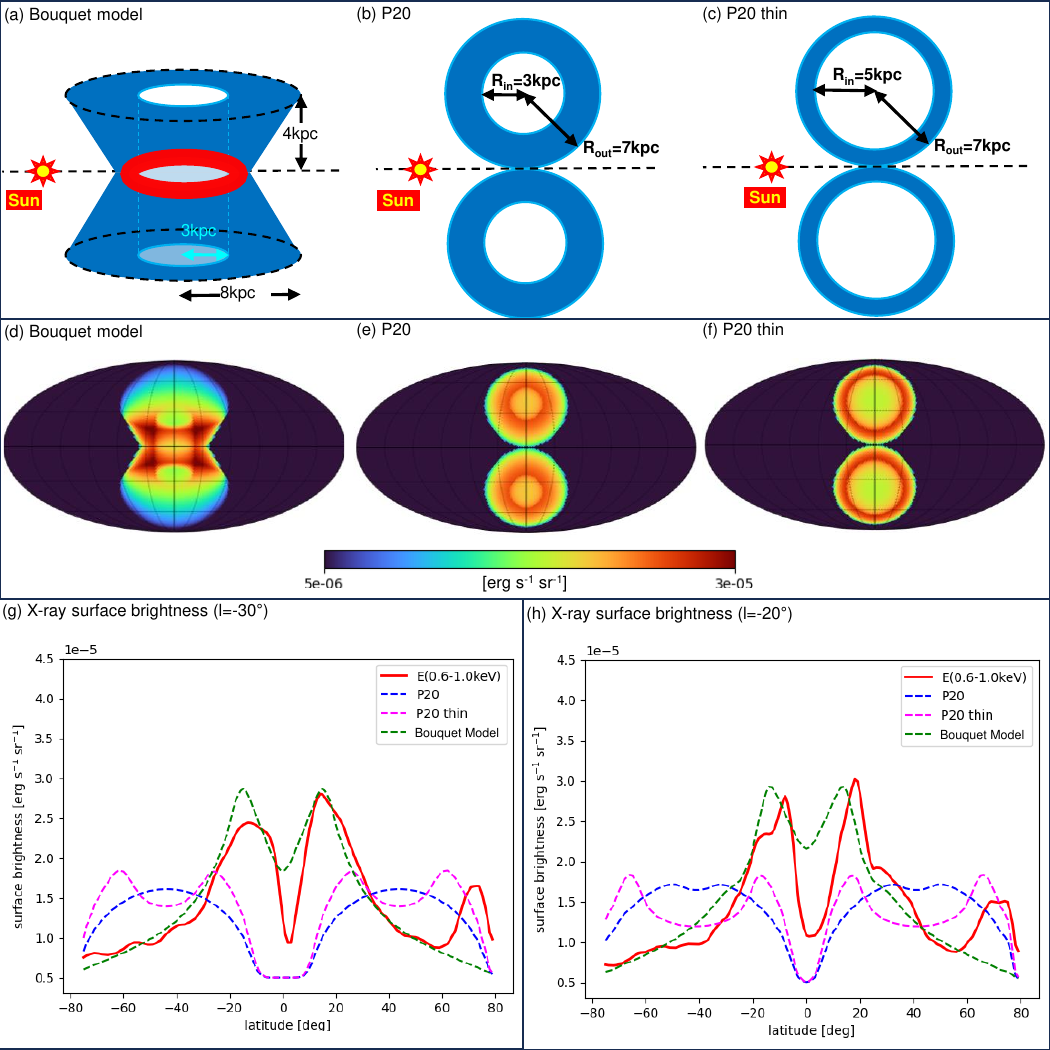}
\end{center}
\caption{{\bf Supplementary Figure 4 $\mid$ The comparison between the ``Bouquet Model'' and the ``{\bf P20}'' Model.} The geometries are described in (a) the simplified ``Bouquet Model'' with upper radius $8$~kpc, the root with 5~kpc Galactic Centric, and the height of 4~kpc. (b) ``{\bf P20}'' Model with the inner radius of 3~kpc and the outer radius of 7~kpc; (c) ``{\bf P20} thin'' Model with the inner radius of 5~kpc and the outer radius of 7~kpc. The expected X-ray emitting maps for 0.6-1.0~keV are presented in (d-f). Positive longitude is influenced by the dust absorption (see Extended Data Figure 2d) so it is not included. Eastern cuts are presented at (g) $l=30^\circ$ and (h) $l=20^\circ$ for the X-ray fluxes at 0.6-1.0~keV for eROSITA measurement (red), the ``Bouquet Model'' (green), ``{\bf P20}'' Model (blue), and ``{\bf P20} thin'' Model (magenta). (h) western cuts at $l=340^\circ$. The ``Bouquet Model'' predicts a decreasing gradient for the X-ray flux from the disc upwards, which provides a clear better fit at low-/mid-latitudes ($\mid b \mid<45^\circ$) and the whole southern sky. In the north side of the North Polar Spur (NPS), the {\bf P20 thin} model is better. In fact, the {\bf P20 thin} model always predicts a large enhancement at high Galactic latitudes, which is only observed in the NPS. }
\label{figSup:XrayEmission}
\end{figure}

\begin{sidewaystable}
\small
\centering
\begin{tabular}{|c|c|c|c|c|c|c|}
\hline \hline
\multicolumn{7}{|l|}{(a) {\bf Seed photon field Modelling}}\\
\hline
\multirow{2}{*}{}  & \multicolumn{2}{c|}{R1-R2} & \multicolumn{2}{c|}{R3-R4} & \multicolumn{2}{c|}{R5-R3}\\
\cline{2-7}
 & T$_{i}$ & $\rho_{E}$ &  T$_{i}$ & $\rho_{E}$ &  T$_{i}$ & $\rho_{E}$ \\
 & [K] & [eV/cm$^{-3}$]& [K] &  [eV/cm$^{-3}$] &  [K] & [eV/cm$^{-3}$] \\
\hline
CMB & 2.73 & 0.24 & 2.73 &  0.24 & 2.73 & 0.24 \\
\hline
FIR  & 33 & 0.08 & 33 & 0.10 & 32 & 0.095 \\
\hline
NIR & 350 & 0.02 & 350 & 0.025 & 300 & 0.025 \\
\hline
Starlight & 4800 & 0.23 & 4750 & 0.31 & 5000 & 0.25 \\
\hline \hline
\multicolumn{7}{|l|}{(b) {\bf SED fitting results}}\\
\hline
\multirow{2}{*}{{\bf Single-Power Law}} & \multicolumn{2}{|c|}{R1-R2} & \multicolumn{2}{|c|}{R3-R4} &  \multicolumn{2}{|c|}{R5-R3}\\
\cline{2-7}
 & Platz+ & Fermi14yr & Platz+ & Fermi14yr  & \multicolumn{2}{|c|}{} \\
\cline{1-5}
Amp [$10^{37}$~eV$^{-1}$] & $4.10\pm0.60$ & $5.10\pm0.60$ & $3.10\pm0.60$ & $2.80\pm0.50$ & \multicolumn{2}{|c|}{} \\
\cline{1-5}
B [~$\mu{\mathrm G}$] & $1.97\pm0.20$ & $1.72\pm0.10$ & $1.40\pm0.20$ & $1.36\pm0.15$ & \multicolumn{2}{|c|}{cannot be fitted} \\
\cline{1-5}
Index  & $-3.40\pm0.06$ & $-3.41\pm0.04$ & $-3.38\pm0.10$ & $-3.40\pm0.06$ & \multicolumn{2}{|c|}{} \\
\hline
{\bf Single-Power Law}+  &  \multicolumn{2}{|c|}{R1-R2} &  \multicolumn{2}{|c|}{R3-R4}&  \multicolumn{2}{|c|}{R5-R3}\\
\cline{2-7}
{\bf exponential cutoff} & Platz+ & Fermi14yr & Platz+ & Fermi14yr & Platz+ & Fermi14yr \\
\hline
Amp [$10^{37}$~eV$^{-1}$]& $10^{+5}_{-3}$ & $11^{+4}_{-3}$ & $3.5^{+1.5}_{-0.8}$ & $3.2^{+1.4}_{-0.7}$ & $33^{+8}_{-4}$ & $63\pm6$ \\
\hline
B [~$\mu{\mathrm G}$] & $1.90\pm0.20$ & $1.81\pm0.13$ & $1.4^{+0.3}_{-0.2}$ & $1.40\pm0.16$ & $1.0^{+0.5}_{-0.4}$ & $0.9\pm0.3$ \\
\hline
Index  & $-3.21\pm0.09$ & $-3.20\pm0.09$ & $-3.33\pm0.13$ & $-3.36^{+0.11}_{-0.08}$ & $-3.00^{+0.3}_{-0.13}$ & $-2.86\pm0.06$ \\
\hline
E$_{cutoff}$ [~TeV] & $0.7^{+0.9}_{-0.4}$ & $0.8^{+0.5}_{-0.3}$ & $11^{+40}_{-9}$ & $5^{+30}_{-4}$ & $14^{+20}_{-9}$ & $4.4^{+1.4}_{-0.9}$ \\
\hline \hline
\multicolumn{7}{|l|}{(c) {\bf Star formation rate for clumps in the inner Galactic disc}}\\
\hline
Longitude & $+33^\circ$ & $+27^\circ$ &   $+16^\circ$ &     $351^\circ$ &  $337^\circ$ &  $330^\circ$  \\
\hline
$\Sigma_{SFR}$[~$M_\odot yr^{-1}$] & $0.22$ & $0.17$ &   $0.15$ &     $0.11$ &  $0.16$ &  $0.12$  \\
\hline \hline
\end{tabular}
\caption{{\bf Supplementary Table 1 $\mid$ Results summary.} (a) seed photon field modelling. We obtain the radiation modelling from Ref. \cite{Popescu17ISRF}{} (ISRF+CMB) and simplify the results with 4 blackbody radiations as seed photon field for SED studies. $T_i$ is the corresponding temperature for the blackbody radiation and $\rho_E$ is the energy density of the photon field. The fitting curve is presented in Supplementary Figure 1-3a. (b) SED fitting results. Two origins of $\gamma-$ray diffuse emissions are considered: Platz+\cite{Platz22} and Fermi14yr\cite{FermiCatalog4FGLDR3}{}. The SED fitting steps are presented in Supplementary Figure 1-3b and the resulting corner maps are in Supplementary Figure 1-3c. The fitting parameters include: ``Amp [$10^{37}$~eV$^{-1}$]'' -- the fitting amplitude, ``B [~$\mu{\mathrm G}$]'' -- the magnetic field strength, ``index'' -- the electron index, and E$_{cutoff}$ [~TeV] -- cutoff energy (only for the fit with exponential cutoff). (c) Star formation rates along the lines of sight with high rate directions in a radius of $3^\circ$. Only the inner Galactic disc where the eROSITA Bubbles are rooted from was considered ($l\lesssim\mid 35^\circ\mid$, see Extended Data Figure 8). The numbers are obtained from Ref.\cite{Elia22}{} based on the observations by Herschel Telescope data.
}
\label{tab:results}
\end{sidewaystable}

\begin{sidewaystable}
\centering
\begin{tabular}{|c|c|c|c|c|c|c|c|c|}
\hline \hline
{\bf $0.408$~GHz} & patch R1 & patch R2 & patch R3 & patch R4 & patch R5 & R1-R2 & R3-R4 & R5-R3\\
\hline
Signal &0.20494 & 0.10305 & 0.13333 & 0.08102 & 0.14737 & 0.10189 & 0.05231 & 0.01404\\
\hline
Err$_{accuracy}$ & 0.01025 & 0.00515 & 0.00667 & 0.00405 & 0.00737 & 0.01147 & 0.00780 & 0.00994\\
\hline
Err$_{sensitivity}$ & 0.00038 & 0.00044 & 0.00033 & 0.00037 & 0.00037 & 0.00058 & 0.00050 & 0.00050\\
\hline
{\bf $1.4$~GHz} & patch R1 & patch R2 & patch R3 & patch R4 & patch R5 & R1-R2 & R3-R4 & R5-R3\\
\hline
Signal &0.09191 & 0.05230 & 0.07704 & 0.06887 & 0.08884 & 0.03961 & 0.00817 & 0.01180\\
\hline
Err$_{accuracy}$ & 0.00460 & 0.00261 & 0.00385 & 0.00344 & 0.00444 & 0.00529 & 0.00517 & 0.00588\\
\hline
Err$_{sensitivity}$ & 0.00022 & 0.00025 & 0.00019 & 0.00022 & 0.00022 & 0.00034 & 0.00029 & 0.00029\\
\hline
{\bf $11$~GHz} & patch R1 & patch R2 & patch R3 & patch R4 & patch R5 & R1-R2 & R3-R4 & R5-R3\\
\hline
Signal & 0.04198 & 0.04094 & null& null & null& 0.00103 & null & null\\
\hline
Err$_{accuracy}$ & 0.00210 & 0.00205 & null& null & null& 0.00293 & null& null\\
\hline
Err$_{sensitivity}$ & 0.00002 & 0.00002 & null & null & null& 0.00003 & null& null\\
\hline
{\bf $13$~GHz} & patch R1 & patch R2 & patch R3 & patch R4 & patch R5 & R1-R2 & R3-R4 & R5-R3\\
\hline
Signal & 0.03908 & 0.03851 & null& null & null& 0.00057 & null & null\\
\hline
Err$_{accuracy}$ & 0.00195 & 0.00193 & null& null & null& 0.00274 & null& null\\
\hline
Err$_{sensitivity}$ & 0.00002 & 0.00002 & null & null & null& 0.00003 & null& null\\
\hline
{\bf $23$~GHz} syn & patch R1 & patch R2 & patch R3 & patch R4 & patch R5 & R1-R2 & R3-R4 & R5-R3\\
\hline
Signal & 0.00172 & 0.00041 & 0.00103 & 0.00045 & 0.00137 & 0.00131 & 0.00058 & 0.00034\\
\hline
Err$_{tot}$ & 0.00044 & 0.00020 & 0.00039 & 0.00025 & 0.00054 & 0.00049 & 0.00046 & 0.00067\\
\hline
{\bf $30$~GHz} syn & patch R1 & patch R2 & patch R3 & patch R4 & patch R5 & R1-R2 & R3-R4 & R5-R3\\
\hline
Signal & 0.00173 & 0.00061 & 0.00084 & 0.00051 & 0.00099 & 0.00112 & 0.00033 & 0.00015\\
\hline
Err$_{accuracy}$ & 0.00009 & 0.00003 & 0.00004 & 0.00003 & 0.00005 & 0.00009 & 0.00005 & 0.00007\\
\hline
Err$_{sensitivity}$ & 0.00009 & 0.00011 & 0.00008 & 0.00009 & 0.00009 & 0.00014 & 0.00012 & 0.00012\\
\hline \hline
\end{tabular}
\caption{{\bf Supplementary Table  2 $\mid$ Radio data for SED analysis.} References for the data survey are listed in the Methods.
Errors from the flux density calibration accuracy and errors from the beam sensitivity are listed with units ~MJy/sr. The patches are listed in Figure 4.
}
\label{tab:sed}
\end{sidewaystable}

\newpage

\clearpage

\bibliographystyle{sn-nature}
\bibliography{Outflows}

\begin{thebibliography}{100}
\expandafter\ifx\csname url\endcsname\relax
  \def\url#1{\burl{#1}}\fi
\expandafter\ifx\csname urlprefix\endcsname\relax\def\urlprefix{URL }\fi
\providecommand{\bibinfo}[2]{#2}
\providecommand{\eprint}[2][]{\url{#2}}
\providecommand{\doi}[1]{\url{https://doi.org/#1}}
\bibcommenthead

\bibitem{Strickland04xHalo}
\bibinfo{author}{{Strickland}, D.~K.}, \bibinfo{author}{{Heckman}, T.~M.},
  \bibinfo{author}{{Colbert}, E. J.~M.}, \bibinfo{author}{{Hoopes}, C.~G.} \&
  \bibinfo{author}{{Weaver}, K.~A.}
\newblock \bibinfo{title}{{A High Spatial Resolution X-Ray and
  H${\ensuremath{\alpha}}$ Study of Hot Gas in the Halos of Star-forming Disk
  Galaxies. I. Spatial and Spectral Properties of the Diffuse X-Ray Emission}}.
\newblock \emph{\bibinfo{journal}{Astrophys. J. S.}}
  \textbf{\bibinfo{volume}{151}}, \bibinfo{pages}{193--236}
  (\bibinfo{year}{2004}).

\bibitem{StricklandHeckman07xHalo}
\bibinfo{author}{{Strickland}, D.~K.} \& \bibinfo{author}{{Heckman}, T.~M.}
\newblock \bibinfo{title}{{Iron Line and Diffuse Hard X-Ray Emission from the
  Starburst Galaxy M82}}.
\newblock \emph{\bibinfo{journal}{Astrophys. J.}}
  \textbf{\bibinfo{volume}{658}}, \bibinfo{pages}{258--281}
  (\bibinfo{year}{2007}).

\bibitem{Krause20ChangesXXII}
\bibinfo{author}{{Krause}, M.} \emph{et~al.}
\newblock \bibinfo{title}{{CHANG-ES. XXII. Coherent magnetic fields in the
  halos of spiral galaxies}}.
\newblock \emph{\bibinfo{journal}{Astron. Astrophys.}}
  \textbf{\bibinfo{volume}{639}}, \bibinfo{pages}{A112} (\bibinfo{year}{2020}).

\bibitem{Krause19halo}
\bibinfo{author}{{Krause}, M.}
\newblock \bibinfo{title}{{Magnetic Fields and Halos in Spiral Galaxies}}.
\newblock \emph{\bibinfo{journal}{Galaxies}} \textbf{\bibinfo{volume}{7}},
  \bibinfo{pages}{54} (\bibinfo{year}{2019}).

\bibitem{Predehl20}
\bibinfo{author}{{Predehl}, P.} \emph{et~al.}
\newblock \bibinfo{title}{{Detection of large-scale X-ray bubbles in the Milky
  Way halo}}.
\newblock \emph{\bibinfo{journal}{Nature}} \textbf{\bibinfo{volume}{588}},
  \bibinfo{pages}{227--231} (\bibinfo{year}{2020}).

\bibitem{DataWMAP1}
\bibinfo{author}{{Hinshaw}, G.} \emph{et~al.}
\newblock \bibinfo{title}{{Nine-year Wilkinson Microwave Anisotropy Probe
  (WMAP) Observations: Cosmological Parameter Results}}.
\newblock \emph{\bibinfo{journal}{Astrophys. J., Suppl. Ser.}}
  \textbf{\bibinfo{volume}{208}}, \bibinfo{pages}{19} (\bibinfo{year}{2013}).

\bibitem{Vidal15}
\bibinfo{author}{{Vidal}, M.}, \bibinfo{author}{{Dickinson}, C.},
  \bibinfo{author}{{Davies}, R.~D.} \& \bibinfo{author}{{Leahy}, J.~P.}
\newblock \bibinfo{title}{{Polarized radio filaments outside the Galactic
  plane}}.
\newblock \emph{\bibinfo{journal}{Mon. Not. R. Astron. Soc.}}
  \textbf{\bibinfo{volume}{452}}, \bibinfo{pages}{656--675}
  (\bibinfo{year}{2015}).

\bibitem{Carretti13Nat}
\bibinfo{author}{{Carretti}, E.} \emph{et~al.}
\newblock \bibinfo{title}{{Giant magnetized outflows from the centre of the
  Milky Way}}.
\newblock \emph{\bibinfo{journal}{Nature}} \textbf{\bibinfo{volume}{493}},
  \bibinfo{pages}{66--69} (\bibinfo{year}{2013}).

\bibitem{Wolleben21}
\bibinfo{author}{{Wolleben}, M.} \emph{et~al.}
\newblock \bibinfo{title}{{The Global Magneto-ionic Medium Survey: A Faraday
  Depth Survey of the Northern Sky Covering 1280-1750 MHz}}.
\newblock \emph{\bibinfo{journal}{Astron. J.}} \textbf{\bibinfo{volume}{162}},
  \bibinfo{pages}{35} (\bibinfo{year}{2021}).

\bibitem{Liu17LHB}
\bibinfo{author}{{Liu}, W.} \emph{et~al.}
\newblock \bibinfo{title}{{The Structure of the Local Hot Bubble}}.
\newblock \emph{\bibinfo{journal}{Astrophys. J.}}
  \textbf{\bibinfo{volume}{834}}, \bibinfo{pages}{33} (\bibinfo{year}{2017}).

\bibitem{Berkhuijsen71SNR}
\bibinfo{author}{{Berkhuijsen}, E.~M.}, \bibinfo{author}{{Haslam}, C.~G.~T.} \&
  \bibinfo{author}{{Salter}, C.~J.}
\newblock \bibinfo{title}{{Are the galactic loops supernova remnants?}}
\newblock \emph{\bibinfo{journal}{Astron. Astrophys.}}
  \textbf{\bibinfo{volume}{14}}, \bibinfo{pages}{252} (\bibinfo{year}{1971}).

\bibitem{Lallement18}
\bibinfo{author}{{Lallement}, R.} \emph{et~al.}
\newblock \bibinfo{title}{{Three-dimensional maps of interstellar dust in the
  Local Arm: using Gaia, 2MASS, and APOGEE-DR14}}.
\newblock \emph{\bibinfo{journal}{Astron. Astrophys.}}
  \textbf{\bibinfo{volume}{616}}, \bibinfo{pages}{A132} (\bibinfo{year}{2018}).

\bibitem{Sofue15mn}
\bibinfo{author}{{Sofue}, Y.}
\newblock \bibinfo{title}{{The North Polar Spur and Aquila Rift}}.
\newblock \emph{\bibinfo{journal}{Mon. Not. R. Astron. Soc.}}
  \textbf{\bibinfo{volume}{447}}, \bibinfo{pages}{3824--3831}
  (\bibinfo{year}{2015}).

\bibitem{Burn66}
\bibinfo{author}{{Burn}, B.~J.}
\newblock \bibinfo{title}{{On the depolarization of discrete radio sources by
  Faraday dispersion}}.
\newblock \emph{\bibinfo{journal}{Mon. Not. R. Astron. Soc.}}
  \textbf{\bibinfo{volume}{133}}, \bibinfo{pages}{67} (\bibinfo{year}{1966}).

\bibitem{Sokoloff98}
\bibinfo{author}{{Sokoloff}, D.~D.} \emph{et~al.}
\newblock \bibinfo{title}{{Depolarization and Faraday effects in galaxies}}.
\newblock \emph{\bibinfo{journal}{Mon. Not. R. Astron. Soc.}}
  \textbf{\bibinfo{volume}{299}}, \bibinfo{pages}{189--206}
  (\bibinfo{year}{1998}).

\bibitem{Lallement22}
\bibinfo{author}{{Lallement}, R.}
\newblock \bibinfo{title}{{North Polar Spur/Loop I: gigantic outskirt of the
  Northern Fermi bubble or nearby hot gas cavity blown by supernovae?}}
\newblock \emph{\bibinfo{journal}{Comptes Rendus Physique}}
  \textbf{\bibinfo{volume}{23}}, \bibinfo{pages}{1--24} (\bibinfo{year}{2023}).

\bibitem{Platz22}
\bibinfo{author}{{Scheel-Platz}, L.~I.} \emph{et~al.}
\newblock \bibinfo{title}{{Multicomponent imaging of the Fermi gamma-ray sky in
  the spatio-spectral domain}}.
\newblock \emph{\bibinfo{journal}{Astron. Astrophys.}}
  \textbf{\bibinfo{volume}{680}}, \bibinfo{pages}{A2} (\bibinfo{year}{2023}).

\bibitem{Dobler10haze}
\bibinfo{author}{{Dobler}, G.}, \bibinfo{author}{{Finkbeiner}, D.~P.},
  \bibinfo{author}{{Cholis}, I.}, \bibinfo{author}{{Slatyer}, T.} \&
  \bibinfo{author}{{Weiner}, N.}
\newblock \bibinfo{title}{{The Fermi Haze: A Gamma-ray Counterpart to the
  Microwave Haze}}.
\newblock \emph{\bibinfo{journal}{Astrophys. J.}}
  \textbf{\bibinfo{volume}{717}}, \bibinfo{pages}{825--842}
  (\bibinfo{year}{2010}).

\bibitem{Planck13haze}
\bibinfo{author}{{Planck Collaboration}} \emph{et~al.}
\newblock \bibinfo{title}{{Planck intermediate results. IX. Detection of the
  Galactic haze with Planck}}.
\newblock \emph{\bibinfo{journal}{Astron. Astrophys.}}
  \textbf{\bibinfo{volume}{554}}, \bibinfo{pages}{A139} (\bibinfo{year}{2013}).

\bibitem{Lacki14}
\bibinfo{author}{{Lacki}, B.~C.}
\newblock \bibinfo{title}{{The Fermi bubbles as starburst wind termination
  shocks.}}
\newblock \emph{\bibinfo{journal}{Mon. Not. R. Astron. Soc.}}
  \textbf{\bibinfo{volume}{444}}, \bibinfo{pages}{L39--L43}
  (\bibinfo{year}{2014}).

\bibitem{Crocker15FermiMultiCMZ}
\bibinfo{author}{{Crocker}, R.~M.}, \bibinfo{author}{{Bicknell}, G.~V.},
  \bibinfo{author}{{Taylor}, A.~M.} \& \bibinfo{author}{{Carretti}, E.}
\newblock \bibinfo{title}{{A Unified Model of the Fermi Bubbles, Microwave
  Haze, and Polarized Radio Lobes: Reverse Shocks in the Galactic
  Center{\textquoteright}s Giant Outflows}}.
\newblock \emph{\bibinfo{journal}{Astrophys. J.}}
  \textbf{\bibinfo{volume}{808}}, \bibinfo{pages}{107} (\bibinfo{year}{2015}).

\bibitem{Su10Fermi}
\bibinfo{author}{{Su}, M.}, \bibinfo{author}{{Slatyer}, T.~R.} \&
  \bibinfo{author}{{Finkbeiner}, D.~P.}
\newblock \bibinfo{title}{{Giant Gamma-ray Bubbles from Fermi-LAT: Active
  Galactic Nucleus Activity or Bipolar Galactic Wind?}}
\newblock \emph{\bibinfo{journal}{Astrophys. J.}}
  \textbf{\bibinfo{volume}{724}}, \bibinfo{pages}{1044--1082}
  (\bibinfo{year}{2010}).

\bibitem{Ackermann14FermiSpect}
\bibinfo{author}{{Ackermann}, M.} \emph{et~al.}
\newblock \bibinfo{title}{{The Spectrum and Morphology of the Fermi Bubbles}}.
\newblock \emph{\bibinfo{journal}{Astrophys. J.}}
  \textbf{\bibinfo{volume}{793}}, \bibinfo{pages}{64} (\bibinfo{year}{2014}).

\bibitem{Remazeilles15Haslam408}
\bibinfo{author}{{Remazeilles}, M.}, \bibinfo{author}{{Dickinson}, C.},
  \bibinfo{author}{{Banday}, A.~J.}, \bibinfo{author}{{Bigot-Sazy}, M.~A.} \&
  \bibinfo{author}{{Ghosh}, T.}
\newblock \bibinfo{title}{{An improved source-subtracted and destriped 408-MHz
  all-sky map}}.
\newblock \emph{\bibinfo{journal}{Mon. Not. R. Astron. Soc.}}
  \textbf{\bibinfo{volume}{451}}, \bibinfo{pages}{4311--4327}
  (\bibinfo{year}{2015}).

\bibitem{DataDRAO1}
\bibinfo{author}{{Wolleben}, M.}, \bibinfo{author}{{Landecker}, T.~L.},
  \bibinfo{author}{{Reich}, W.} \& \bibinfo{author}{{Wielebinski}, R.}
\newblock \bibinfo{title}{{An absolutely calibrated survey of polarized
  emission from the northern sky at 1.4 GHz. Observations and data reduction}}.
\newblock \emph{\bibinfo{journal}{Astron. Astrophys.}}
  \textbf{\bibinfo{volume}{448}}, \bibinfo{pages}{411--424}
  (\bibinfo{year}{2006}).

\bibitem{DataQUIJOTE}
\bibinfo{author}{{Rubi{\~n}o-Mart{\'\i}n}, J.~A.} \emph{et~al.}
\newblock \bibinfo{title}{{QUIJOTE scientific results - IV. A northern sky
  survey in intensity and polarization at 10-20 GHz with the multifrequency
  instrument}}.
\newblock \emph{\bibinfo{journal}{Mon. Not. R. Astron. Soc.}}
  \textbf{\bibinfo{volume}{519}}, \bibinfo{pages}{3383--3431}
  (\bibinfo{year}{2023}).

\bibitem{WMAP14synchro}
\bibinfo{author}{{Fuskeland}, U.}, \bibinfo{author}{{Wehus}, I.~K.},
  \bibinfo{author}{{Eriksen}, H.~K.} \& \bibinfo{author}{{N{\ae}ss}, S.~K.}
\newblock \bibinfo{title}{{Spatial Variations in the Spectral Index of
  Polarized Synchrotron Emission in the 9 yr WMAP Sky Maps}}.
\newblock \emph{\bibinfo{journal}{Astrophys. J.}}
  \textbf{\bibinfo{volume}{790}}, \bibinfo{pages}{104} (\bibinfo{year}{2014}).

\bibitem{DataPLANCKCommander}
\bibinfo{author}{{Planck Collaboration}} \emph{et~al.}
\newblock \bibinfo{title}{{Planck 2018 results. IV. Diffuse component
  separation}}.
\newblock \emph{\bibinfo{journal}{Astron. Astrophys.}}
  \textbf{\bibinfo{volume}{641}}, \bibinfo{pages}{A4} (\bibinfo{year}{2020}).

\bibitem{Fermi21DataInstrument}
\bibinfo{author}{{Ajello}, M.} \emph{et~al.}
\newblock \bibinfo{title}{{Fermi Large Area Telescope Performance after 10
  Years of Operation}}.
\newblock \emph{\bibinfo{journal}{Astrophys. J. S.}}
  \textbf{\bibinfo{volume}{256}}, \bibinfo{pages}{12} (\bibinfo{year}{2021}).

\bibitem{kataoka13}
\bibinfo{author}{{Kataoka}, J.} \emph{et~al.}
\newblock \bibinfo{title}{{Suzaku Observations of the Diffuse X-Ray Emission
  across the Fermi Bubbles' Edges}}.
\newblock \emph{\bibinfo{journal}{Astrophys. J.}}
  \textbf{\bibinfo{volume}{779}}, \bibinfo{pages}{57} (\bibinfo{year}{2013}).

\bibitem{Locatelli24neDensityMW}
\bibinfo{author}{{Locatelli}, N.} \emph{et~al.}
\newblock \bibinfo{title}{{The warm-hot circumgalactic medium of the Milky Way
  as seen by eROSITA}}.
\newblock \emph{\bibinfo{journal}{Astron. Astrophys.}}
  \textbf{\bibinfo{volume}{681}}, \bibinfo{pages}{A78} (\bibinfo{year}{2024}).

\bibitem{Elia22}
\bibinfo{author}{{Elia}, D.} \emph{et~al.}
\newblock \bibinfo{title}{{The Star Formation Rate of the Milky Way as Seen by
  Herschel}}.
\newblock \emph{\bibinfo{journal}{Astrophys. J.}}
  \textbf{\bibinfo{volume}{941}}, \bibinfo{pages}{162} (\bibinfo{year}{2022}).

\bibitem{Chevalier85SF}
\bibinfo{author}{{Chevalier}, R.~A.} \& \bibinfo{author}{{Clegg}, A.~W.}
\newblock \bibinfo{title}{{Wind from a starburst galaxy nucleus}}.
\newblock \emph{\bibinfo{journal}{Nature}} \textbf{\bibinfo{volume}{317}},
  \bibinfo{pages}{44--45} (\bibinfo{year}{1985}).

\bibitem{Heckman00}
\bibinfo{author}{{Heckman}, T.~M.}, \bibinfo{author}{{Lehnert}, M.~D.},
  \bibinfo{author}{{Strickland}, D.~K.} \& \bibinfo{author}{{Armus}, L.}
\newblock \bibinfo{title}{{Absorption-Line Probes of Gas and Dust in Galactic
  Superwinds}}.
\newblock \emph{\bibinfo{journal}{Astrophys. J., Suppl. Ser.}}
  \textbf{\bibinfo{volume}{129}}, \bibinfo{pages}{493--516}
  (\bibinfo{year}{2000}).

\bibitem{Veilleux05}
\bibinfo{author}{{Veilleux}, S.}, \bibinfo{author}{{Cecil}, G.} \&
  \bibinfo{author}{{Bland-Hawthorn}, J.}
\newblock \bibinfo{title}{{Galactic Winds}}.
\newblock \emph{\bibinfo{journal}{Annu. Rev. Astron. Astrophys.}}
  \textbf{\bibinfo{volume}{43}}, \bibinfo{pages}{769--826}
  (\bibinfo{year}{2005}).

\bibitem{Strickland09}
\bibinfo{author}{{Strickland}, D.~K.} \& \bibinfo{author}{{Heckman}, T.~M.}
\newblock \bibinfo{title}{{Supernova Feedback Efficiency and Mass Loading in
  the Starburst and Galactic Superwind Exemplar M82}}.
\newblock \emph{\bibinfo{journal}{Astrophys. J.}}
  \textbf{\bibinfo{volume}{697}}, \bibinfo{pages}{2030--2056}
  (\bibinfo{year}{2009}).

\bibitem{Vink14_shockMA}
\bibinfo{author}{{Vink}, J.} \& \bibinfo{author}{{Yamazaki}, R.}
\newblock \bibinfo{title}{{A Critical Shock Mach Number for Particle
  Acceleration in the Absence of Pre-existing Cosmic Rays: $M=\sqrt5$}}.
\newblock \emph{\bibinfo{journal}{Astrophys. J.}}
  \textbf{\bibinfo{volume}{780}}, \bibinfo{pages}{125} (\bibinfo{year}{2014}).

\bibitem{Guo18_shockMA}
\bibinfo{author}{{Guo}, X.}, \bibinfo{author}{{Sironi}, L.} \&
  \bibinfo{author}{{Narayan}, R.}
\newblock \bibinfo{title}{{Electron Heating in Low Mach Number Perpendicular
  Shocks. II. Dependence on the Pre-shock Conditions}}.
\newblock \emph{\bibinfo{journal}{Astrophys. J.}}
  \textbf{\bibinfo{volume}{858}}, \bibinfo{pages}{95} (\bibinfo{year}{2018}).

\bibitem{Nguyen22SFringHDsimu}
\bibinfo{author}{{Nguyen}, D.~D.} \& \bibinfo{author}{{Thompson}, T.~A.}
\newblock \bibinfo{title}{{Galactic Winds and Bubbles from Nuclear Starburst
  Rings}}.
\newblock \emph{\bibinfo{journal}{Astrophys. J. Lett.}}
  \textbf{\bibinfo{volume}{935}}, \bibinfo{pages}{L24} (\bibinfo{year}{2022}).

\bibitem{MarascoFraternali11}
\bibinfo{author}{{Marasco}, A.} \& \bibinfo{author}{{Fraternali}, F.}
\newblock \bibinfo{title}{{Modelling the H I halo of the Milky Way}}.
\newblock \emph{\bibinfo{journal}{Astron. Astrophys.}}
  \textbf{\bibinfo{volume}{525}}, \bibinfo{pages}{A134} (\bibinfo{year}{2011}).

\bibitem{Faerman17}
\bibinfo{author}{{Faerman}, Y.}, \bibinfo{author}{{Sternberg}, A.} \&
  \bibinfo{author}{{McKee}, C.~F.}
\newblock \bibinfo{title}{{Massive Warm/Hot Galaxy Coronae as Probed by
  UV/X-Ray Oxygen Absorption and Emission. I. Basic Model}}.
\newblock \emph{\bibinfo{journal}{Astrophys. J.}}
  \textbf{\bibinfo{volume}{835}}, \bibinfo{pages}{52} (\bibinfo{year}{2017}).

\bibitem{Sancisi2001}
\bibinfo{author}{{Sancisi}, R.}, \bibinfo{author}{{Fraternali}, F.},
  \bibinfo{author}{{Oosterloo}, T.} \& \bibinfo{author}{{van Moorsel}, G.}
\newblock \bibinfo{editor}{{Funes}, J.~G.} \& \bibinfo{editor}{{Corsini},
  E.~M.} (eds) \emph{\bibinfo{title}{{The Vertical Structure and Kinematics of
  HI in Spiral Galaxies}}}.
\newblock (eds \bibinfo{editor}{{Funes}, J.~G.} \& \bibinfo{editor}{{Corsini},
  E.~M.}) \emph{\bibinfo{booktitle}{Galaxy Disks and Disk Galaxies}}, Vol.
  \bibinfo{volume}{230} of \emph{\bibinfo{series}{Astronomical Society of the
  Pacific Conference Series}}, \bibinfo{pages}{111--118}
  (\bibinfo{year}{2001}).
\newblock \eprint{astro-ph/0010407}.

\bibitem{Marasco13}
\bibinfo{author}{{Marasco}, A.}, \bibinfo{author}{{Marinacci}, F.} \&
  \bibinfo{author}{{Fraternali}, F.}
\newblock \bibinfo{title}{{On the origin of the warm-hot absorbers in the Milky
  Way's halo}}.
\newblock \emph{\bibinfo{journal}{Mon. Not. R. Astron. Soc.}}
  \textbf{\bibinfo{volume}{433}}, \bibinfo{pages}{1634--1647}
  (\bibinfo{year}{2013}).

\bibitem{Meliani24BhaloSimulation}
\bibinfo{author}{{Meliani}, Z.} \emph{et~al.}
\newblock \bibinfo{title}{{The galactic bubbles of starburst galaxies. The
  influence of galactic large-scale magnetic fields}}.
\newblock \emph{\bibinfo{journal}{Astron. Astrophys.}}
  \textbf{\bibinfo{volume}{683}}, \bibinfo{pages}{A178} (\bibinfo{year}{2024}).

\bibitem{Sofue16mn}
\bibinfo{author}{{Sofue}, Y.} \emph{et~al.}
\newblock \bibinfo{title}{{Galactic Centre hypershell model for the North Polar
  Spurs}}.
\newblock \emph{\bibinfo{journal}{Mon. Not. R. Astron. Soc.}}
  \textbf{\bibinfo{volume}{459}}, \bibinfo{pages}{108--120}
  (\bibinfo{year}{2016}).

\bibitem{Sofue21}
\bibinfo{author}{{Sofue}, Y.} \& \bibinfo{author}{{Kataoka}, J.}
\newblock \bibinfo{title}{{Interaction of the galactic-centre super bubbles
  with the gaseous disc}}.
\newblock \emph{\bibinfo{journal}{Mon. Not. R. Astron. Soc.}}
  \textbf{\bibinfo{volume}{506}}, \bibinfo{pages}{2170--2180}
  (\bibinfo{year}{2021}).

\bibitem{Yang2022NatAs}
\bibinfo{author}{{Yang}, H. Y.~K.}, \bibinfo{author}{{Ruszkowski}, M.} \&
  \bibinfo{author}{{Zweibel}, E.~G.}
\newblock \bibinfo{title}{{Fermi and eROSITA bubbles as relics of the past
  activity of the Galaxy's central black hole}}.
\newblock \emph{\bibinfo{journal}{Nat. Astron.}} \textbf{\bibinfo{volume}{6}},
  \bibinfo{pages}{584--591} (\bibinfo{year}{2022}).

\bibitem{Mou23NC}
\bibinfo{author}{{Mou}, G.} \emph{et~al.}
\newblock \bibinfo{title}{{Asymmetric eROSITA bubbles as the evidence of a
  circumgalactic medium wind}}.
\newblock \emph{\bibinfo{journal}{Nat. Commun.}} \textbf{\bibinfo{volume}{14}},
  \bibinfo{pages}{781} (\bibinfo{year}{2023}).

\bibitem{Sarkar19}
\bibinfo{author}{{Sarkar}, K.~C.}
\newblock \bibinfo{title}{{Possible connection between the asymmetry of the
  North Polar Spur and Loop I and Fermi bubbles}}.
\newblock \emph{\bibinfo{journal}{Mon. Not. R. Astron. Soc.}}
  \textbf{\bibinfo{volume}{482}}, \bibinfo{pages}{4813--4823}
  (\bibinfo{year}{2019}).

\bibitem{DataROSAT}
\bibinfo{author}{{Snowden}, S.~L.} \emph{et~al.}
\newblock \bibinfo{title}{{ROSAT Survey Diffuse X-Ray Background Maps. II.}}
\newblock \emph{\bibinfo{journal}{Astrophys. J.}}
  \textbf{\bibinfo{volume}{485}}, \bibinfo{pages}{125--135}
  (\bibinfo{year}{1997}).

\bibitem{Maconi23DustSimulation}
\bibinfo{author}{{Maconi}, E.} \emph{et~al.}
\newblock \bibinfo{title}{{Modelling Local Bubble analogs: synthetic dust
  polarization maps}}.
\newblock \emph{\bibinfo{journal}{Mon. Not. R. Astron. Soc.}}
  \textbf{\bibinfo{volume}{523}}, \bibinfo{pages}{5995--6010}
  (\bibinfo{year}{2023}).

\bibitem{LocalismFrisch11araa}
\bibinfo{author}{{Frisch}, P.~C.}, \bibinfo{author}{{Redfield}, S.} \&
  \bibinfo{author}{{Slavin}, J.~D.}
\newblock \bibinfo{title}{{The Interstellar Medium Surrounding the Sun}}.
\newblock \emph{\bibinfo{journal}{Annu. Rev. Astron. Astrophys.}}
  \textbf{\bibinfo{volume}{49}}, \bibinfo{pages}{237--279}
  (\bibinfo{year}{2011}).

\bibitem{Yeung23LHB}
\bibinfo{author}{{Yeung}, M.~C.~H.} \emph{et~al.}
\newblock \bibinfo{title}{{SRG/eROSITA X-ray shadowing study of giant molecular
  clouds}}.
\newblock \emph{\bibinfo{journal}{Astron. Astrophys.}}
  \textbf{\bibinfo{volume}{676}}, \bibinfo{pages}{A3} (\bibinfo{year}{2023}).

\bibitem{Beck15}
\bibinfo{author}{{Beck}, R.}
\newblock \bibinfo{title}{{Magnetic fields in spiral galaxies}}.
\newblock \emph{\bibinfo{journal}{Astron. Astrophys. Rev.}}
  \textbf{\bibinfo{volume}{24}}, \bibinfo{pages}{4} (\bibinfo{year}{2015}).

\bibitem{WMAPdata}
\bibinfo{author}{{Bennett}, C.~L.} \emph{et~al.}
\newblock \bibinfo{title}{{Nine-year Wilkinson Microwave Anisotropy Probe
  (WMAP) Observations: Final Maps and Results}}.
\newblock \emph{\bibinfo{journal}{Astrophys. J., Suppl. Ser.}}
  \textbf{\bibinfo{volume}{208}}, \bibinfo{pages}{20} (\bibinfo{year}{2013}).

\bibitem{Ehle93}
\bibinfo{author}{{Ehle}, M.} \& \bibinfo{author}{{Beck}, R.}
\newblock \bibinfo{title}{{Ionized gas and intrinsic magnetic fields in the
  spiral galaxy NGC 6946.}}
\newblock \emph{\bibinfo{journal}{Astron. Astrophys.}}
  \textbf{\bibinfo{volume}{273}}, \bibinfo{pages}{45--64}
  (\bibinfo{year}{1993}).

\bibitem{Armstrong95}
\bibinfo{author}{{Armstrong}, J.~W.}, \bibinfo{author}{{Rickett}, B.~J.} \&
  \bibinfo{author}{{Spangler}, S.~R.}
\newblock \bibinfo{title}{{Electron Density Power Spectrum in the Local
  Interstellar Medium}}.
\newblock \emph{\bibinfo{journal}{Astrophys. J.}}
  \textbf{\bibinfo{volume}{443}}, \bibinfo{pages}{209} (\bibinfo{year}{1995}).

\bibitem{Chep2010}
\bibinfo{author}{{Chepurnov}, A.} \& \bibinfo{author}{{Lazarian}, A.}
\newblock \bibinfo{title}{{Extending the Big Power Law in the Sky with
  Turbulence Spectra from Wisconsin H{$\alpha$} Mapper Data}}.
\newblock \emph{\bibinfo{journal}{Astrophys. J.}}
  \textbf{\bibinfo{volume}{710}}, \bibinfo{pages}{853--858}
  (\bibinfo{year}{2010}).

\bibitem{ymw16ne}
\bibinfo{author}{{Yao}, J.~M.}, \bibinfo{author}{{Manchester}, R.~N.} \&
  \bibinfo{author}{{Wang}, N.}
\newblock \bibinfo{title}{{A New Electron-density Model for Estimation of
  Pulsar and FRB Distances}}.
\newblock \emph{\bibinfo{journal}{Astrophys. J.}}
  \textbf{\bibinfo{volume}{835}}, \bibinfo{pages}{29} (\bibinfo{year}{2017}).

\bibitem{JF12b_turbB}
\bibinfo{author}{{Jansson}, R.} \& \bibinfo{author}{{Farrar}, G.~R.}
\newblock \bibinfo{title}{{The Galactic Magnetic Field}}.
\newblock \emph{\bibinfo{journal}{Astrophys. J. Lett.}}
  \textbf{\bibinfo{volume}{761}}, \bibinfo{pages}{L11} (\bibinfo{year}{2012}).

\bibitem{PLANCK16XLIIBfield}
\bibinfo{author}{{Planck Collaboration}} \emph{et~al.}
\newblock \bibinfo{title}{{Planck intermediate results. XLII. Large-scale
  Galactic magnetic fields}}.
\newblock \emph{\bibinfo{journal}{Astron. Astrophys.}}
  \textbf{\bibinfo{volume}{596}}, \bibinfo{pages}{A103} (\bibinfo{year}{2016}).

\bibitem{DataWMAP2}
\bibinfo{author}{{Bennett}, C.~L.} \emph{et~al.}
\newblock \bibinfo{title}{{Nine-year Wilkinson Microwave Anisotropy Probe
  (WMAP) Observations: Final Maps and Results}}.
\newblock \emph{\bibinfo{journal}{Astrophys. J., Suppl. Ser.}}
  \textbf{\bibinfo{volume}{208}}, \bibinfo{pages}{20} (\bibinfo{year}{2013}).

\bibitem{Carretti19MN}
\bibinfo{author}{{Carretti}, E.} \emph{et~al.}
\newblock \bibinfo{title}{{S-band Polarization All-Sky Survey (S-PASS): survey
  description and maps}}.
\newblock \emph{\bibinfo{journal}{Mon. Not. R. Astron. Soc.}}
  \textbf{\bibinfo{volume}{489}}, \bibinfo{pages}{2330--2354}
  (\bibinfo{year}{2019}).

\bibitem{DataDRAO2}
\bibinfo{author}{{Testori}, J.~C.}, \bibinfo{author}{{Reich}, P.} \&
  \bibinfo{author}{{Reich}, W.}
\newblock \bibinfo{title}{{A fully sampled {\ensuremath{\lambda}}21 cm linear
  polarization survey of the southern sky}}.
\newblock \emph{\bibinfo{journal}{Astron. Astrophys.}}
  \textbf{\bibinfo{volume}{484}}, \bibinfo{pages}{733--742}
  (\bibinfo{year}{2008}).

\bibitem{Iacobelli13LOFAR}
\bibinfo{author}{{Iacobelli}, M.} \emph{et~al.}
\newblock \bibinfo{title}{{Studying Galactic interstellar turbulence through
  fluctuations in synchrotron emission. First LOFAR Galactic foreground
  detection}}.
\newblock \emph{\bibinfo{journal}{Astron. Astrophys.}}
  \textbf{\bibinfo{volume}{558}}, \bibinfo{pages}{A72} (\bibinfo{year}{2013}).

\bibitem{HAWC2017}
\bibinfo{author}{{Abeysekara}, A.~U.} \emph{et~al.}
\newblock \bibinfo{title}{{Extended gamma-ray sources around pulsars constrain
  the origin of the positron flux at Earth}}.
\newblock \emph{\bibinfo{journal}{Science}} \textbf{\bibinfo{volume}{358}},
  \bibinfo{pages}{911--914} (\bibinfo{year}{2017}).

\bibitem{FermiCatalog4FGLDR3}
\bibinfo{author}{{Abdollahi}, S.} \emph{et~al.}
\newblock \bibinfo{title}{{Incremental Fermi Large Area Telescope Fourth Source
  Catalog}}.
\newblock \emph{\bibinfo{journal}{Astrophys. J. S.}}
  \textbf{\bibinfo{volume}{260}}, \bibinfo{pages}{53} (\bibinfo{year}{2022}).

\bibitem{Fermipy_Wood17}
\emph{\bibinfo{title}{{Fermipy: An open-source Python package for analysis of
  Fermi-LAT Data}}}, Vol. \bibinfo{volume}{301} of
  \emph{\bibinfo{series}{International Cosmic Ray Conference}}.
\newblock \eprint{1707.09551}.

\bibitem{Popescu17ISRF}
\bibinfo{author}{{Popescu}, C.~C.} \emph{et~al.}
\newblock \bibinfo{title}{{A radiation transfer model for the Milky Way: I.
  Radiation fields and application to high-energy astrophysics}}.
\newblock \emph{\bibinfo{journal}{Mon. Not. R. Astron. Soc.}}
  \textbf{\bibinfo{volume}{470}}, \bibinfo{pages}{2539--2558}
  (\bibinfo{year}{2017}).

\bibitem{naima}
\bibinfo{author}{{Zabalza}, V.}
\newblock \bibinfo{title}{naima: a python package for inference of relativistic
  particle energy distributions from observed nonthermal spectra}.
\newblock \emph{\bibinfo{journal}{Proc.~of International Cosmic Ray Conference
  2015}} \bibinfo{pages}{922} (\bibinfo{year}{2015}).

\bibitem{Aharonian10syn}
\bibinfo{author}{{Aharonian}, F.~A.}, \bibinfo{author}{{Kelner}, S.~R.} \&
  \bibinfo{author}{{Prosekin}, A.~Y.}
\newblock \bibinfo{title}{{Angular, spectral, and time distributions of highest
  energy protons and associated secondary gamma rays and neutrinos propagating
  through extragalactic magnetic and radiation fields}}.
\newblock \emph{\bibinfo{journal}{Phys. Rev. D}} \textbf{\bibinfo{volume}{82}},
  \bibinfo{pages}{043002} (\bibinfo{year}{2010}).

\bibitem{Khangulyan14IC}
\bibinfo{author}{{Khangulyan}, D.}, \bibinfo{author}{{Aharonian}, F.~A.} \&
  \bibinfo{author}{{Kelner}, S.~R.}
\newblock \bibinfo{title}{{Simple Analytical Approximations for Treatment of
  Inverse Compton Scattering of Relativistic Electrons in the Blackbody
  Radiation Field}}.
\newblock \emph{\bibinfo{journal}{Astrophys. J.}}
  \textbf{\bibinfo{volume}{783}}, \bibinfo{pages}{100} (\bibinfo{year}{2014}).

\bibitem{Zirakashvili07AAexpcutoff}
\bibinfo{author}{{Zirakashvili}, V.~N.} \& \bibinfo{author}{{Aharonian}, F.}
\newblock \bibinfo{title}{{Analytical solutions for energy spectra of electrons
  accelerated by nonrelativistic shock-waves in shell type supernova
  remnants}}.
\newblock \emph{\bibinfo{journal}{Astron. Astrophys.}}
  \textbf{\bibinfo{volume}{465}}, \bibinfo{pages}{695--702}
  (\bibinfo{year}{2007}).

\bibitem{Fermi_Instrument12}
\bibinfo{author}{{Ackermann}, M.} \emph{et~al.}
\newblock \bibinfo{title}{{In-flight measurement of the absolute energy scale
  of the Fermi Large Area Telescope}}.
\newblock \emph{\bibinfo{journal}{Astroparticle Physics}}
  \textbf{\bibinfo{volume}{35}}, \bibinfo{pages}{346--353}
  (\bibinfo{year}{2012}).

\bibitem{Blumenthal70}
\bibinfo{author}{{Blumenthal}, G.~R.} \& \bibinfo{author}{{Gould}, R.~J.}
\newblock \bibinfo{title}{{Bremsstrahlung, Synchrotron Radiation, and Compton
  Scattering of High-Energy Electrons Traversing Dilute Gases}}.
\newblock \emph{\bibinfo{journal}{Reviews of Modern Physics}}
  \textbf{\bibinfo{volume}{42}}, \bibinfo{pages}{237--271}
  (\bibinfo{year}{1970}).

\bibitem{Heesen16}
\bibinfo{author}{{Heesen}, V.}, \bibinfo{author}{{Dettmar}, R.-J.},
  \bibinfo{author}{{Krause}, M.}, \bibinfo{author}{{Beck}, R.} \&
  \bibinfo{author}{{Stein}, Y.}
\newblock \bibinfo{title}{{Advective and diffusive cosmic ray transport in
  galactic haloes}}.
\newblock \emph{\bibinfo{journal}{Mon. Not. R. Astron. Soc.}}
  \textbf{\bibinfo{volume}{458}}, \bibinfo{pages}{332--353}
  (\bibinfo{year}{2016}).

\bibitem{Heesen21halorev}
\bibinfo{author}{{Heesen}, V.}
\newblock \bibinfo{title}{{The radio continuum perspective on cosmic-ray
  transport in external galaxies}}.
\newblock \emph{\bibinfo{journal}{Astrophy. S. S.}}
  \textbf{\bibinfo{volume}{366}}, \bibinfo{pages}{117} (\bibinfo{year}{2021}).

\bibitem{SNRrate1Reed05}
\bibinfo{author}{{Reed}, B.~C.}
\newblock \bibinfo{title}{{New Estimates of the Solar-Neighborhood Massive Star
  Birthrate and the Galactic Supernova Rate}}.
\newblock \emph{\bibinfo{journal}{Astron. J.}} \textbf{\bibinfo{volume}{130}},
  \bibinfo{pages}{1652--1657} (\bibinfo{year}{2005}).

\bibitem{SNRrate2Diehl06}
\bibinfo{author}{{Diehl}, R.} \emph{et~al.}
\newblock \bibinfo{title}{{Radioactive $^{26}$Al from massive stars in the
  Galaxy}}.
\newblock \emph{\bibinfo{journal}{Nature}} \textbf{\bibinfo{volume}{439}},
  \bibinfo{pages}{45--47} (\bibinfo{year}{2006}).

\bibitem{SNRrate3Li11}
\bibinfo{author}{{Li}, W.} \emph{et~al.}
\newblock \bibinfo{title}{{Nearby supernova rates from the Lick Observatory
  Supernova Search - III. The rate-size relation, and the rates as a function
  of galaxy Hubble type and colour}}.
\newblock \emph{\bibinfo{journal}{Mon. Not. R. Astron. Soc.}}
  \textbf{\bibinfo{volume}{412}}, \bibinfo{pages}{1473--1507}
  (\bibinfo{year}{2011}).

\bibitem{SNRrate4Rozwadowska21}
\bibinfo{author}{{Rozwadowska}, K.}, \bibinfo{author}{{Vissani}, F.} \&
  \bibinfo{author}{{Cappellaro}, E.}
\newblock \bibinfo{title}{{On the rate of core collapse supernovae in the milky
  way}}.
\newblock \emph{\bibinfo{journal}{New Astro.}} \textbf{\bibinfo{volume}{83}},
  \bibinfo{pages}{101498} (\bibinfo{year}{2021}).

\bibitem{SNRrate5Adams13}
\bibinfo{author}{{Adams}, S.~M.}, \bibinfo{author}{{Kochanek}, C.~S.},
  \bibinfo{author}{{Beacom}, J.~F.}, \bibinfo{author}{{Vagins}, M.~R.} \&
  \bibinfo{author}{{Stanek}, K.~Z.}
\newblock \bibinfo{title}{{Observing the Next Galactic Supernova}}.
\newblock \emph{\bibinfo{journal}{Astrophys. J.}}
  \textbf{\bibinfo{volume}{778}}, \bibinfo{pages}{164} (\bibinfo{year}{2013}).

\bibitem{Poznanski13SNRenergy}
\bibinfo{author}{{Poznanski}, D.}
\newblock \bibinfo{title}{{An emerging coherent picture of red supergiant
  supernova explosions}}.
\newblock \emph{\bibinfo{journal}{Mon. Not. R. Astron. Soc.}}
  \textbf{\bibinfo{volume}{436}}, \bibinfo{pages}{3224--3230}
  (\bibinfo{year}{2013}).

\bibitem{Inoue15}
\bibinfo{author}{{Inoue}, Y.} \emph{et~al.}
\newblock \bibinfo{title}{{Metal enrichment in the Fermi bubbles as a probe of
  their origin}}.
\newblock \emph{\bibinfo{journal}{Publ. Astron. Soc. Jpn.}}
  \textbf{\bibinfo{volume}{67}}, \bibinfo{pages}{56} (\bibinfo{year}{2015}).

\bibitem{LaRocca20}
\bibinfo{author}{{LaRocca}, D.~M.} \emph{et~al.}
\newblock \bibinfo{title}{{An Analysis of the North Polar Spur Using HaloSat}}.
\newblock \emph{\bibinfo{journal}{Astrophys. J.}}
  \textbf{\bibinfo{volume}{904}}, \bibinfo{pages}{54} (\bibinfo{year}{2020}).

\bibitem{Gupta23NA_SFerass}
\bibinfo{author}{{Gupta}, A.}, \bibinfo{author}{{Mathur}, S.},
  \bibinfo{author}{{Kingsbury}, J.}, \bibinfo{author}{{Das}, S.} \&
  \bibinfo{author}{{Krongold}, Y.}
\newblock \bibinfo{title}{{Thermal and chemical properties of the eROSITA
  bubbles from Suzaku observations}}.
\newblock \emph{\bibinfo{journal}{Nat. Astron.}}  (\bibinfo{year}{2023}).

\bibitem{FoxRichter19massoutflow}
\bibinfo{author}{{Fox}, A.~J.} \emph{et~al.}
\newblock \bibinfo{title}{{The Mass Inflow and Outflow Rates of the Milky
  Way}}.
\newblock \emph{\bibinfo{journal}{Astrophys. J.}}
  \textbf{\bibinfo{volume}{884}}, \bibinfo{pages}{53} (\bibinfo{year}{2019}).

\bibitem{Haslam74}
\bibinfo{author}{{Haslam}, C.~G.~T.}
\newblock \bibinfo{title}{{NOD2 A General System of Analysis for
  Radioastronomy}}.
\newblock \emph{\bibinfo{journal}{Astron. Astrophys. Suppl.}}
  \textbf{\bibinfo{volume}{15}}, \bibinfo{pages}{333} (\bibinfo{year}{1974}).

\bibitem{Haslam82}
\bibinfo{author}{{Haslam}, C.~G.~T.}, \bibinfo{author}{{Salter}, C.~J.},
  \bibinfo{author}{{Stoffel}, H.} \& \bibinfo{author}{{Wilson}, W.~E.}
\newblock \bibinfo{title}{{A 408 MHz all-sky continuum survey. II. The atlas of
  contour maps.}}
\newblock \emph{\bibinfo{journal}{Astronomy and Astrophysics, Suppl. Ser.}}
  \textbf{\bibinfo{volume}{47}}, \bibinfo{pages}{1--143}
  (\bibinfo{year}{1982}).

\bibitem{SPASS_Carretti19}
\bibinfo{author}{{Carretti}, E.} \emph{et~al.}
\newblock \bibinfo{title}{{S-band Polarization All-Sky Survey (S-PASS): survey
  description and maps}}.
\newblock \emph{\bibinfo{journal}{Mon. Not. R. Astron. Soc.}}
  \textbf{\bibinfo{volume}{489}}, \bibinfo{pages}{2330--2354}
  (\bibinfo{year}{2019}).

\bibitem{PLANCK18diffuseSepa}
\bibinfo{author}{{Planck Collaboration}} \emph{et~al.}
\newblock \bibinfo{title}{{Planck 2018 results. IV. Diffuse component
  separation}}.
\newblock \emph{\bibinfo{journal}{Astron. Astrophys.}}
  \textbf{\bibinfo{volume}{641}}, \bibinfo{pages}{A4} (\bibinfo{year}{2020}).

\bibitem{PLANCK18dustPol}
\bibinfo{author}{{Planck Collaboration}} \emph{et~al.}
\newblock \bibinfo{title}{{Planck 2018 results. XI. Polarized dust
  foregrounds}}.
\newblock \emph{\bibinfo{journal}{Astron. Astrophys.}}
  \textbf{\bibinfo{volume}{641}}, \bibinfo{pages}{A11} (\bibinfo{year}{2020}).

\bibitem{Python}
\bibinfo{author}{Van~Rossum, G.} \& \bibinfo{author}{Drake, F.~L.}
\newblock \emph{\bibinfo{title}{Python 3 Reference Manual}}
  (\bibinfo{publisher}{CreateSpace}, \bibinfo{address}{Scotts Valley, CA},
  \bibinfo{year}{2009}).

\bibitem{numpy1}
\bibinfo{author}{{van der Walt}, S.}, \bibinfo{author}{{Colbert}, S.~C.} \&
  \bibinfo{author}{{Varoquaux}, G.}
\newblock \bibinfo{title}{{The NumPy Array: A Structure for Efficient Numerical
  Computation}}.
\newblock \emph{\bibinfo{journal}{Computing in Science and Engineering}}
  \textbf{\bibinfo{volume}{13}}, \bibinfo{pages}{22--30}
  (\bibinfo{year}{2011}).

\bibitem{numpy2}
\bibinfo{author}{Harris, C.~R.} \emph{et~al.}
\newblock \bibinfo{title}{Array programming with {NumPy}}.
\newblock \emph{\bibinfo{journal}{Nature}} \textbf{\bibinfo{volume}{585}},
  \bibinfo{pages}{357--362} (\bibinfo{year}{2020}).
\newblock \urlprefix\url{https://doi.org/10.1038/s41586-020-2649-2}.

\bibitem{healpy1}
\bibinfo{author}{{G{\'o}rski}, K.~M.} \emph{et~al.}
\newblock \bibinfo{title}{{HEALPix: A Framework for High-Resolution
  Discretization and Fast Analysis of Data Distributed on the Sphere}}.
\newblock \emph{\bibinfo{journal}{Astrophys. J.}}
  \textbf{\bibinfo{volume}{622}}, \bibinfo{pages}{759--771}
  (\bibinfo{year}{2005}).

\bibitem{healpy2}
\bibinfo{author}{Zonca, A.} \emph{et~al.}
\newblock \bibinfo{title}{healpy: equal area pixelization and spherical
  harmonics transforms for data on the sphere in python}.
\newblock \emph{\bibinfo{journal}{Journal of Open Source Software}}
  \textbf{\bibinfo{volume}{4}}, \bibinfo{pages}{1298} (\bibinfo{year}{2019}).
\newblock \urlprefix\url{https://doi.org/10.21105/joss.01298}.

\bibitem{Astropy13}
\bibinfo{author}{{Astropy Collaboration}} \emph{et~al.}
\newblock \bibinfo{title}{{Astropy: A community Python package for astronomy}}.
\newblock \emph{\bibinfo{journal}{Astron. Astrophys.}}
  \textbf{\bibinfo{volume}{558}}, \bibinfo{pages}{A33} (\bibinfo{year}{2013}).

\bibitem{Astropy18}
\bibinfo{author}{{Astropy Collaboration}} \emph{et~al.}
\newblock \bibinfo{title}{{The Astropy Project: Building an Open-science
  Project and Status of the v2.0 Core Package}}.
\newblock \emph{\bibinfo{journal}{Astron. J.}} \textbf{\bibinfo{volume}{156}},
  \bibinfo{pages}{123} (\bibinfo{year}{2018}).

\bibitem{Astropy22}
\bibinfo{author}{{Astropy Collaboration}} \emph{et~al.}
\newblock \bibinfo{title}{{The Astropy Project: Sustaining and Growing a
  Community-oriented Open-source Project and the Latest Major Release (v5.0) of
  the Core Package}}.
\newblock \emph{\bibinfo{journal}{Astrophys. J.}}
  \textbf{\bibinfo{volume}{935}}, \bibinfo{pages}{167} (\bibinfo{year}{2022}).

\bibitem{CRPropa}
\bibinfo{author}{{Alves Batista}, R.} \emph{et~al.}
\newblock \bibinfo{title}{{CRPropa 3.2 - an advanced framework for high-energy
  particle propagation in extragalactic and galactic spaces}}.
\newblock \emph{\bibinfo{journal}{JCAP}} \textbf{\bibinfo{volume}{2022}},
  \bibinfo{pages}{035} (\bibinfo{year}{2022}).

\bibitem{Jupyter}
\bibinfo{author}{Kluyver, T.} \emph{et~al.}
\newblock \bibinfo{editor}{Loizides, F.} \& \bibinfo{editor}{Scmidt, B.} (eds)
  \emph{\bibinfo{title}{Jupyter notebooks ? a publishing format for
  reproducible computational workflows}}.
\newblock (eds \bibinfo{editor}{Loizides, F.} \& \bibinfo{editor}{Scmidt, B.})
  \emph{\bibinfo{booktitle}{Positioning and Power in Academic Publishing:
  Players, Agents and Agendas}}, \bibinfo{pages}{87--90}
  (\bibinfo{publisher}{IOS Press}, \bibinfo{year}{2016}).
\newblock \urlprefix\url{https://eprints.soton.ac.uk/403913/}.

\bibitem{Matplotlib}
\bibinfo{author}{{Hunter}, J.~D.}
\newblock \bibinfo{title}{{Matplotlib: A 2D Graphics Environment}}.
\newblock \emph{\bibinfo{journal}{Computing in Science and Engineering}}
  \textbf{\bibinfo{volume}{9}}, \bibinfo{pages}{90--95} (\bibinfo{year}{2007}).

\bibitem{ds9Joye03}
\bibinfo{author}{{Joye}, W.~A.} \& \bibinfo{author}{{Mandel}, E.}
\newblock \bibinfo{editor}{{Payne}, H.~E.}, \bibinfo{editor}{{Jedrzejewski},
  R.~I.} \& \bibinfo{editor}{{Hook}, R.~N.} (eds) \emph{\bibinfo{title}{{New
  Features of SAOImage DS9}}}.
\newblock (eds \bibinfo{editor}{{Payne}, H.~E.},
  \bibinfo{editor}{{Jedrzejewski}, R.~I.} \& \bibinfo{editor}{{Hook}, R.~N.})
  \emph{\bibinfo{booktitle}{Astronomical Data Analysis Software and Systems
  XII}}, Vol. \bibinfo{volume}{295} of \emph{\bibinfo{series}{Astronomical
  Society of the Pacific Conference Series}}, \bibinfo{pages}{489}
  (\bibinfo{year}{2003}).

\bibitem{DatastockertReich86}
\bibinfo{author}{{Reich}, P.} \& \bibinfo{author}{{Reich}, W.}
\newblock \bibinfo{title}{{A radio continuum survey of the northern sky at 1420
  MHz. II}}.
\newblock \emph{\bibinfo{journal}{Astron. Astrophys. Suppl.}}
  \textbf{\bibinfo{volume}{63}}, \bibinfo{pages}{205} (\bibinfo{year}{1986}).

\bibitem{DatastockertReich01}
\bibinfo{author}{{Reich}, P.}, \bibinfo{author}{{Testori}, J.~C.} \&
  \bibinfo{author}{{Reich}, W.}
\newblock \bibinfo{title}{{A radio continuum survey of the southern sky at 1420
  MHz. The atlas of contour maps}}.
\newblock \emph{\bibinfo{journal}{Astron. Astrophys. Suppl.}}
  \textbf{\bibinfo{volume}{376}}, \bibinfo{pages}{861--877}
  (\bibinfo{year}{2001}).

\bibitem{MillerBregman15}
\bibinfo{author}{{Miller}, M.~J.} \& \bibinfo{author}{{Bregman}, J.~N.}
\newblock \bibinfo{title}{{Constraining the Milky Way's Hot Gas Halo with O VII
  and O VIII Emission Lines}}.
\newblock \emph{\bibinfo{journal}{Astrophys. J.}}
  \textbf{\bibinfo{volume}{800}}, \bibinfo{pages}{14} (\bibinfo{year}{2015}).

\bibitem{Bland-Hawthorn03}
\bibinfo{author}{{Bland-Hawthorn}, J.} \& \bibinfo{author}{{Cohen}, M.}
\newblock \bibinfo{title}{{The Large-Scale Bipolar Wind in the Galactic
  Center}}.
\newblock \emph{\bibinfo{journal}{Astrophys. J.}}
  \textbf{\bibinfo{volume}{582}}, \bibinfo{pages}{246--256}
  (\bibinfo{year}{2003}).

\bibitem{Ponti19}
\bibinfo{author}{{Ponti}, G.} \emph{et~al.}
\newblock \bibinfo{title}{{An X-ray chimney extending hundreds of parsecs above
  and below the Galactic Centre}}.
\newblock \emph{\bibinfo{journal}{Nature}} \textbf{\bibinfo{volume}{567}},
  \bibinfo{pages}{347--350} (\bibinfo{year}{2019}).

\end{thebibliography}

\end{document}